\documentclass[prd,twocolumn,english,showpacs,superscriptaddress,amsmath,amssymb,floatfix,nofootinbib]{revtex4-1}
\makeatletter
\def\@fnsymbol#1{%
 \ensuremath{%
  \ifcase#1\or
   *\or                        \dagger                   \or
   \ddagger                \or \mathsection              \or
   \mathparagraph\or
   **\or                       \dagger\dagger            \or
   \ddagger\ddagger        \or \mathsection \mathsection \or
   \mathparagraph\mathparagraph\or
   *{*}*\ignorespaces      \or \dagger\dagger\dagger     \or
   \ddagger\ddagger\ddagger\or \mathsection \mathsection \mathsection \or
   \mathparagraph\mathparagraph\mathparagraph\or
  \else
   \@ctrerr
  \fi
 }%
}%
\makeatother

\usepackage{tikz}  

\usepackage{graphicx}
\usepackage[normalem]{ulem}
\usepackage[caption=false]{subfig}
\usepackage{float}
\usepackage{soul,xcolor}
\usepackage{nicefrac}
\usepackage[colorlinks = true,linkcolor = blue, urlcolor  = blue,citecolor = red,anchorcolor = blue]{hyperref}

\definecolor{darkgreen}{RGB}{50,205,50}
\newcommand{\ipmxx}{\mbox{\textsc{IPM-xFitter}}}
\newcommand{\ipmx}{\mbox{\textsc{IPMx}}}
\let\oldcite\cite
\renewcommand{\cite}[1]{\mbox{\oldcite{#1}}}  %
\newcommand{\xfitter}{\textsc{xFitter}}

\newcommand{\orcid}[1]{\,\href{https://orcid.org/#1}{\includegraphics[width=9pt]{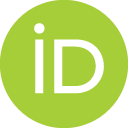}}\,}

\newcommand{\orcidHA}{0000-0002-2017-7706} %
\newcommand{\orcidMS}{0000-0002-2127-6933} %
\newcommand{\orcidHK}{0000-0001-7078-7177} %

\newcommand{\orcidSA}{0000-0001-5450-0447} %
\newcommand{\orcidFG}{0000-0002-8506-274X} %
\newcommand{\orcidAG}{0000-0002-8553-7338} %
\newcommand{\orcidAL}{0000-0001-7968-5388} %
\newcommand{\orcidFO}{0000-0001-6799-2436} %
\newcommand{\orcidOZ}{0000-0003-3783-6330} %

\begin{document}

\setstcolor{blue}
\hyphenpenalty=10000
\title {QCD analysis of pion fragmentation functions  in the \xfitter{} framework}

\def\ipm{\affiliation{School of Particles and Accelerators, Institute for Research in Fundamental     Sciences (IPM), P.O.Box 19395-5531, Tehran, Iran}}
\def\ustm{\affiliation{Department of Physics, University of Science and Technology of Mazandaran,     P.O.Box 48518-78195, Behshahr, Iran}}
\def\smu{\affiliation{Department of Physics, Southern Methodist University,
    Dallas, TX 75275-0175, U.S.A.}}
\def\desy{\affiliation{Deutsches Elektronen-Synchrotron (DESY), Notkestrasse 85, D-22607 Hamburg, Germany}}
\def\dubna{\affiliation{Joint Institute for Nuclear Research, Joliot-Curie 6, Dubna, Moscow region, Russia, 141980}}
\def\maxPlanck{\affiliation{Max-Planck-Institut f\"ur Physik, F\"ohringer Ring 6, D-80805 M\"unchen, Germany}}{}
\def\oxford{\affiliation{Particle Physics, Denys Wilkinson Bdg, Keble Road, University of Oxford, OX1 3RH Oxford, UK}}
\def\rome{\affiliation{University of Rome Tor Vergata and INFN, Sezione di Roma 2, Via della Ricerca Scientifica 1,00133 Roma, Italy}}
\def\cracow{\affiliation{T. Kosciuszko Cracow University of Technology, PL-30-084, Cracow, Poland}}
\def\hamburg{\affiliation{Hamburg University, II. Institute for Theoretical Physics, Luruper Chaussee 149, D-22761 Hamburg,Germany}}

\def\cern{\affiliation{CERN, CH-1211 Geneva 23, Switzerland}}

\author{Hamed Abdolmaleki\orcid{\orcidHA}} 
\email{Abdolmaleki@ipm.ir}
\ipm{}  

\author{Maryam Soleymaninia\orcid{\orcidMS}}
\email{Maryam\_Soleymaninia@ipm.ir}
\ipm{}

\author{Hamzeh Khanpour\orcid{\orcidHK}}  
\email{Hamzeh.Khanpour@cern.ch}
\ustm{}
\ipm{}

\author{Simone Amoroso\orcid{\orcidSA}}
\email{Simone.Amoroso@cern.ch}
\desy{}

\author{Francesco~Giuli\orcid{\orcidFG}}
\email{francesco.giuli@roma2.infn.it}
\rome{}

\author{Alexander~Glazov\orcid{\orcidAG}}
\email{alexandre.glazov@desy.de}
\desy{}

\author{Agnieszka~Luszczak\orcid{\orcidAL}}
\email{agnieszka.luszczak@desy.de}
\cracow{}

\author{Fredrick~Olness\orcid{\orcidFO}}
\email{olness@smu.edu}
\smu{}

\author{Oleksandr~Zenaiev\orcid{\orcidOZ}}
\email{oleksandr.zenaiev@cern.ch}
\cern{}

\collaboration{\xfitter{} Developers' Team:}

\date{\today}
\begin{abstract}
We present the first open-source analysis of
fragmentation functions (FFs) of charged pions (entitled \ipmxx{})
performed at 
next-to-leading order (NLO) and next-to-next-to-leading order (NNLO) 
accuracy in perturbative QCD using the \xfitter{} framework. 
This study incorporates a comprehensive
and up-to-date set of pion production data from single-inclusive 
annihilation (SIA) processes, together with the most recent measurements of inclusive 
cross-sections of single pion by the BELLE collaboration.
The determination of pion FFs along with their theoretical uncertainties is performed in 
the Zero-Mass Variable-Flavor Number Scheme (ZM-VFNS).
We also present comparisons of our FFs set with recent fits from the literature.
The resulting NLO and NNLO pion FFs provide valuable insights 
for applications in present and future high-energy analysis of pion final state processes.
\\[10pt]
\vspace*{1.5cm}
\vspace*{0.8cm}
\end{abstract}

\maketitle
\tableofcontents{}

\def\figChiPlot{
\begin{figure*}[b]
\centering
\includegraphics[width=0.60\textwidth]{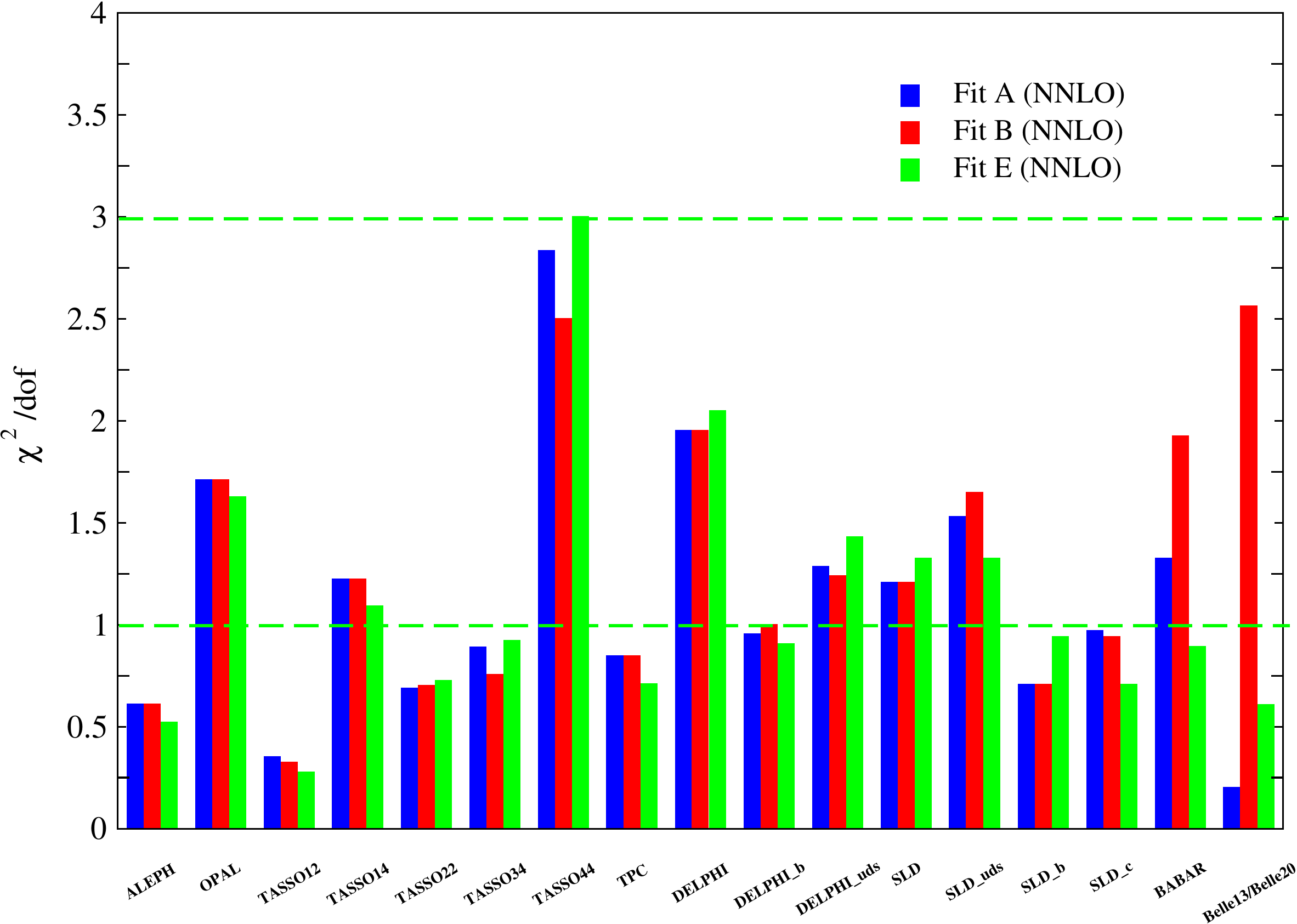}	
\caption{The $\chi^2/dof$ for each individual experiment for Fits~A, B, and~E.
The data used in the Fit~A are in blue,   Fit~B is red, and  Fit~E is green.
The numerical results for all fits (A,B,C,D,E) are listed in Table~\ref{tab-SIA-chi2}.
For reference, (green) guide lines are shown for a $\chi^2/dof$ of 1 and 3. 
\label{fig:chi2-all}}
\end{figure*}
}%

\def\figNNLO{
\begin{figure*}[tbh]
	\includegraphics[width=0.33\textwidth]{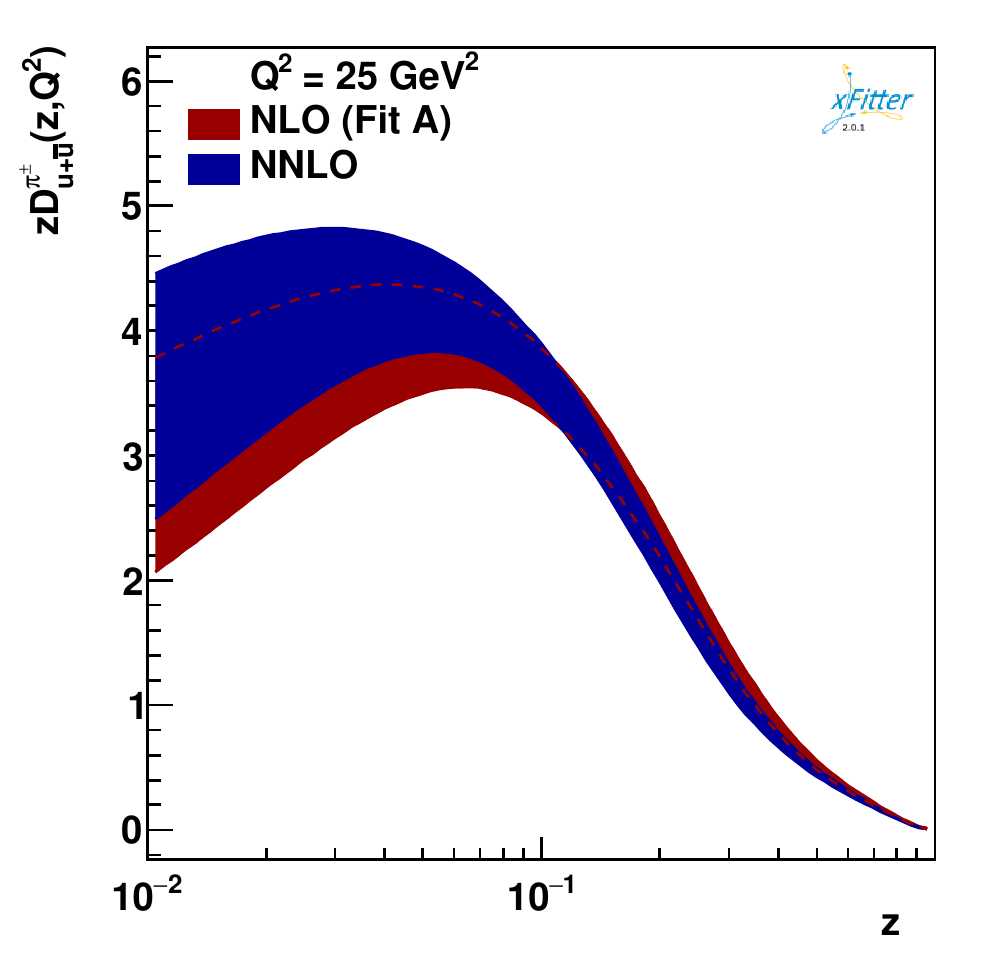}
	\includegraphics[width=0.33\textwidth]{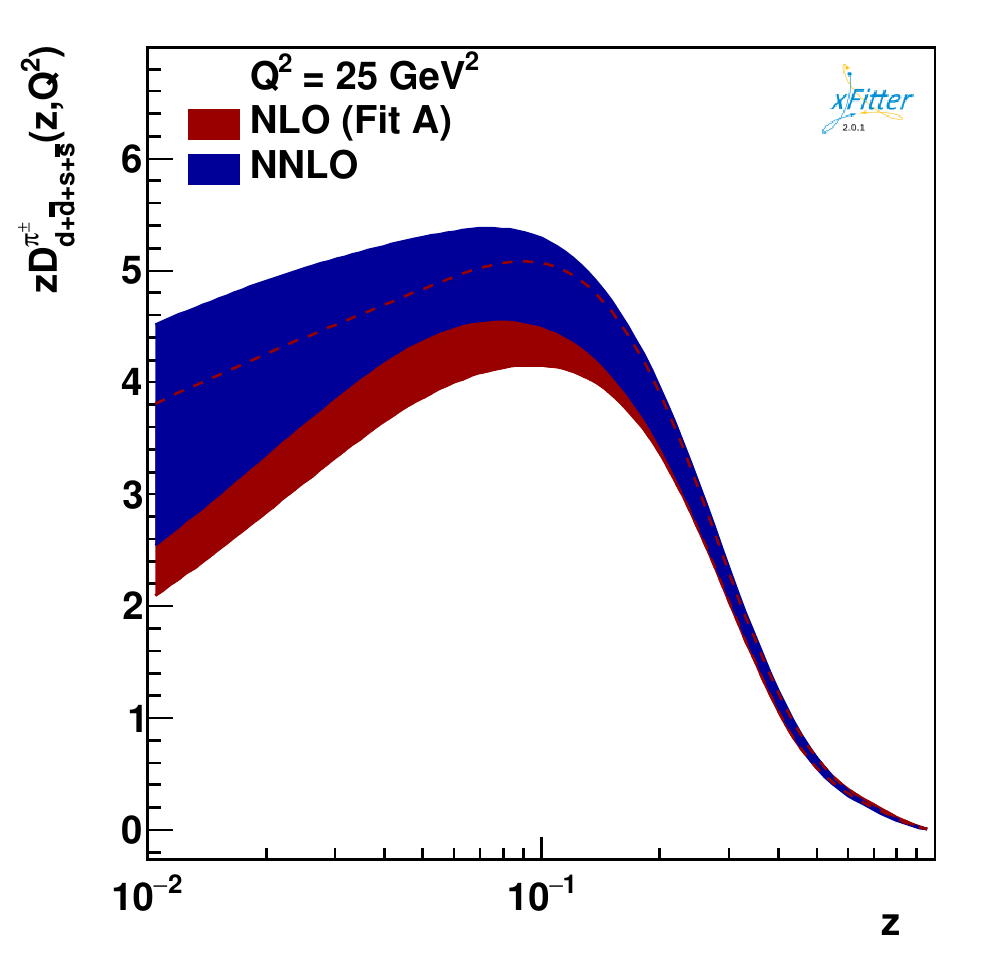}
	\includegraphics[width=0.33\textwidth]{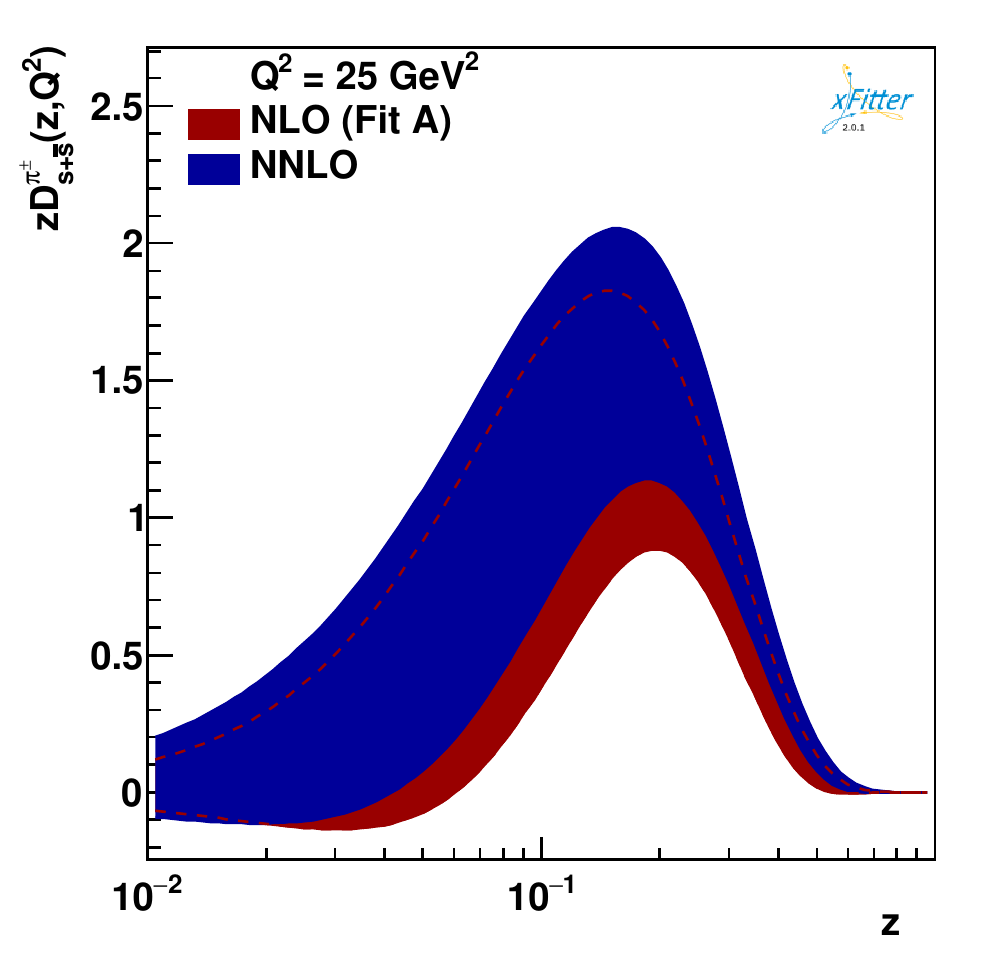}
	\\
	\includegraphics[width=0.33\textwidth]{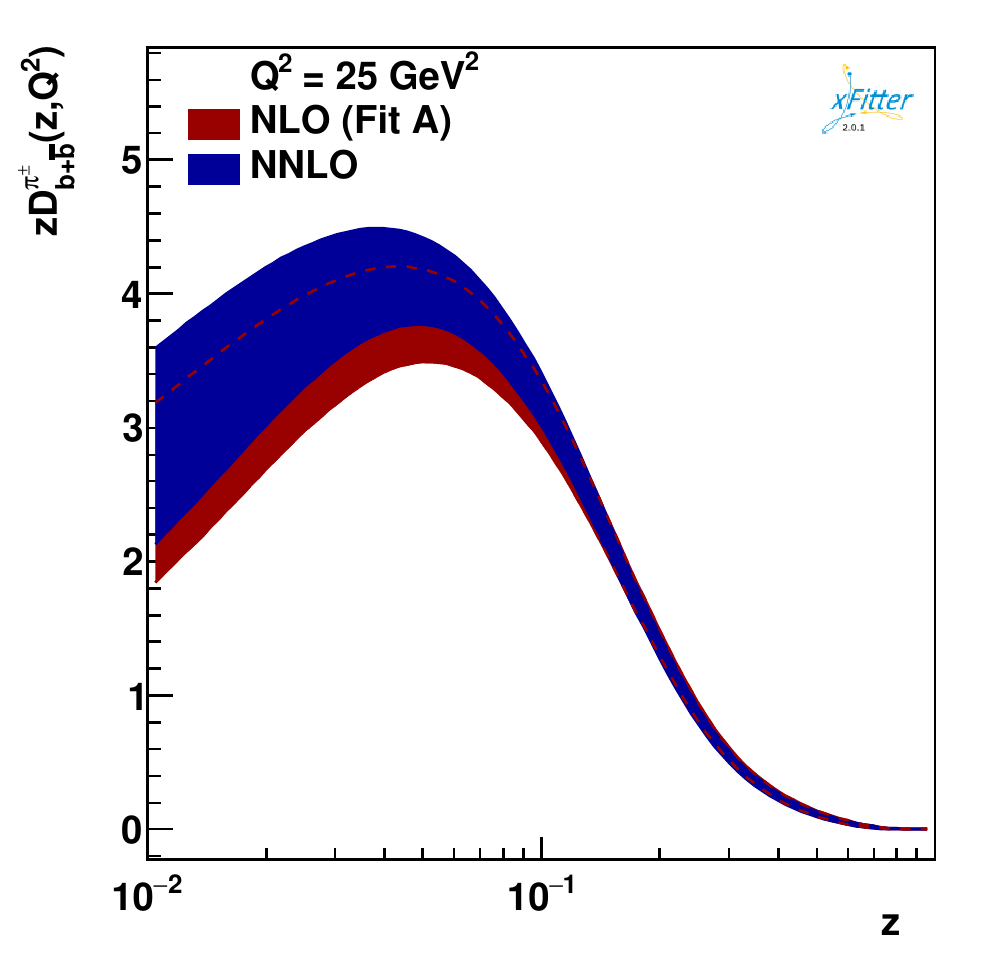}
	\includegraphics[width=0.33\textwidth]{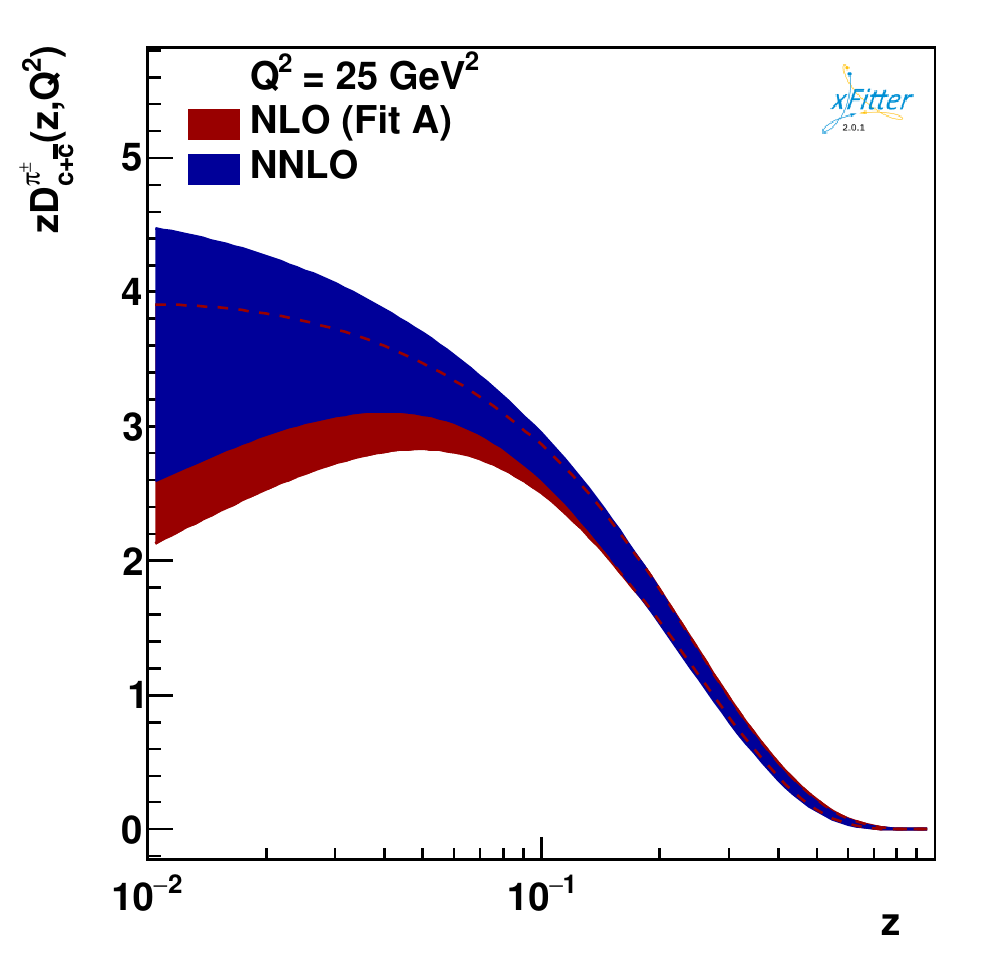}	
	\includegraphics[width=0.33\textwidth]{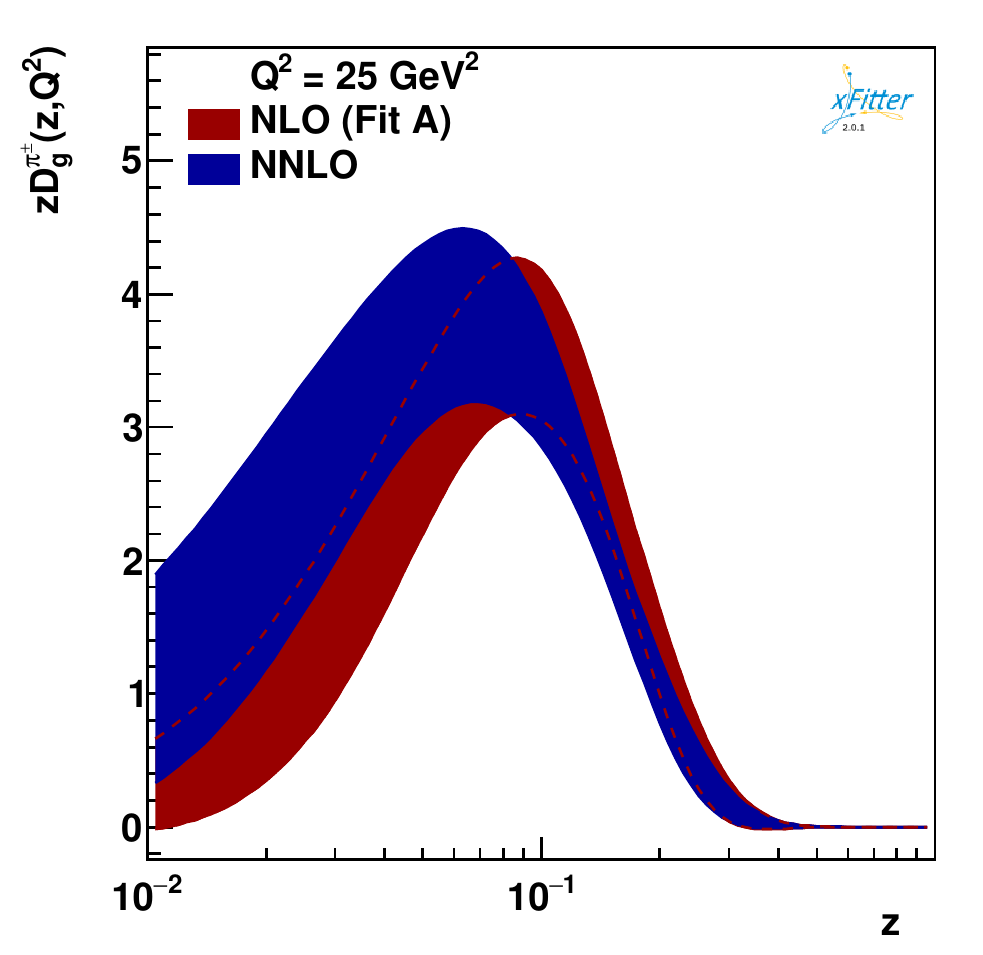}
	\\		
	\caption{A comparison of the NLO and NNLO results for Fit~A at $Q_0^2=25$~GeV${}^2$. 
	As the results for up and down are identical, we choose to also display 
	the $d^+ {+} s^+$ combination.
	Note that we are plotting the sum $\pi^+ {+} \pi^-$.
		\label{fig:FitA-1}}
\end{figure*}
}%

\def\figAB{
\begin{figure*}[tbh]
	\includegraphics[width=0.33\textwidth]{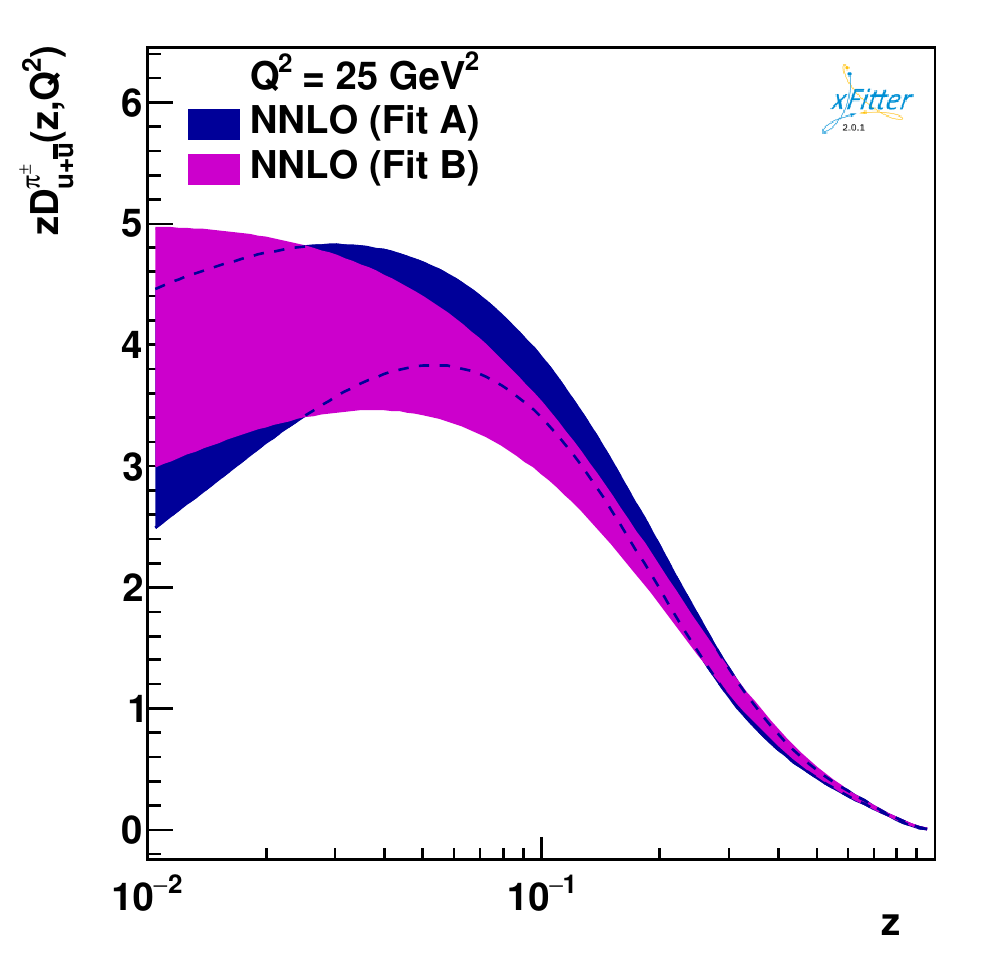}
	\includegraphics[width=0.33\textwidth]{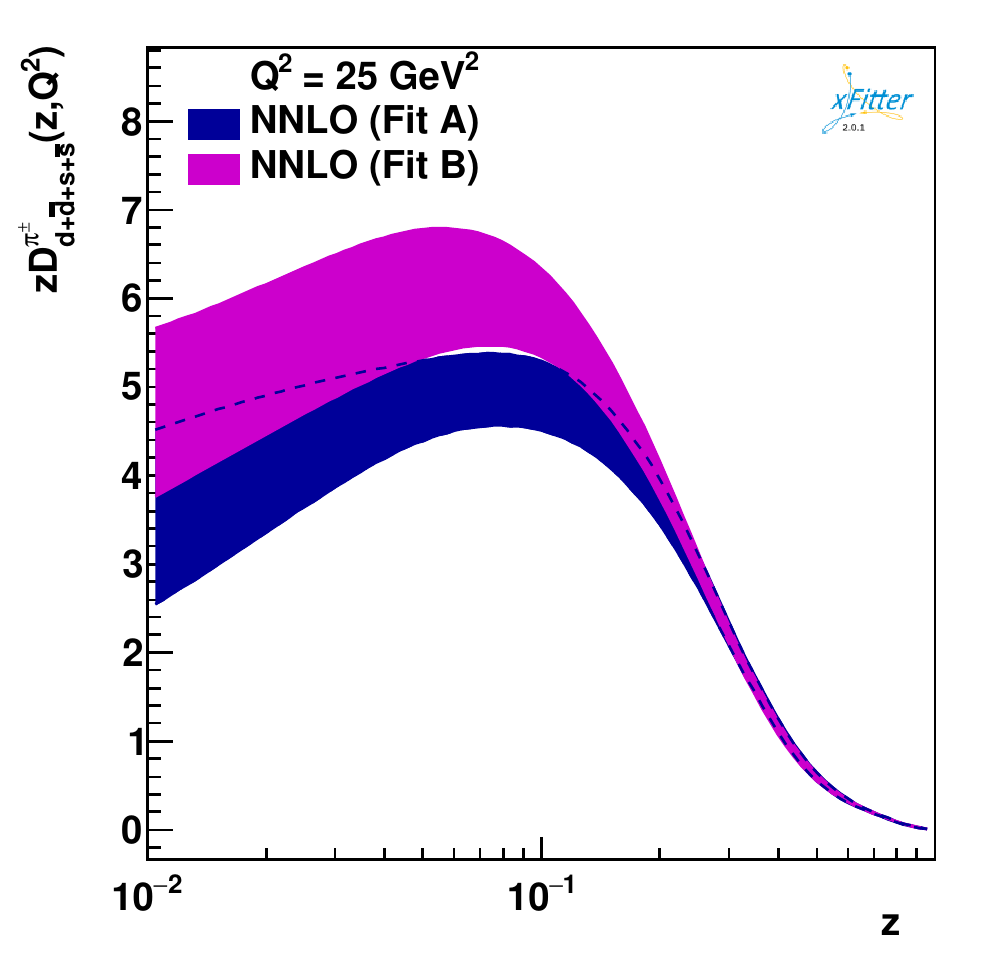}
	\includegraphics[width=0.33\textwidth]{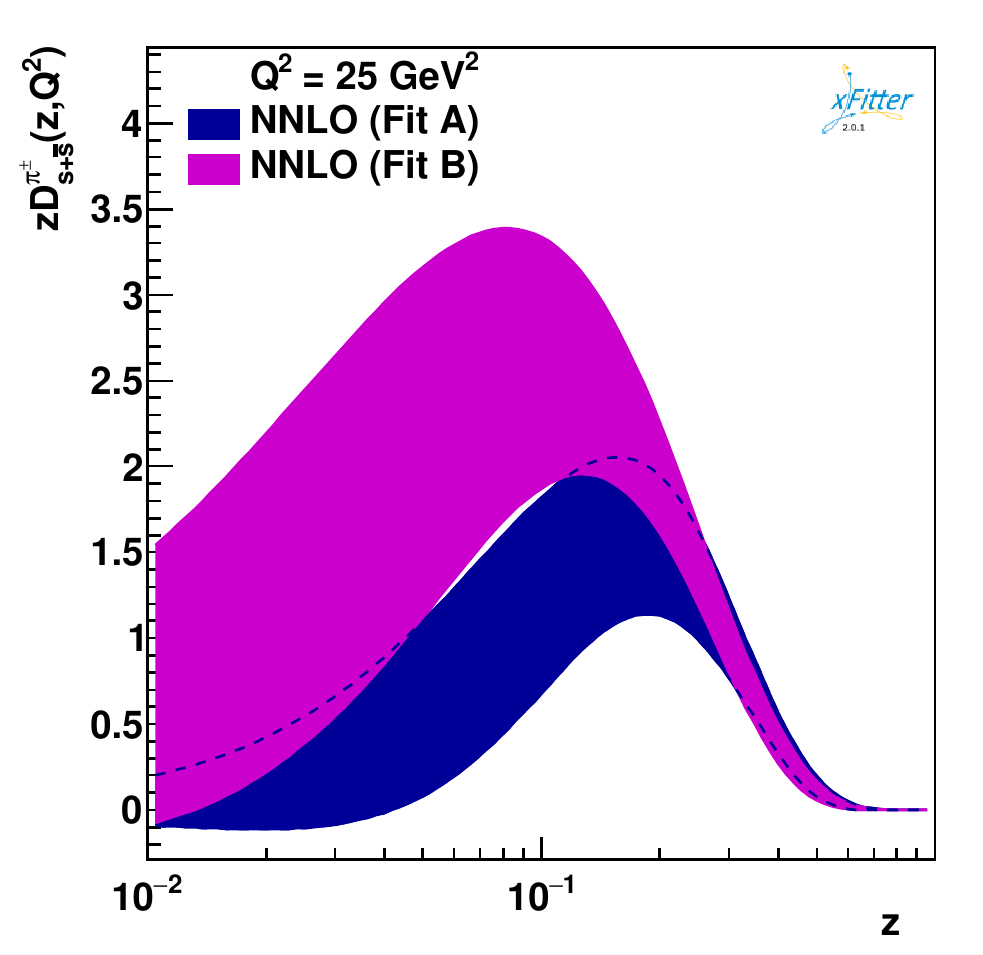}
	\\
	\includegraphics[width=0.33\textwidth]{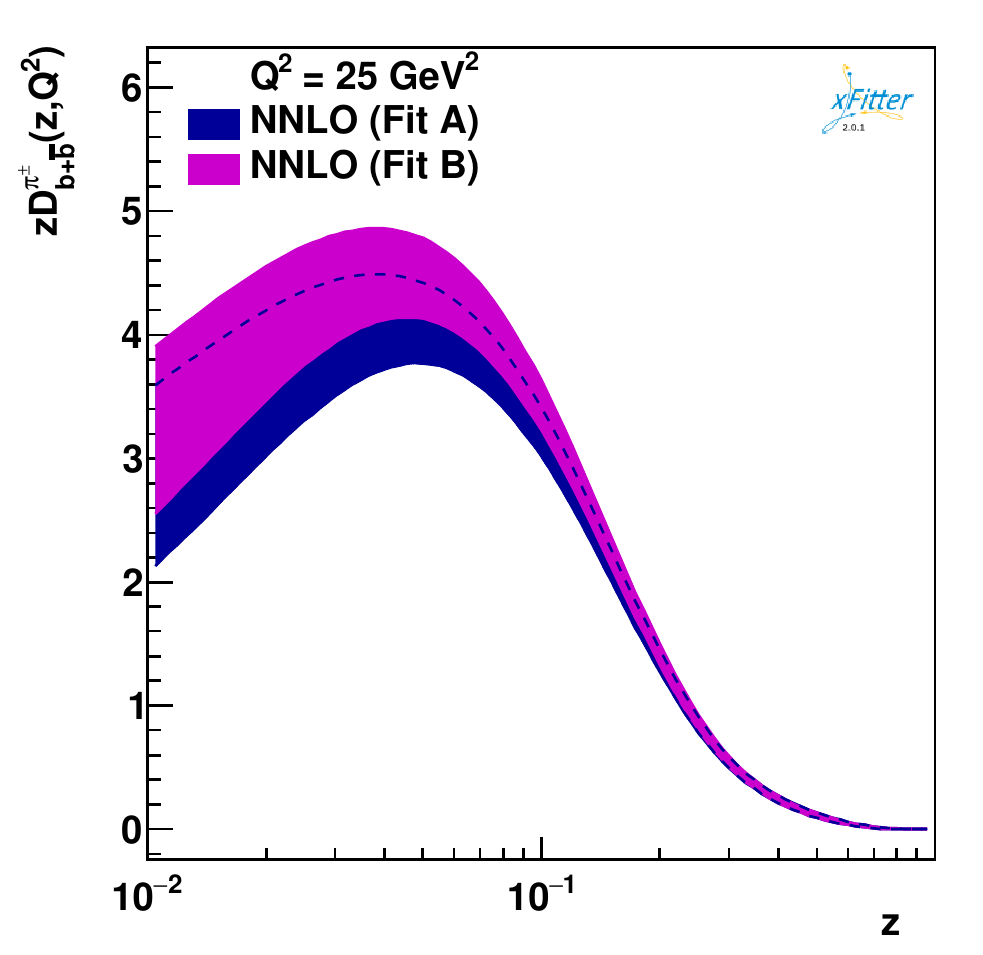}
	\includegraphics[width=0.33\textwidth]{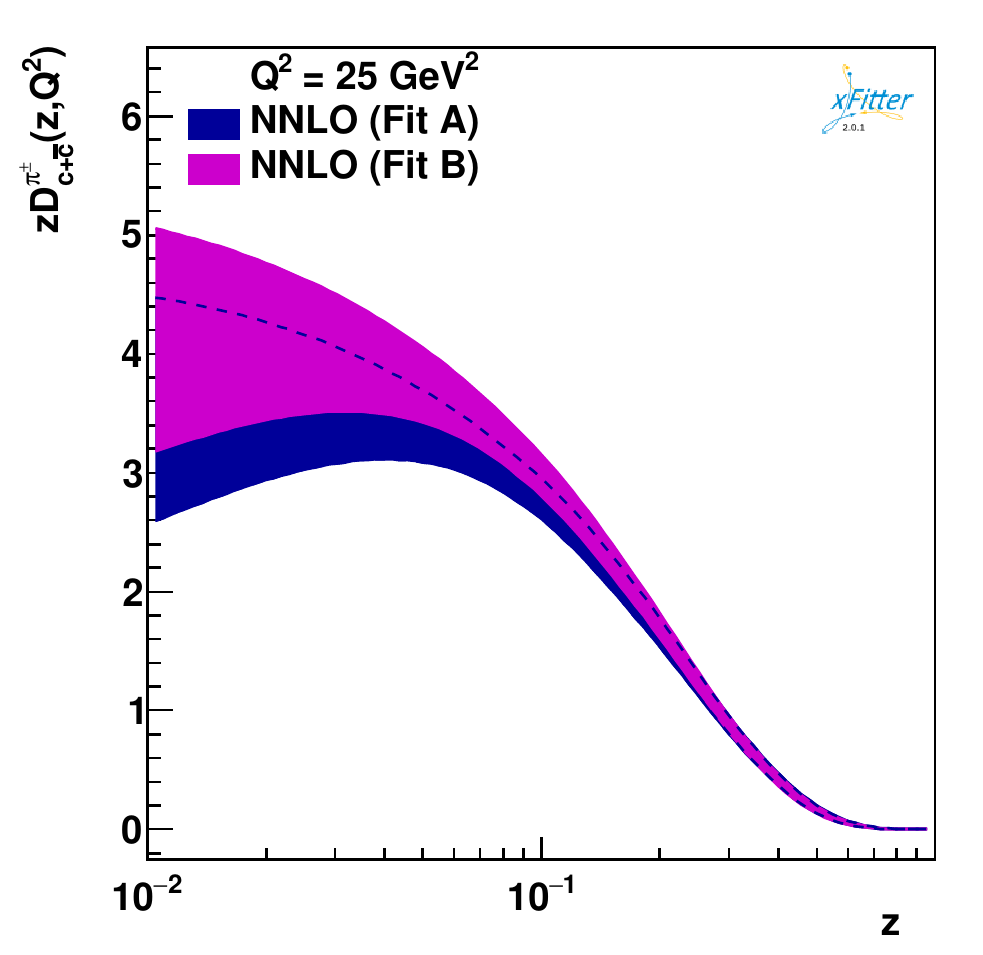}	
	\includegraphics[width=0.33\textwidth]{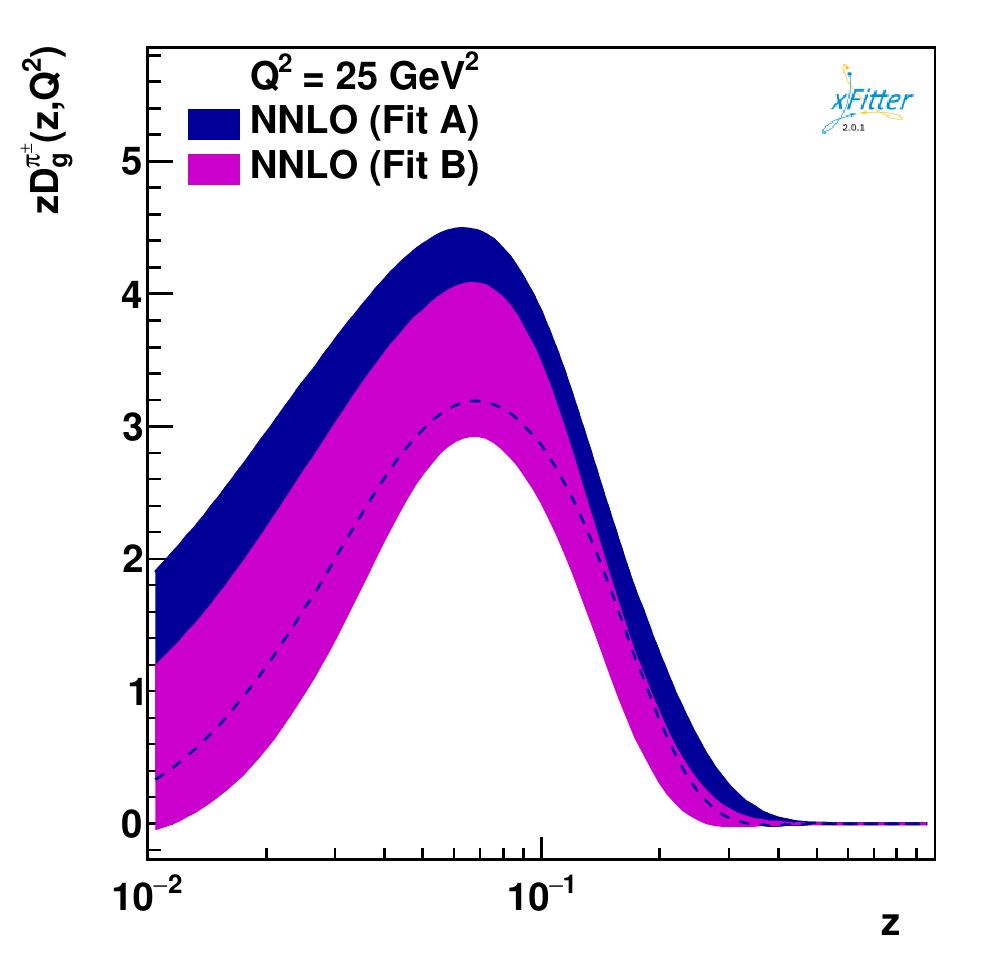}
	\\		
	\caption{A comparison of the NNLO results for
	Fit~A (with BELLE13) and Fit~B (with BELLE20) analyses at $Q_0^2=25$~GeV${}^2$. 
		\label{fig:Belle20-1}}
\end{figure*}
}%

\def\figABC{
\begin{figure*}[htp]
	\includegraphics[width=0.33\textwidth]{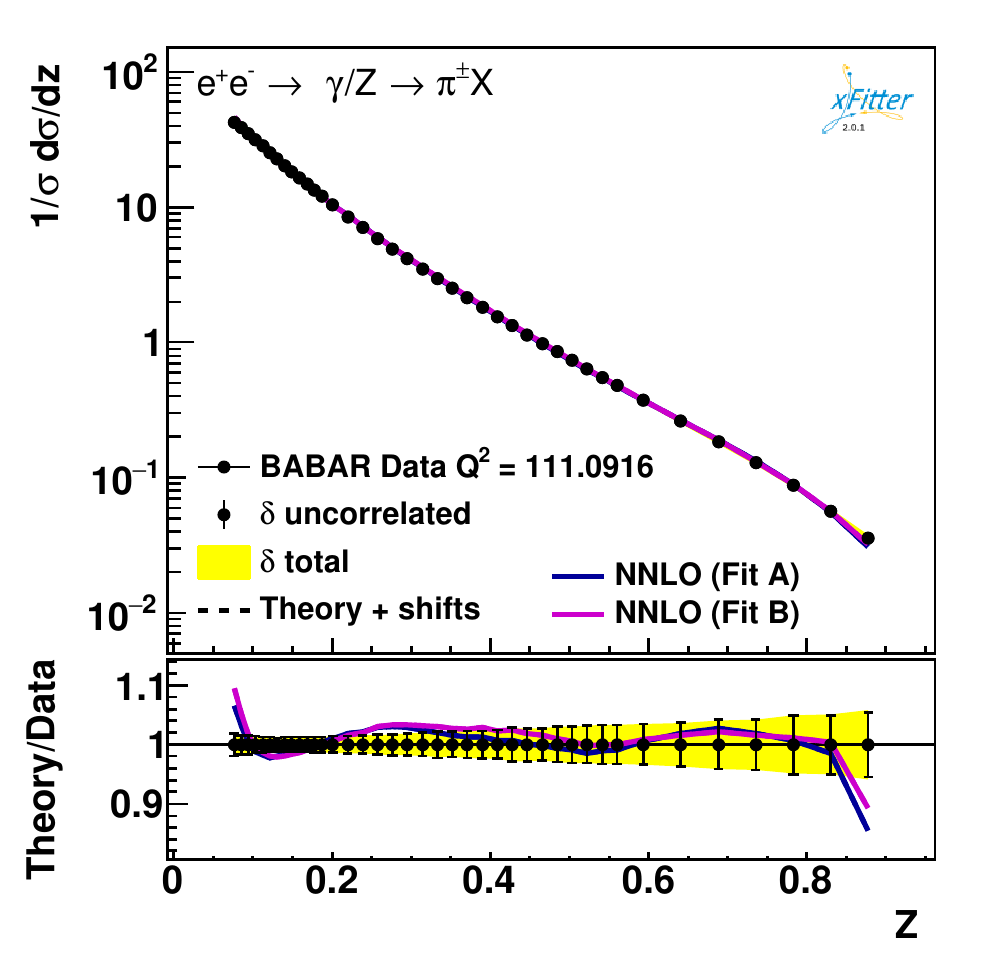}
	\includegraphics[width=0.33\textwidth]{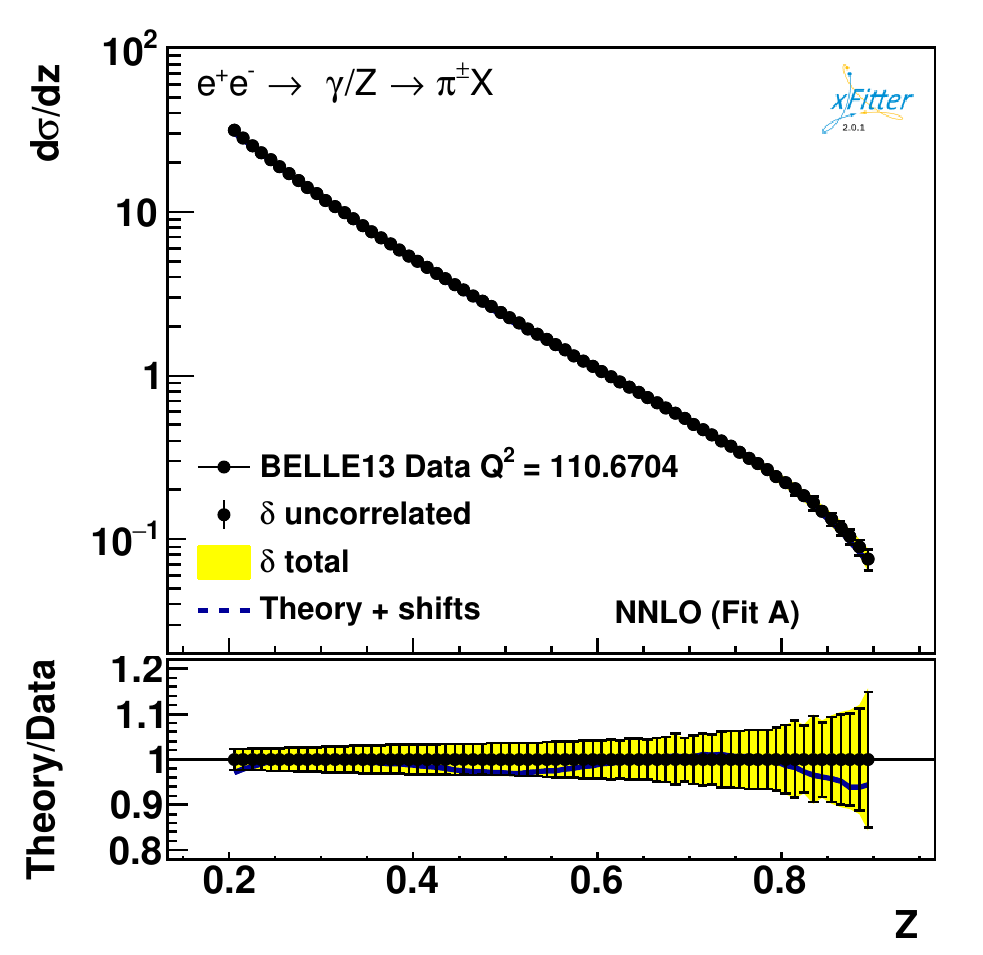}
	\includegraphics[width=0.33\textwidth]{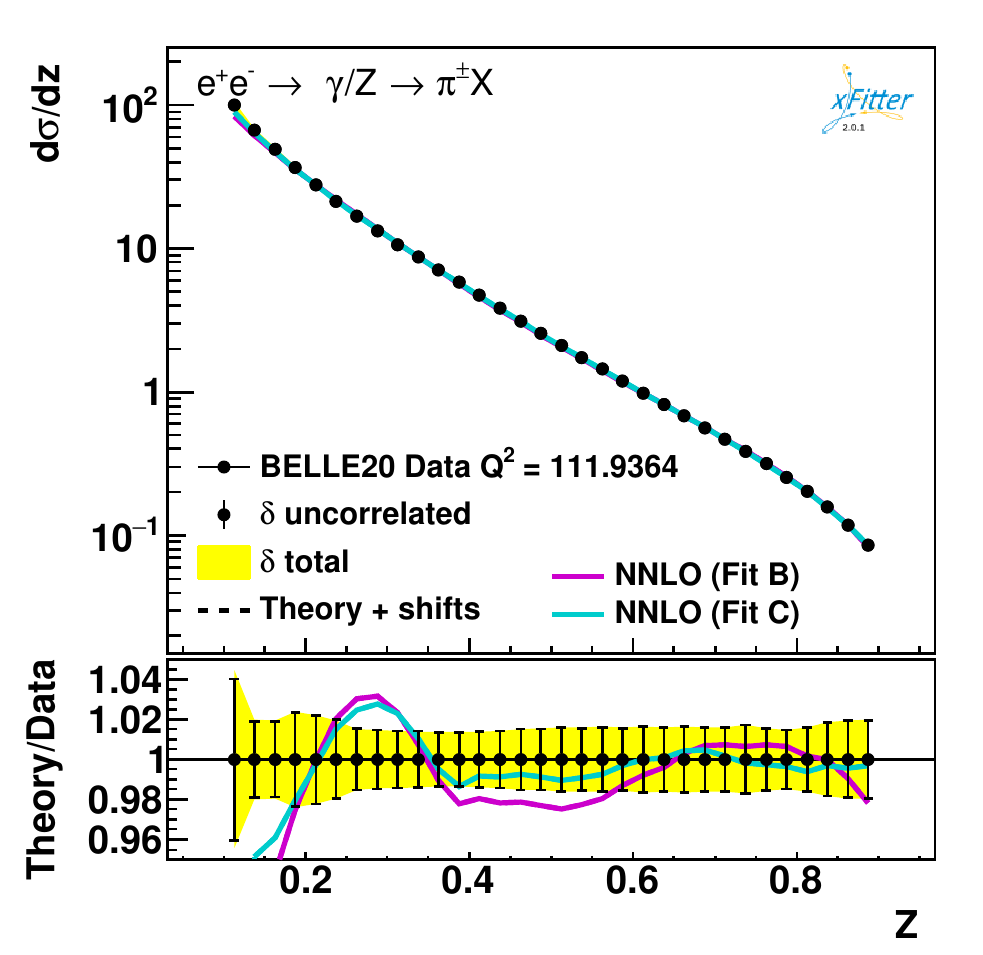}
	\caption{A comparison of BELLE13, BELLE20, and BaBar for Fits~A, B, and C.
	(Not all data sets are included in each fit.)
	}
	\label{fig:Compare-with-SIA-abc}
	\end{figure*}
}%

\def\figDE{
\begin{figure*}[htp]
	\includegraphics[width=0.33\textwidth]{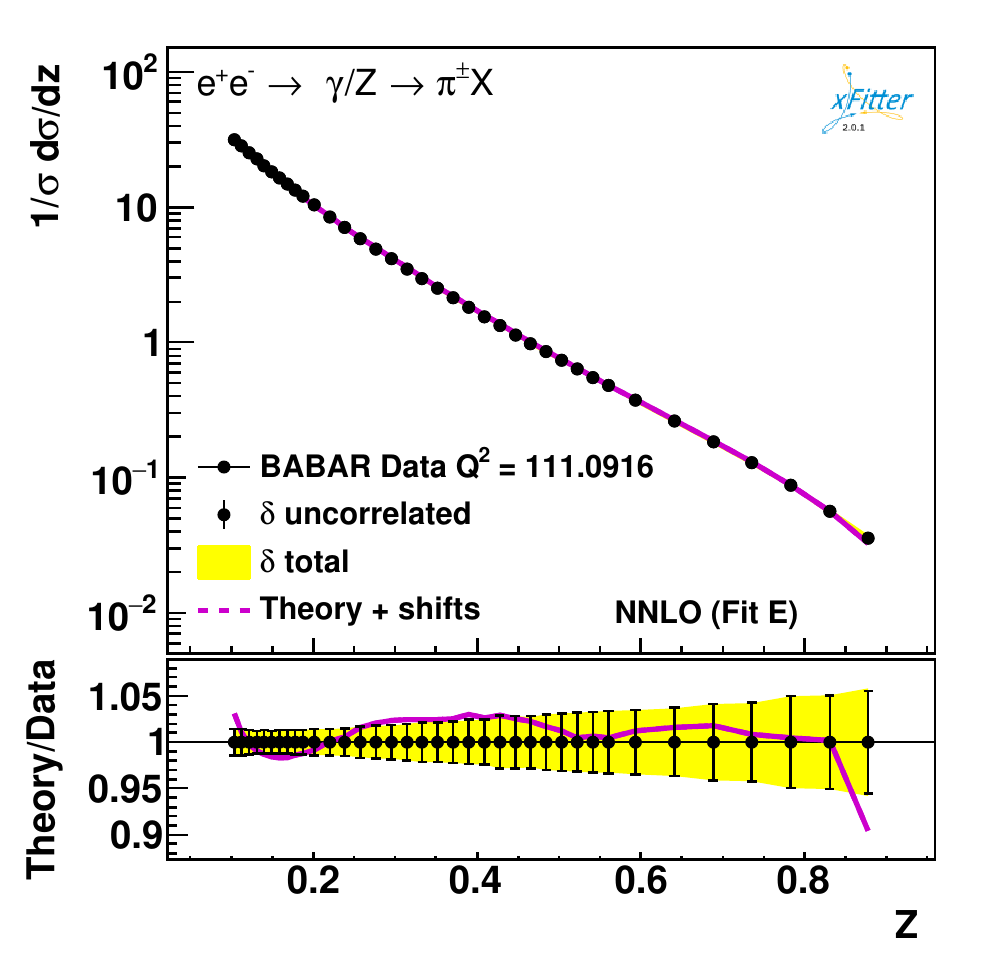}
	\includegraphics[width=0.33\textwidth]{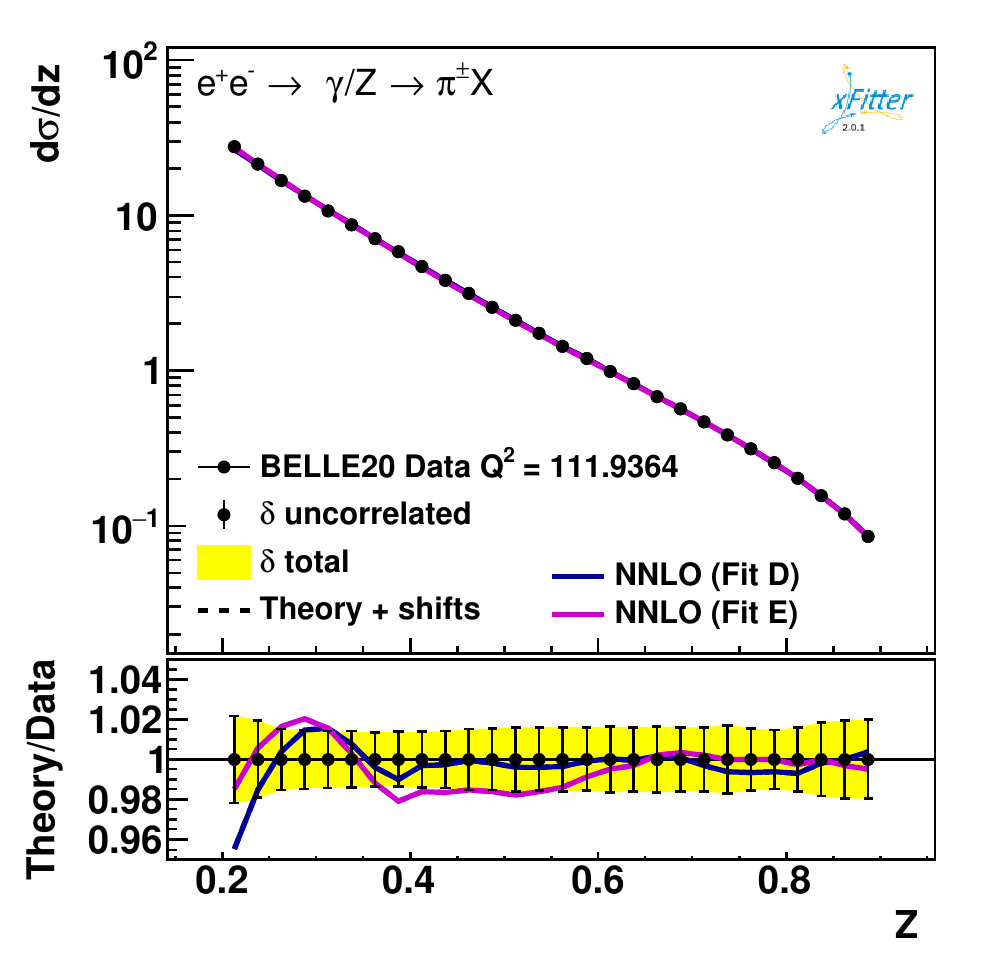}
	\caption{A comparison of  BELLE20 and BaBar for Fits~D and E.
	(BaBar is not included in Fit~D.)}
	\label{fig:Compare-with-SIA-de}
\end{figure*}
}%

\def\figBElowQ{
\begin{figure*}[tbh]
	\begin{center}
		\includegraphics[width=0.33\textwidth]{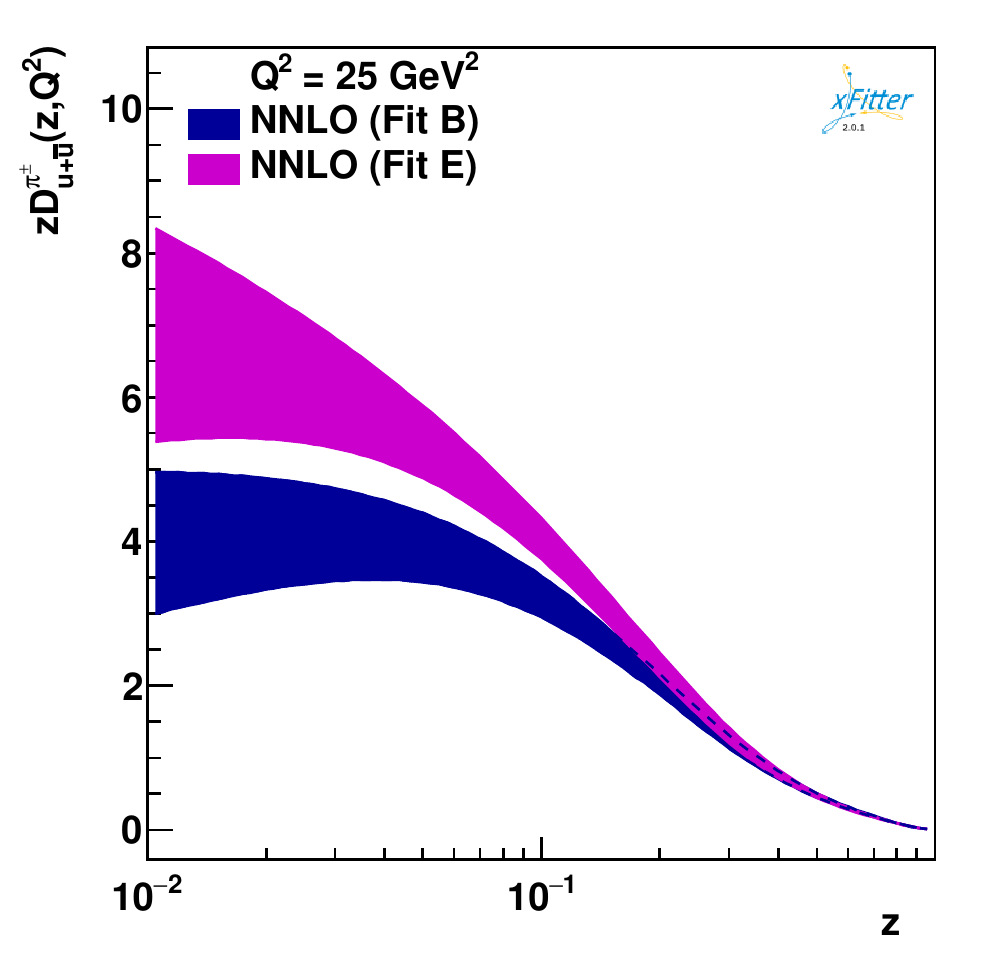}
		\includegraphics[width=0.33\textwidth]{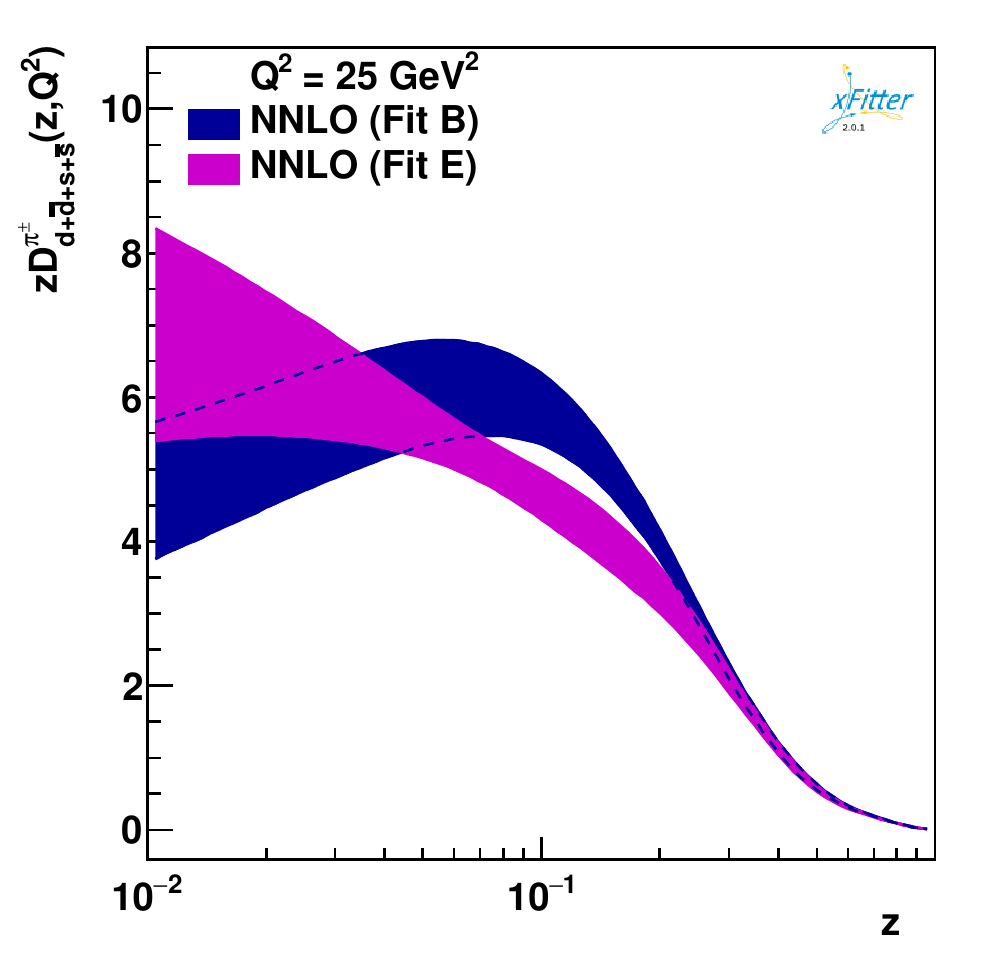}
		\includegraphics[width=0.33\textwidth]{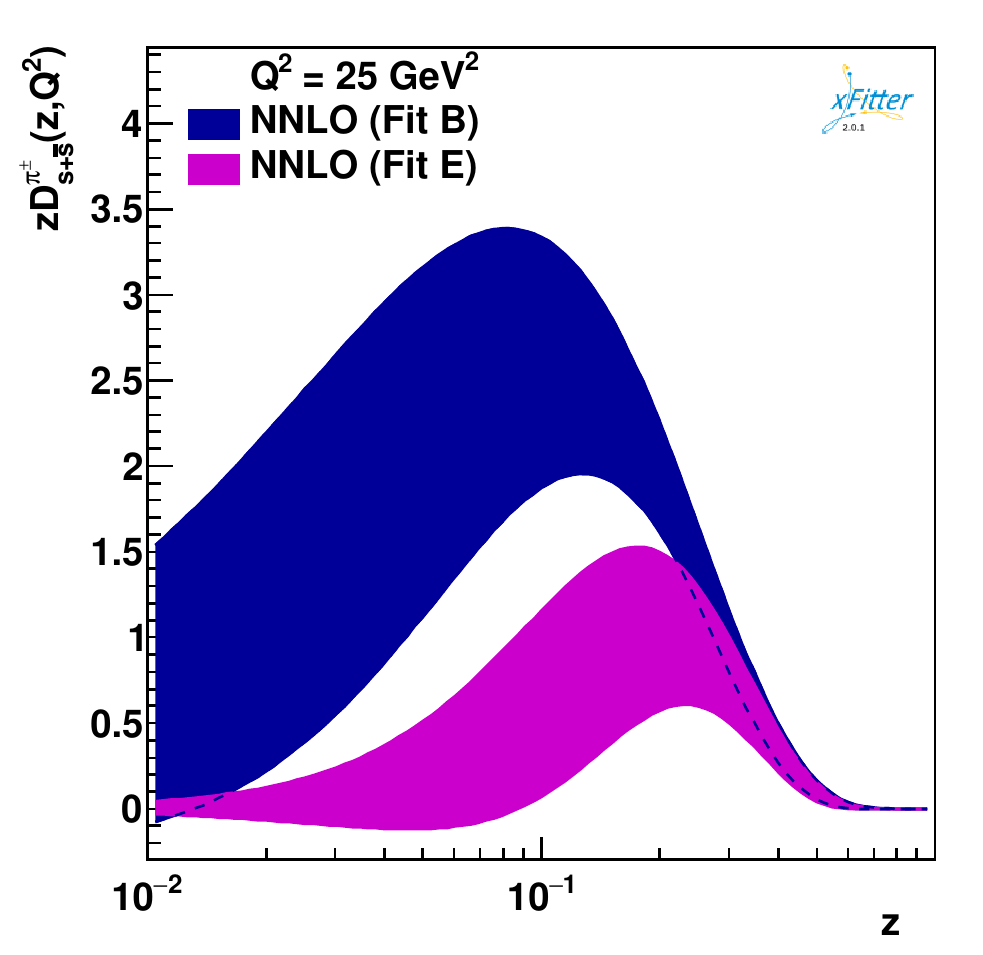}
		\\
		\includegraphics[width=0.33\textwidth]{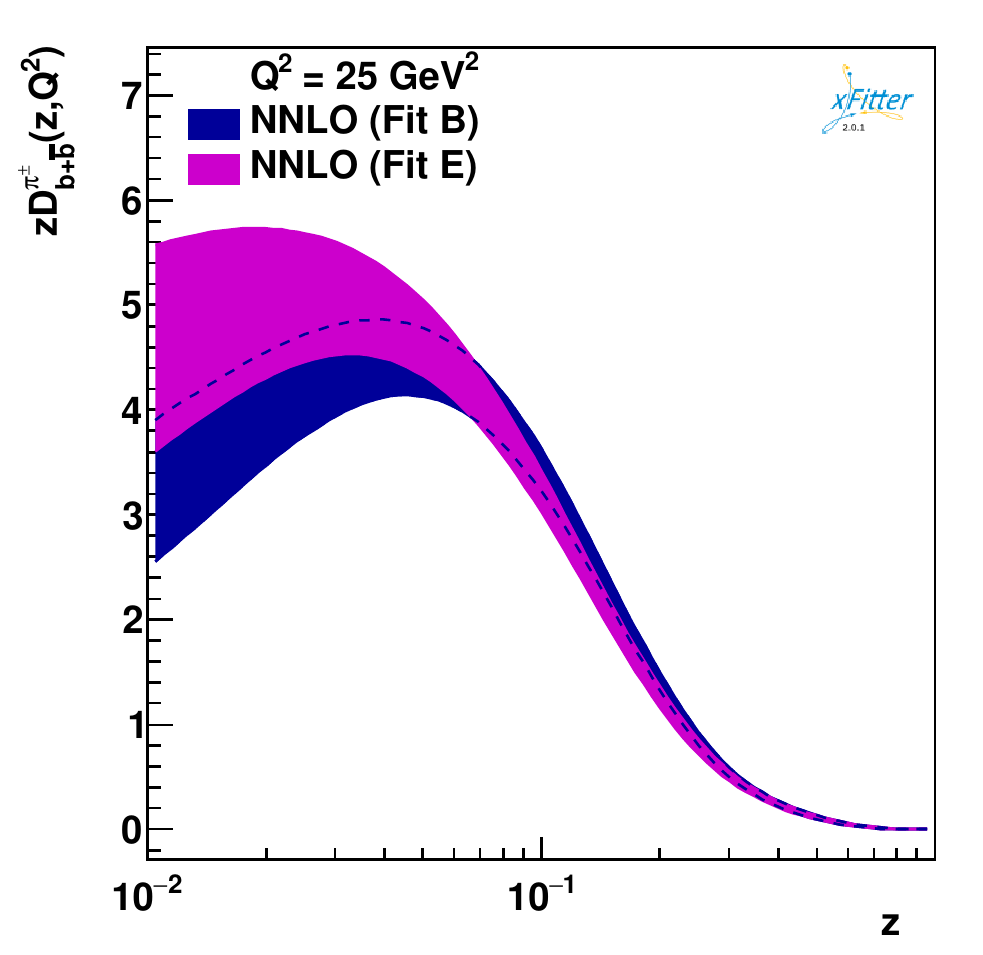}
        \includegraphics[width=0.33\textwidth]{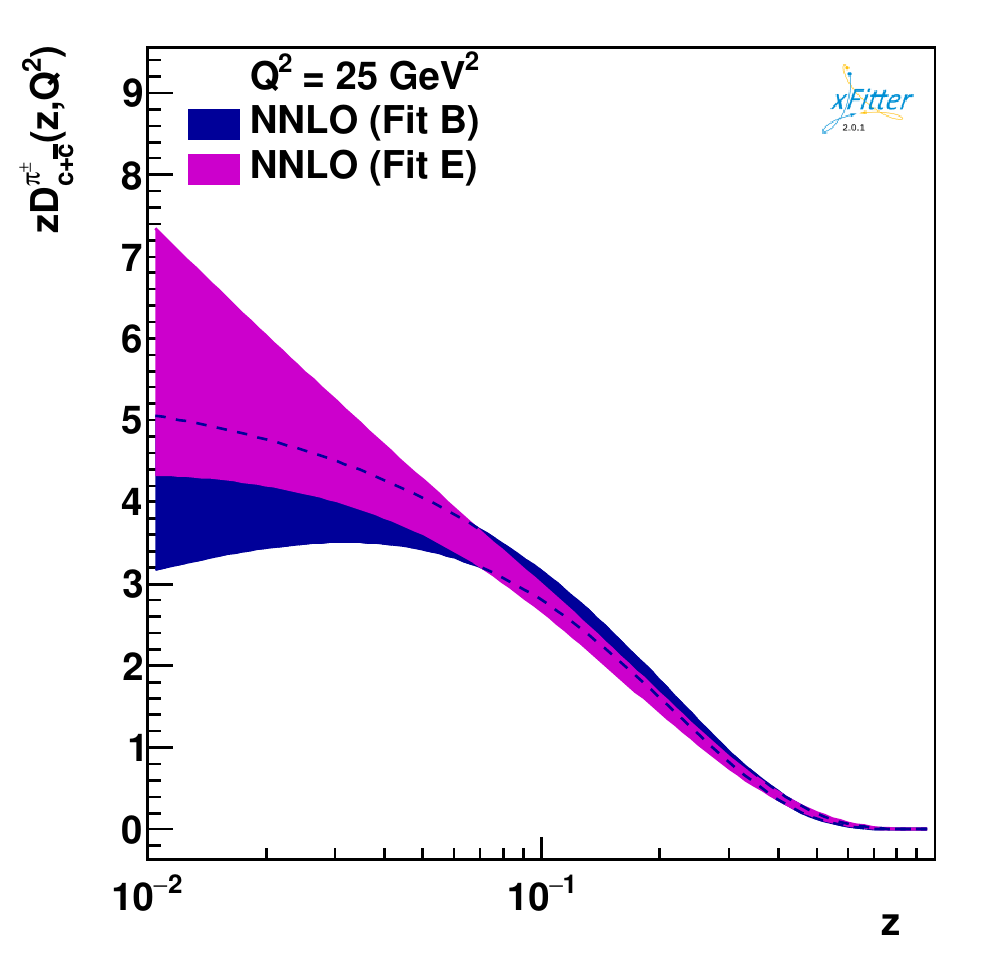}	
        \includegraphics[width=0.33\textwidth]{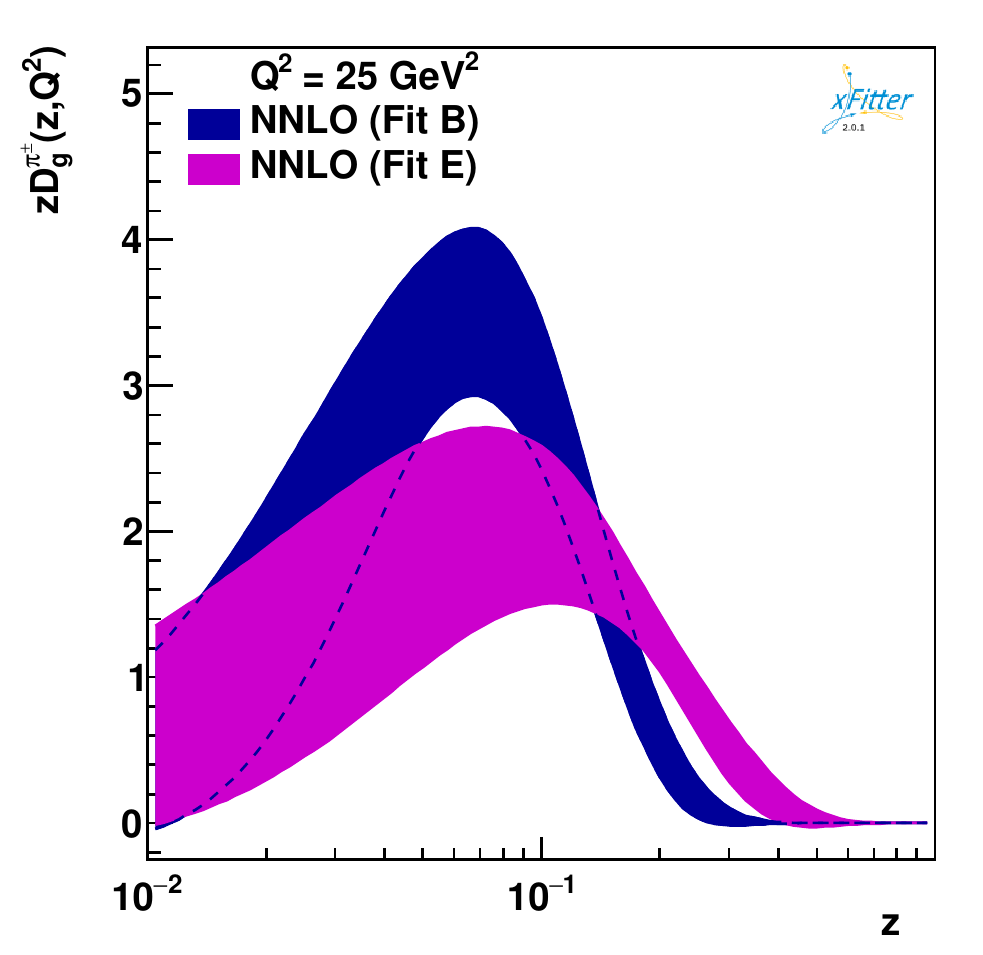}
\end{center}
\caption{
The comparison of the charged pion FFs determined for Fit~B and Fit~E
at $Q_0^2=25$~GeV${}^2$.  as a function of $z$.
Both fits include BELLE20 and BaBar data.
Fit~B has no cuts in $z$, 
while Fit~E imposed a $z{>}0.2$ cut on BELLE20 
and a $z{>}0.1$ cut on BaBar. 
\label{fig:FFs-BE}}
\end{figure*}
}%

\def\figBEhighQ{
\begin{figure*}[tbh]
	\begin{center}
		\includegraphics[width=0.33\textwidth]{./Plots/Belle20-cut/q2_8317_pdf_u+ubar-BE}
		\includegraphics[width=0.33\textwidth]{./Plots/Belle20-cut/q2_8317_pdf_dos}
		\includegraphics[width=0.33\textwidth]{./Plots/Belle20-cut/q2_8317_pdf_s+bars-BE}
		\\
		\includegraphics[width=0.33\textwidth]{./Plots/Belle20-cut/q2_8317_pdf_b+bbar-BE}
        \includegraphics[width=0.33\textwidth]{./Plots/Belle20-cut/q2_8317_pdf_c+cbar-BE}	
        \includegraphics[width=0.33\textwidth]{./Plots/Belle20-cut/q2_8317_pdf_g-BE}
\end{center}
\caption{
The comparison of the charged pion FFs determined for Fit~B and Fit~E
at $Q^2=8317$~GeV$^2$ as a function of $z$.
Both fits include BELLE20 and BaBar data.
Fit~B has no cuts in $z$, 
while Fit~E imposed a $z{>}0.2$ cut on BELLE20 
and a $z{>}0.1$ cut on BaBar. 
\label{fig:FFs-BE2}}
\end{figure*}
}%

\def\figScale{
\begin{figure*}[tbh]	
	\includegraphics[width=0.33\textwidth]{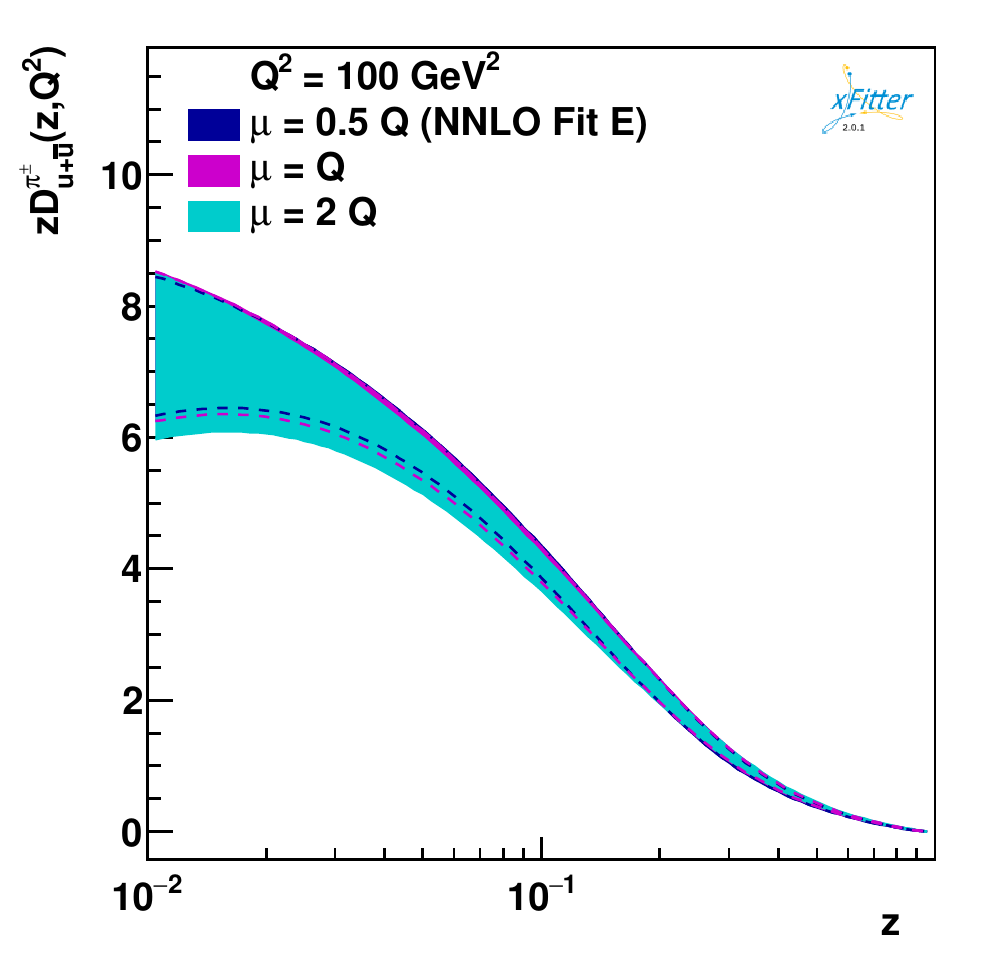}
	\includegraphics[width=0.33\textwidth]{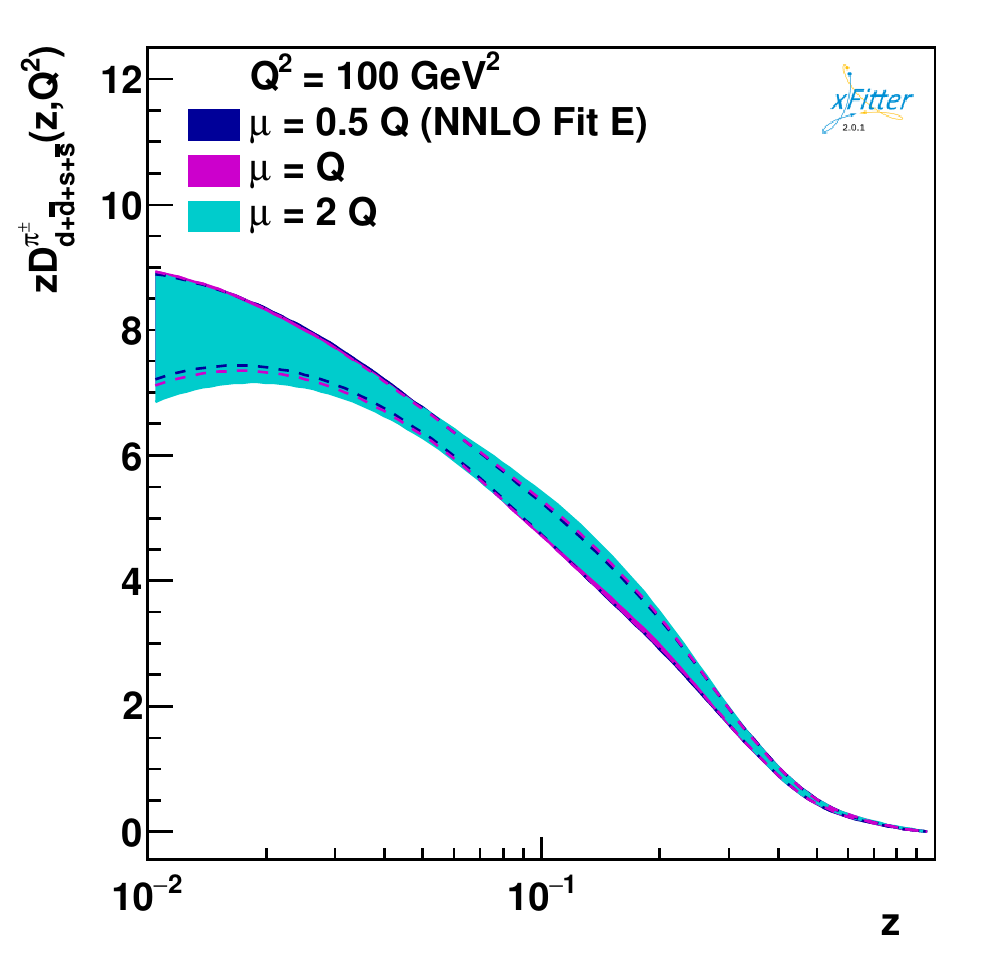}
	\includegraphics[width=0.33\textwidth]{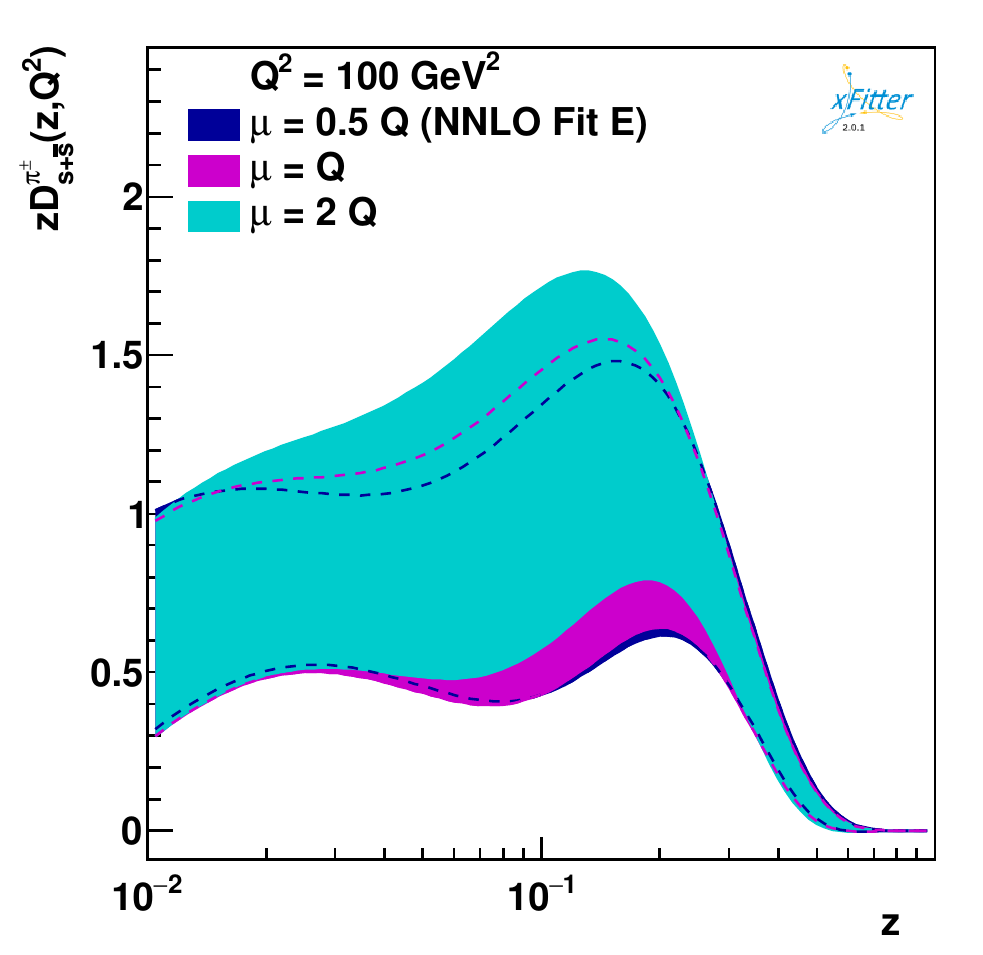}
	\\
	\includegraphics[width=0.33\textwidth]{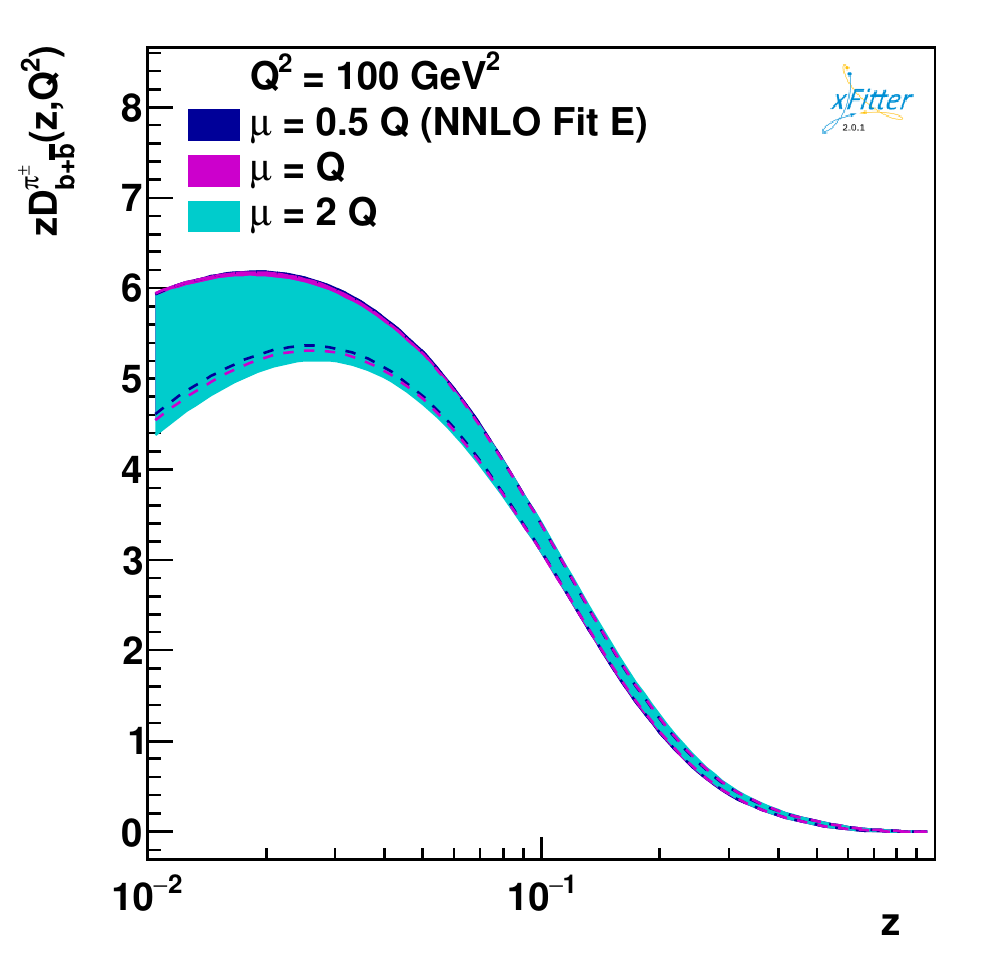}
	\includegraphics[width=0.33\textwidth]{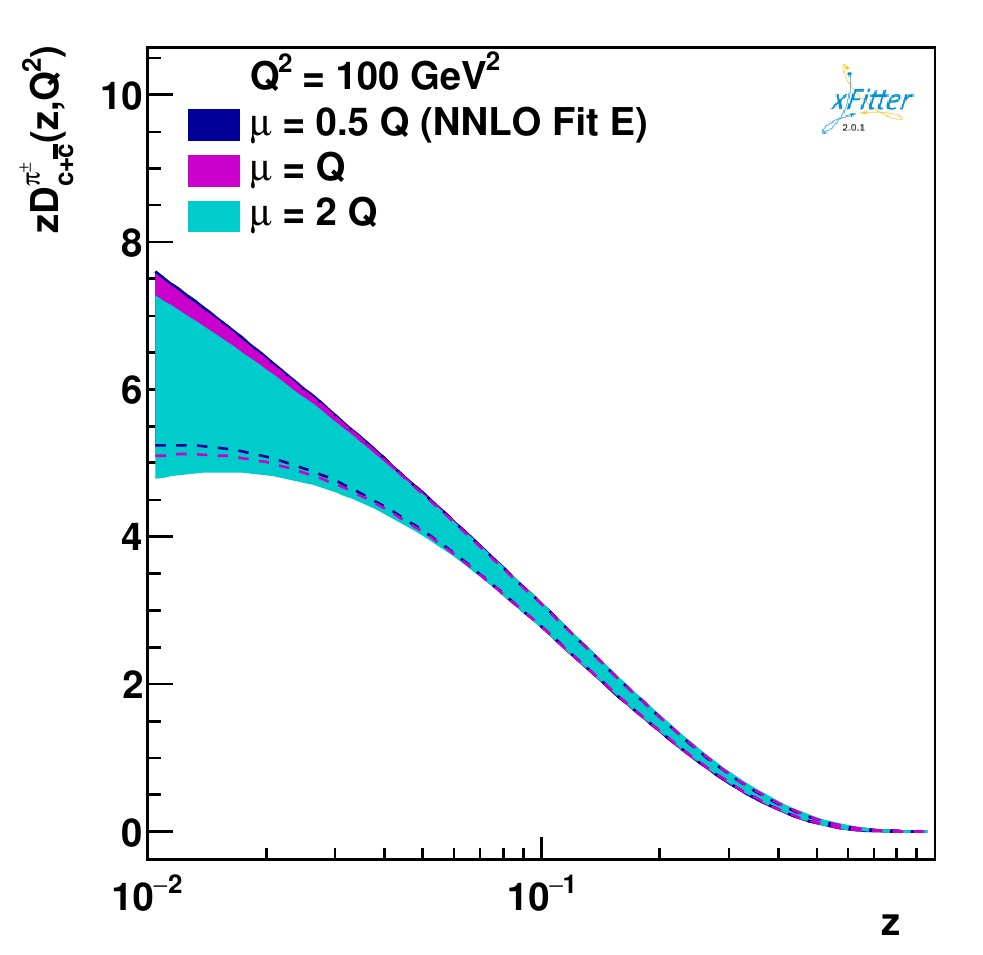}	
	\includegraphics[width=0.33\textwidth]{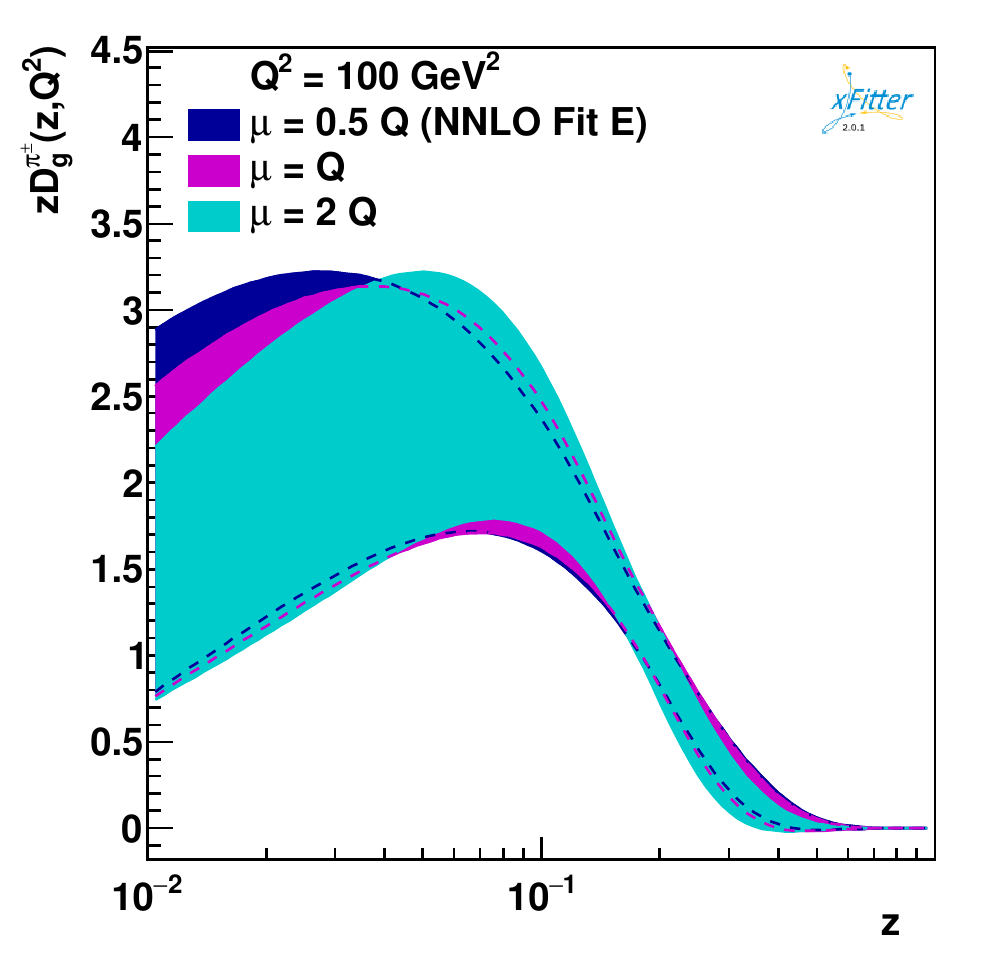}
	\caption{The NNLO charged pion FFs extracted of Fit~E for different choices of the scale $\mu$ as function of $z$. 
	We choose $Q^2=100$~GeV${}^2$ and display the values $\mu=\{5, 10, 20\}$~GeV. 
\label{fig:FFs-scale}}
\end{figure*}
}%

\def\figCompLitLowQ{
\begin{figure*}[tbh]
	\includegraphics[width=0.33\textwidth]{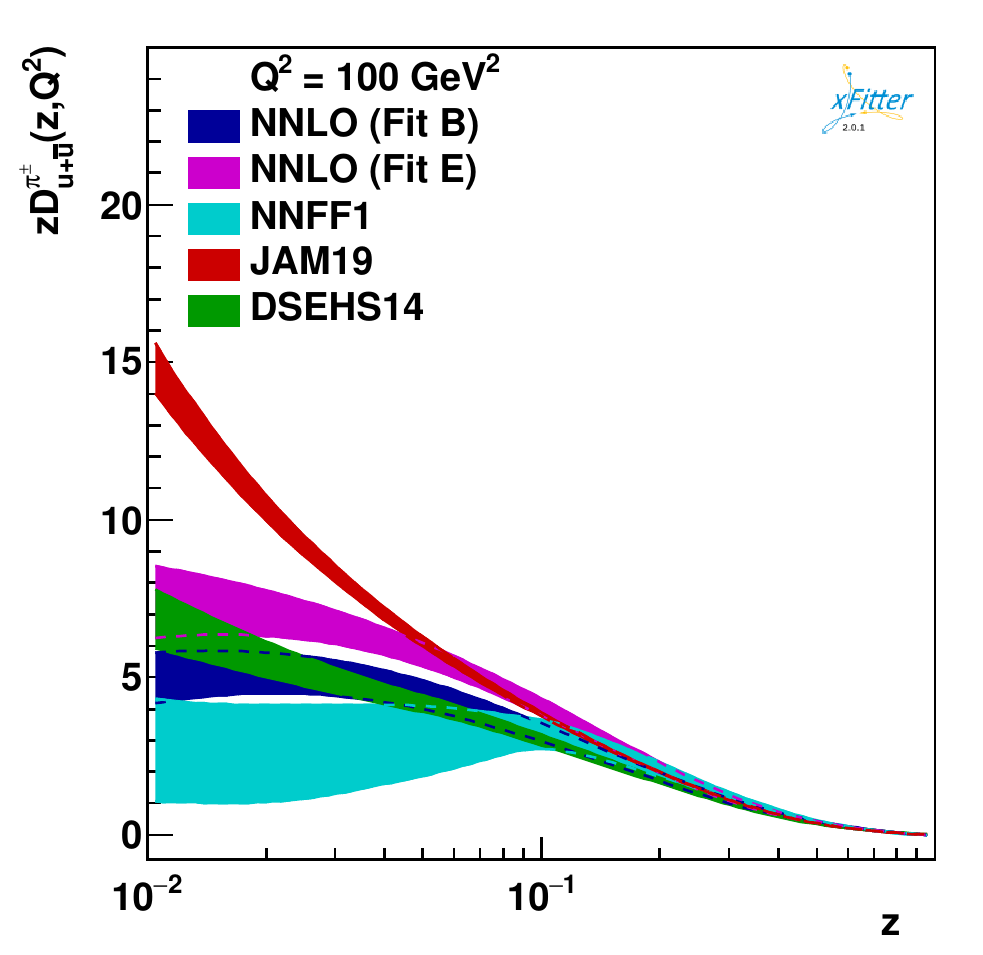}
	\includegraphics[width=0.33\textwidth]{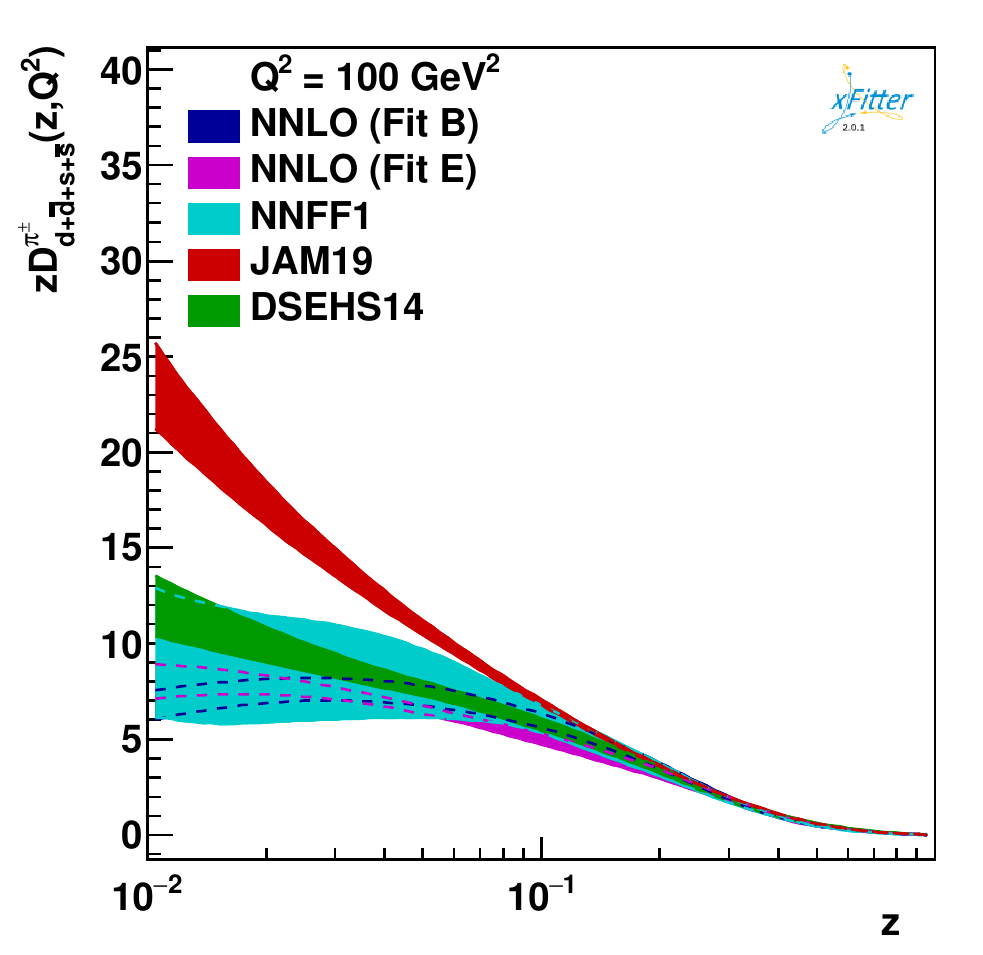}
	\includegraphics[width=0.33\textwidth]{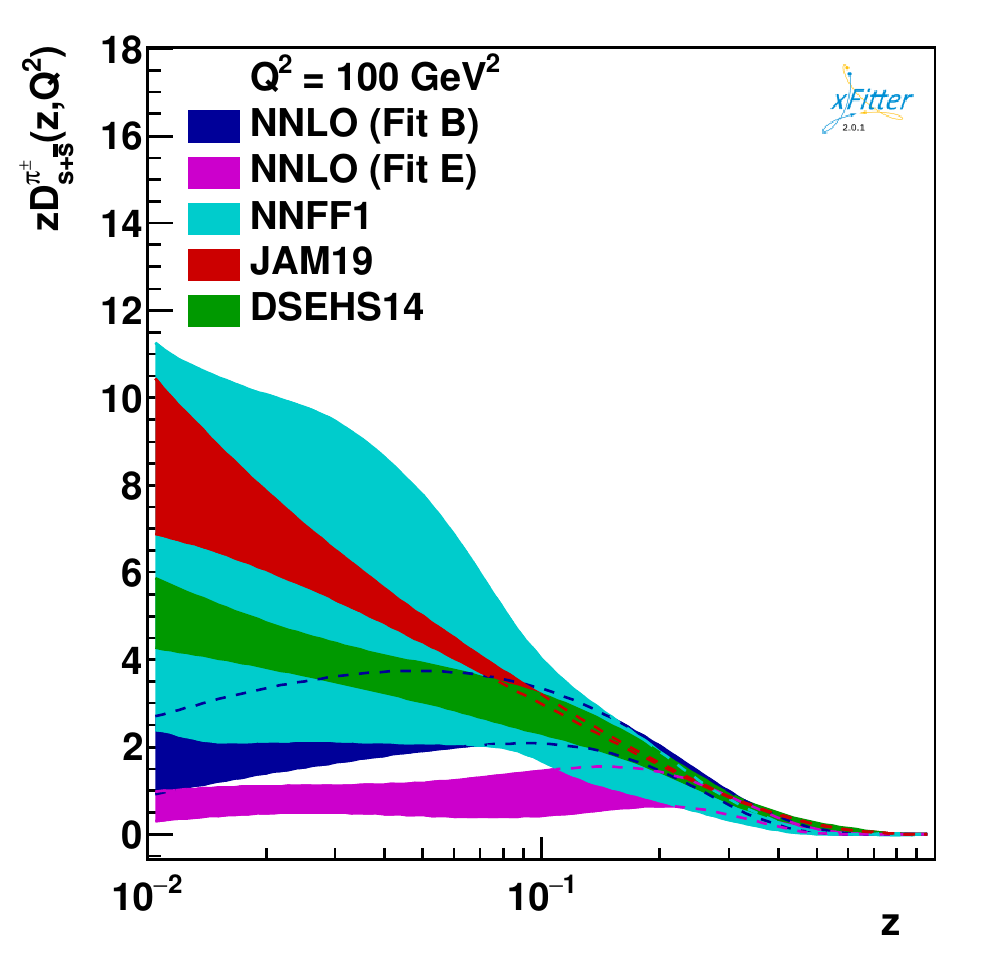}
	\\
	\includegraphics[width=0.33\textwidth]{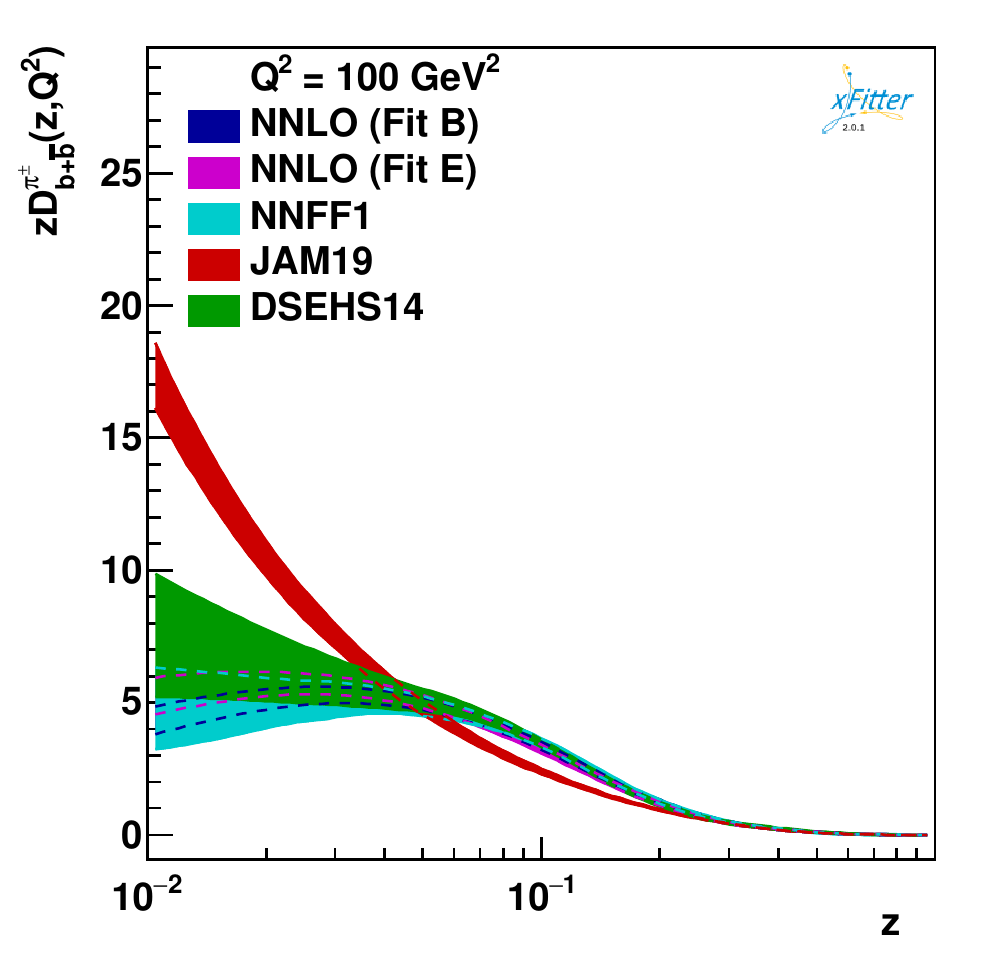}
	\includegraphics[width=0.33\textwidth]{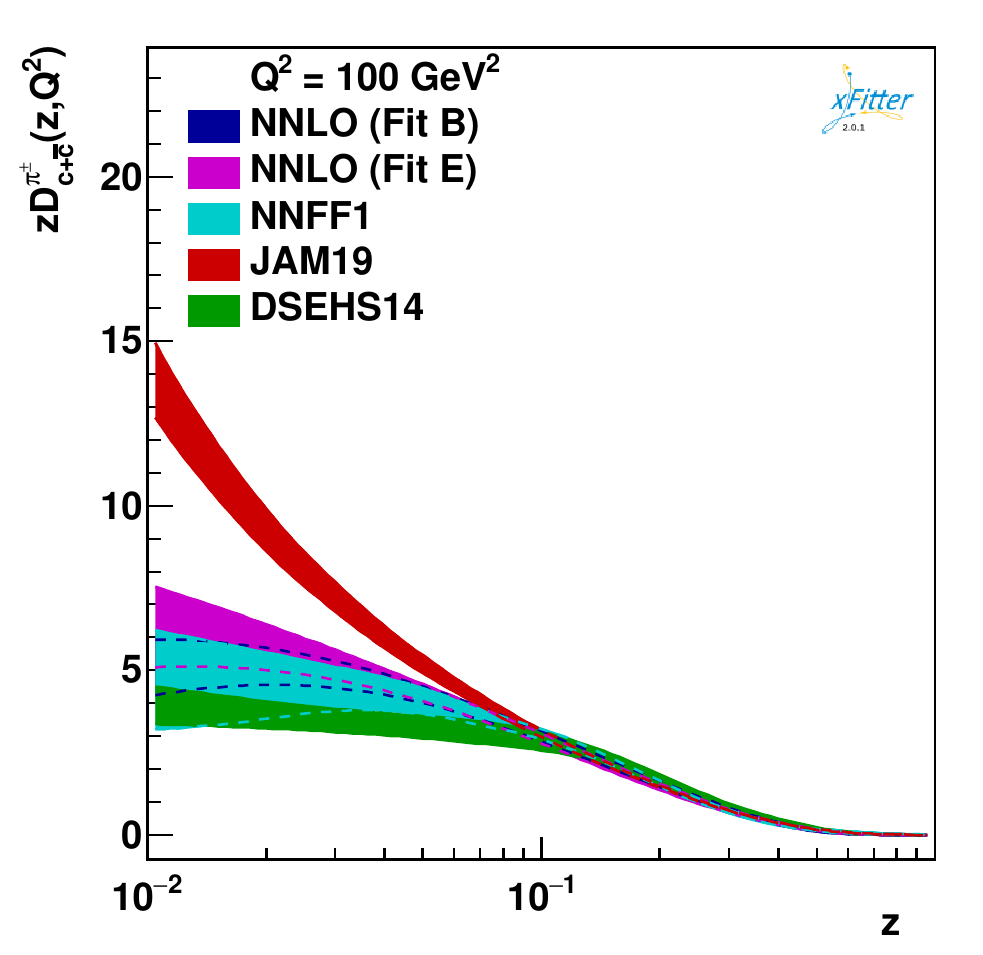}
	\includegraphics[width=0.33\textwidth]{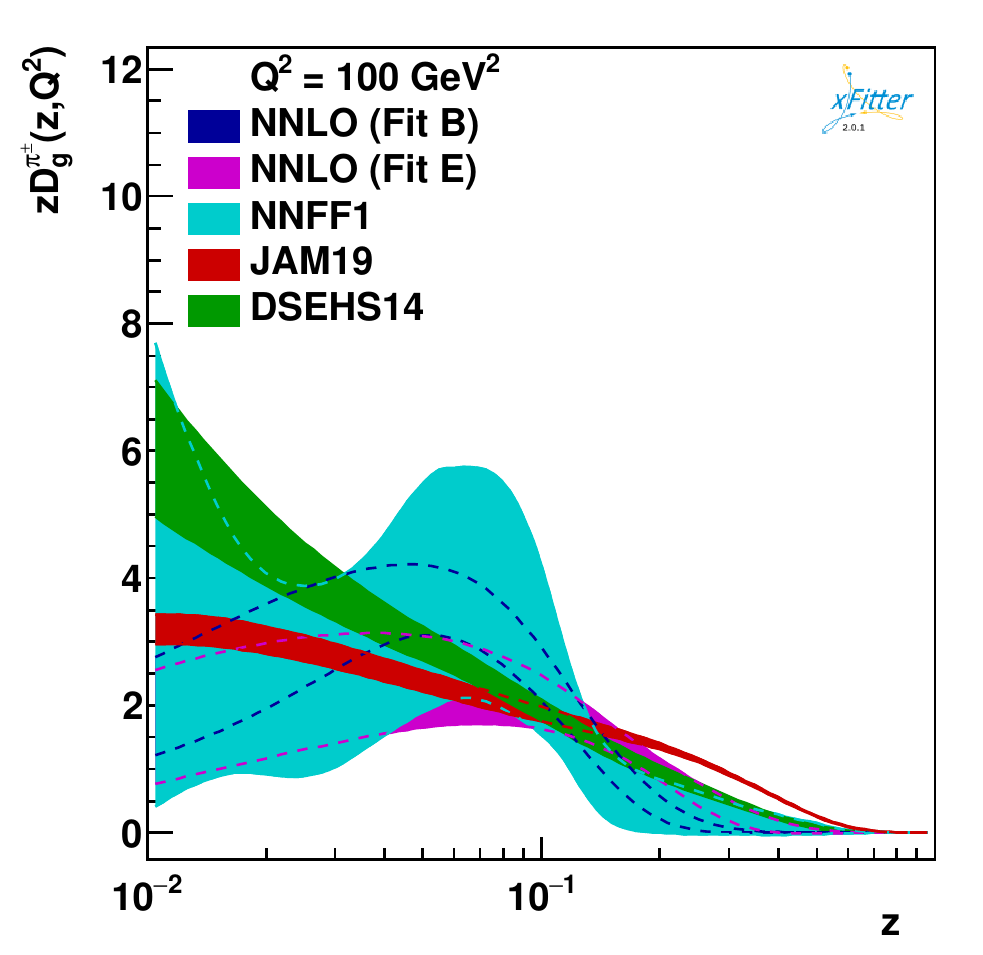}
	\caption{A comparison of our    preferred Fit~E [\ipmx{}]  as well as Fit~B  for 
	charged pion FFs  ($\pi^+ {+} \pi^-$) at NNLO
	with results from the literature at $Q^2=100$~GeV${}^2$.
	We display  
	{\tt NNFF1}~\cite{Bertone:2017tyb} at NNLO,
	{\tt JAM19}~\cite{Sato:2019yez} at NLO, 
	{\tt DSEHS}~\cite{deFlorian:2014xna} at NLO, 
 with their uncertainties at $Q^2 = 100$~GeV$^2$. 
 Note,  discretion is necessary when interpreting the very low~$z$ region  as
 the extrapolation of the FF grids extends beyond the
 region fitted in the individual analyses. 
For example, the JAM19 focus was on SIDIS in the region $z\gtrsim 0.2$,
and NNFF1 used a lower kinematic cut of $z_{min} {=} 0.02$ for $Q{=}M_Z$ and $0.075$ for $Q<M_Z$.
While Fit~E is our preferred fit, we also display Fit~B to highlight the impact of the low~$z$ cuts.
\label{fig:FFs-other-1}}
\end{figure*}
}%

\def\figCompLitHighQ{
\begin{figure*}[tbh]
	\includegraphics[width=0.33\textwidth]{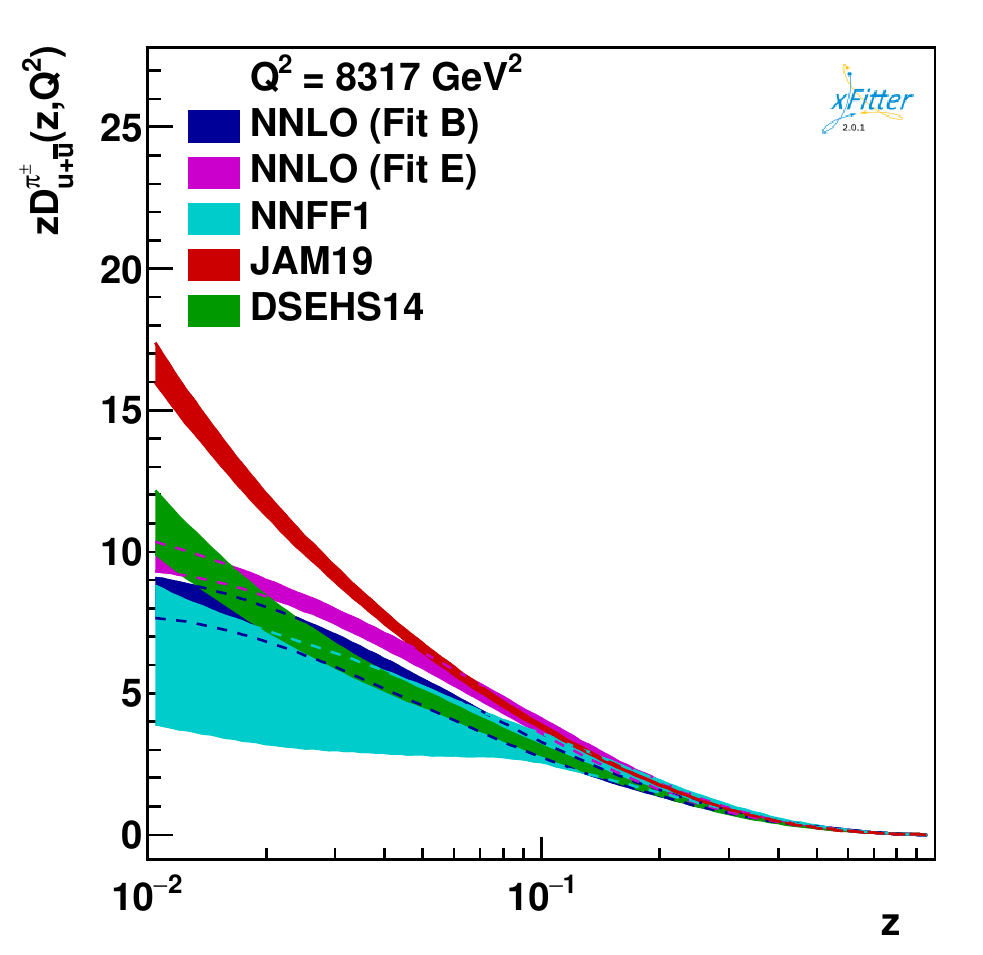}
\includegraphics[width=0.33\textwidth]{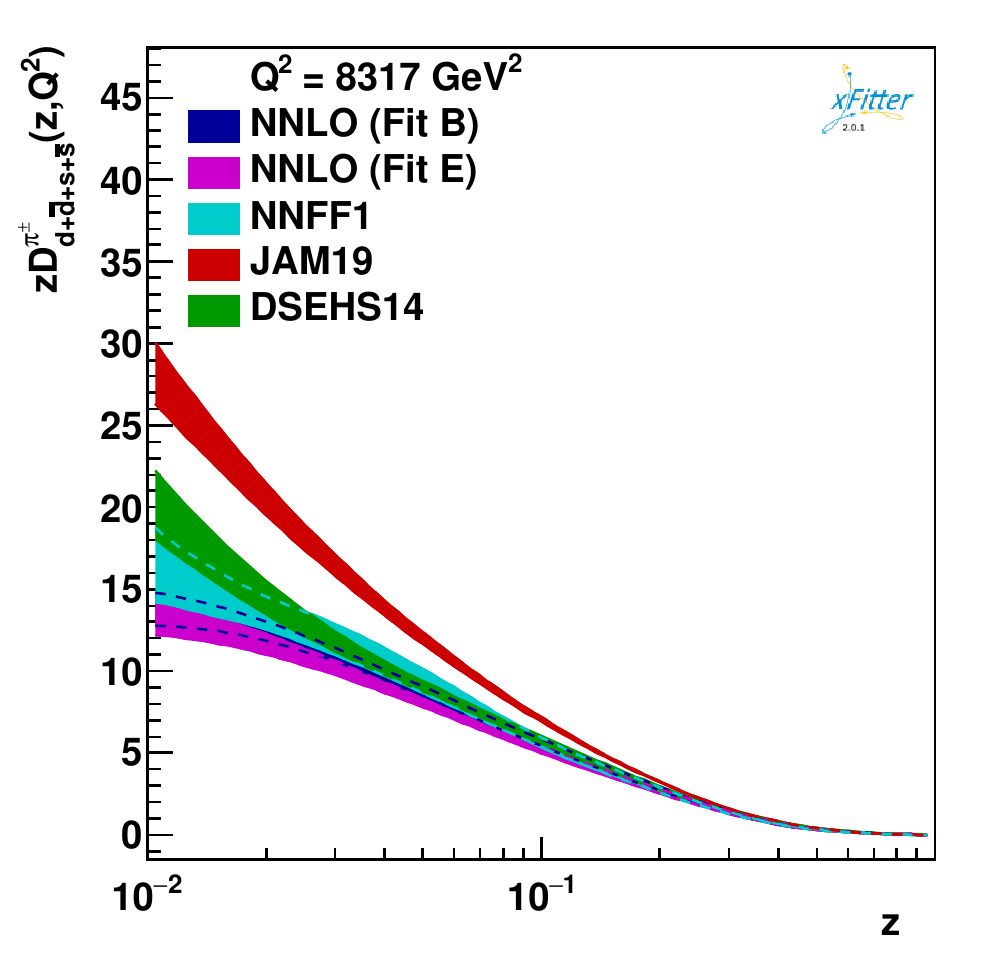}
\includegraphics[width=0.33\textwidth]{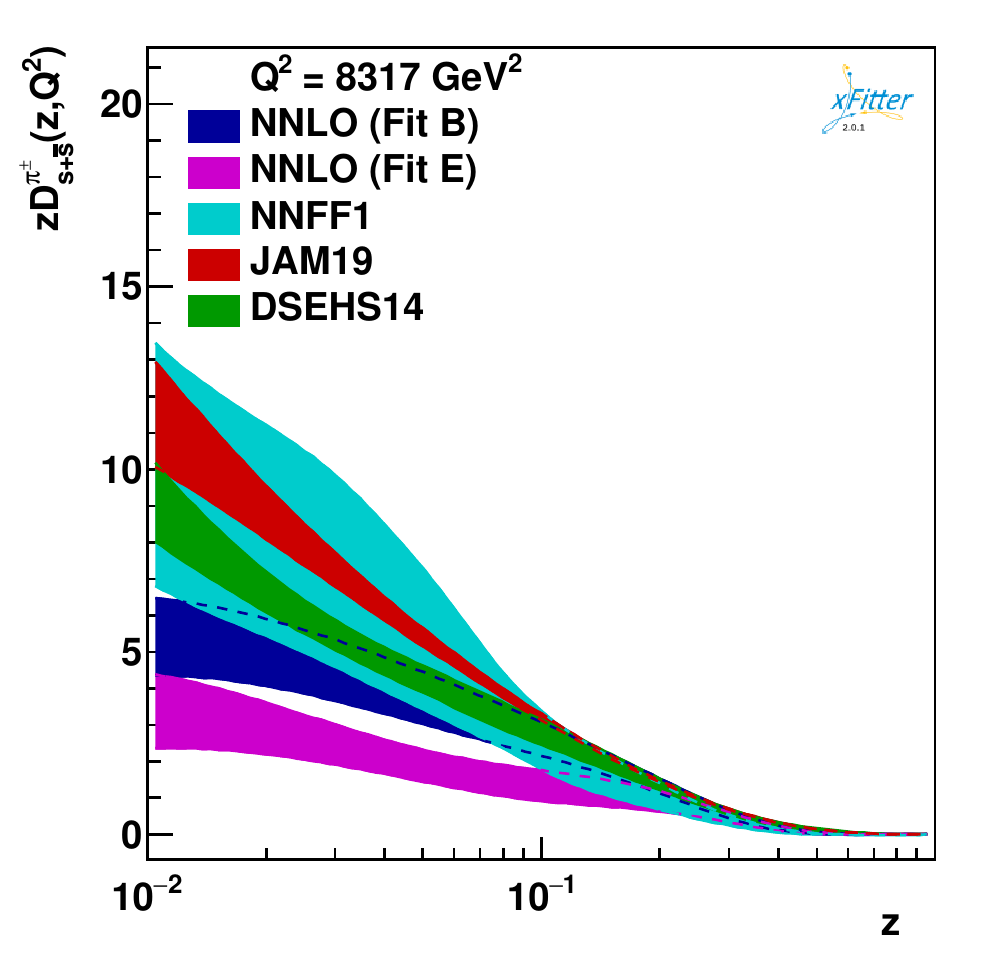}
\\
\includegraphics[width=0.33\textwidth]{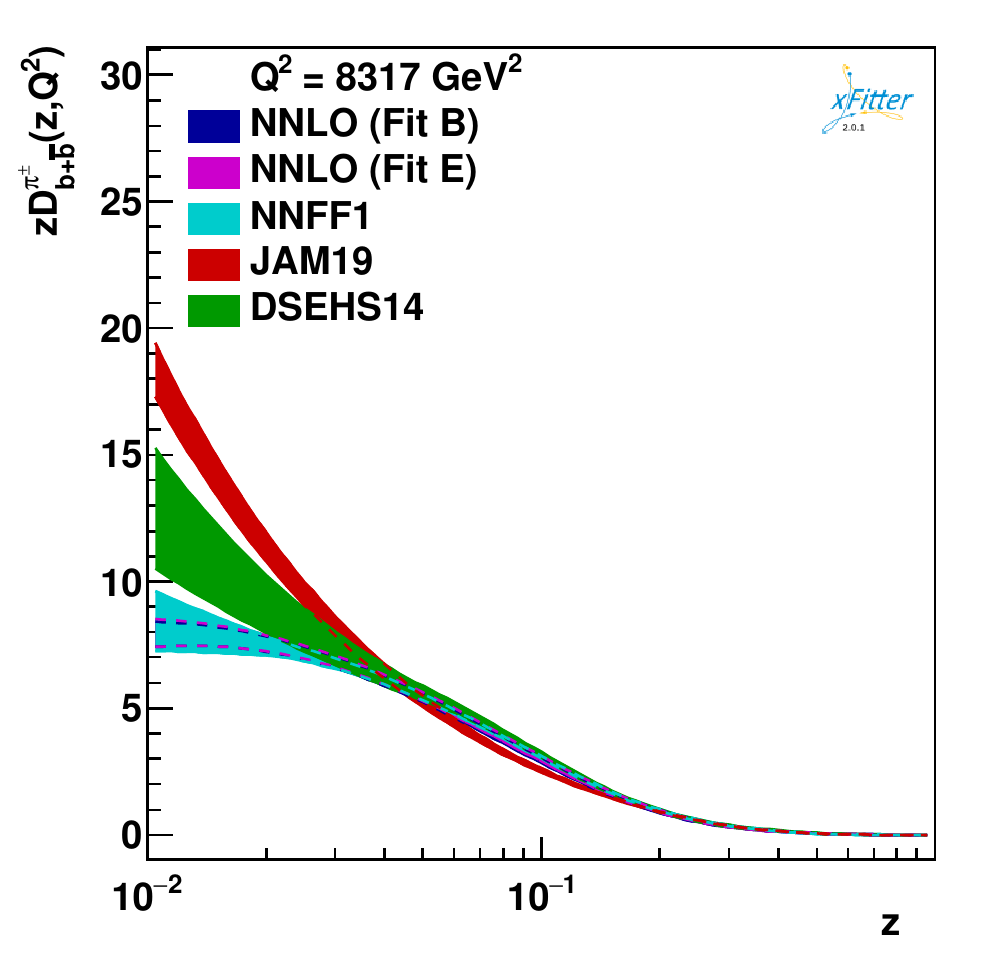}
\includegraphics[width=0.33\textwidth]{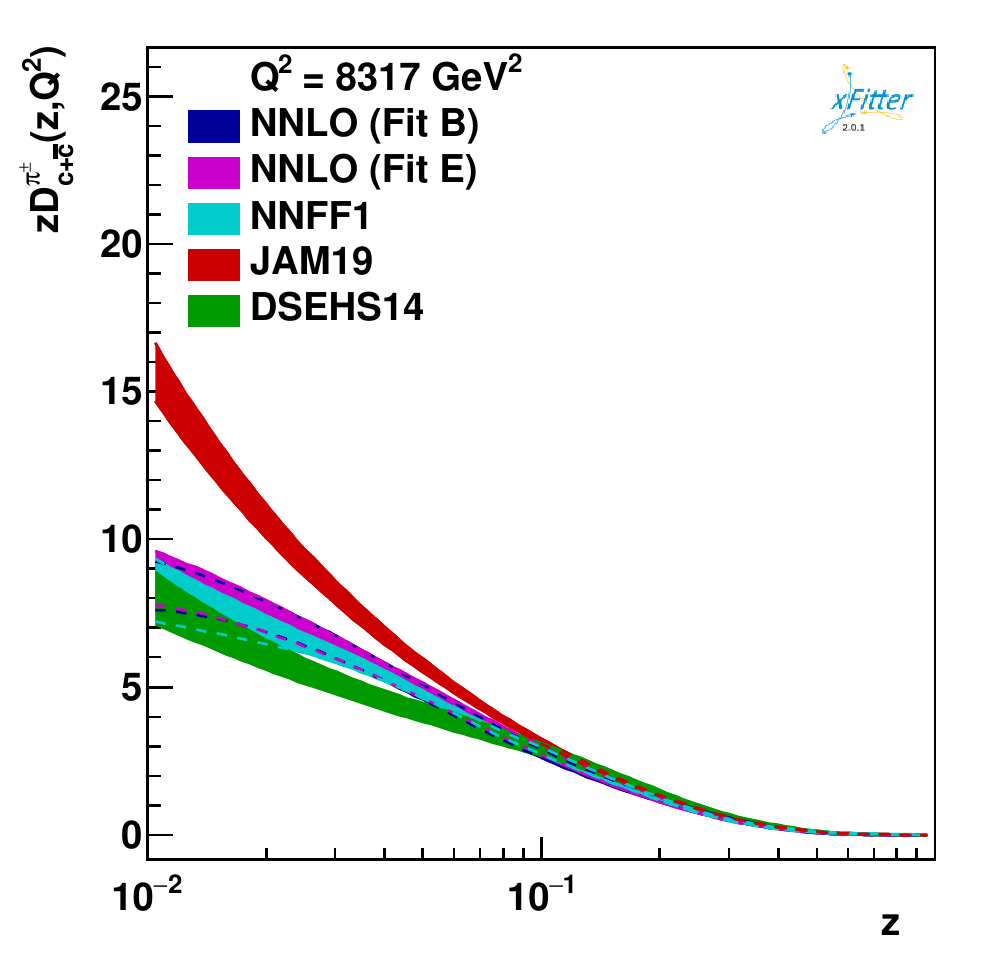}
\includegraphics[width=0.33\textwidth]{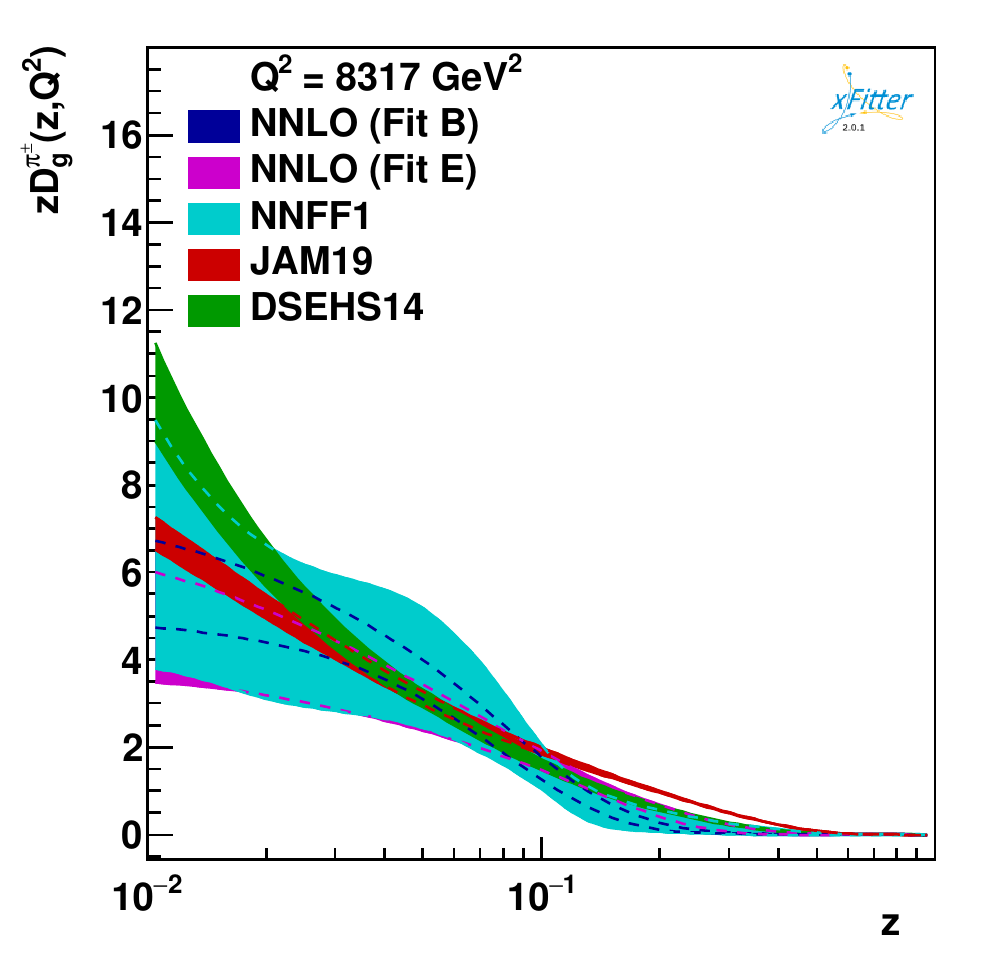}
\caption{  
Same as Fig.~\ref{fig:FFs-other-1} but   for $Q^2 = M_Z^2$. 
\label{fig:FFs-other-2}}
\end{figure*}
}%

\def\dataFigi{
\begin{figure*}[!ptb]
	\includegraphics[width=0.33\textwidth]{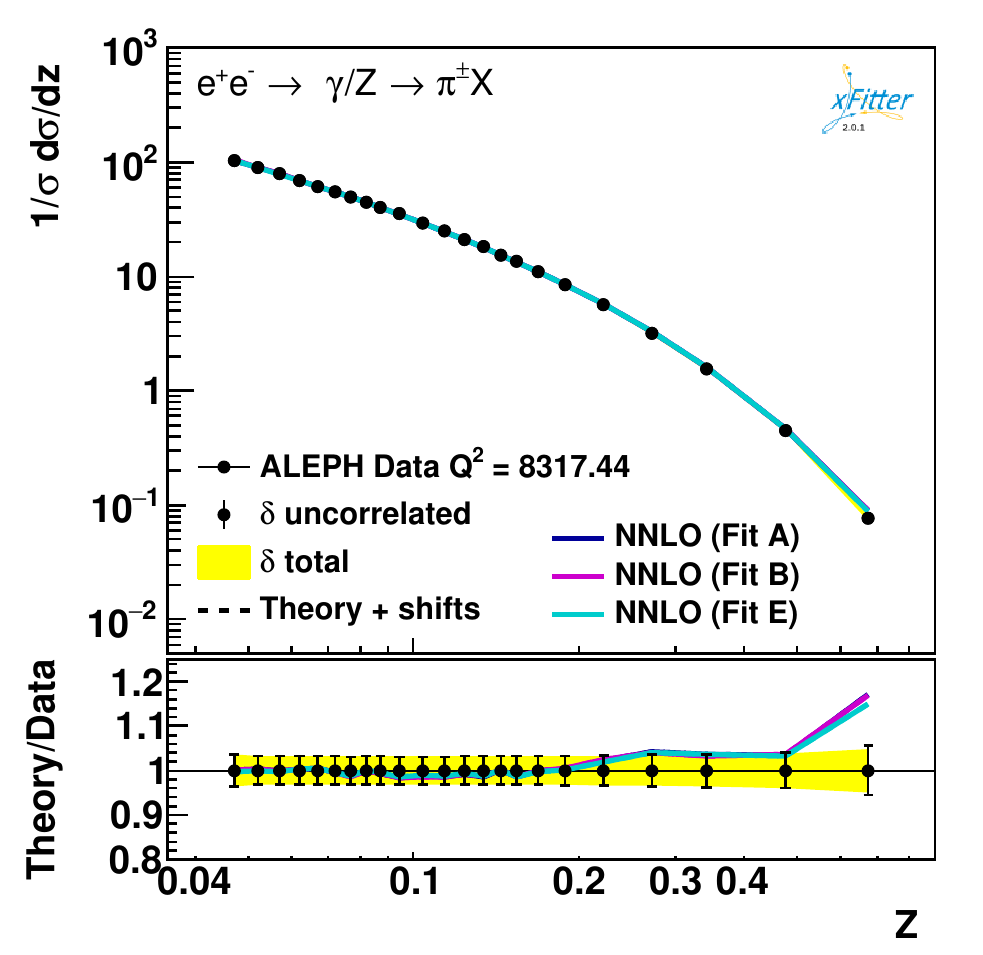}
	\includegraphics[width=0.33\textwidth]{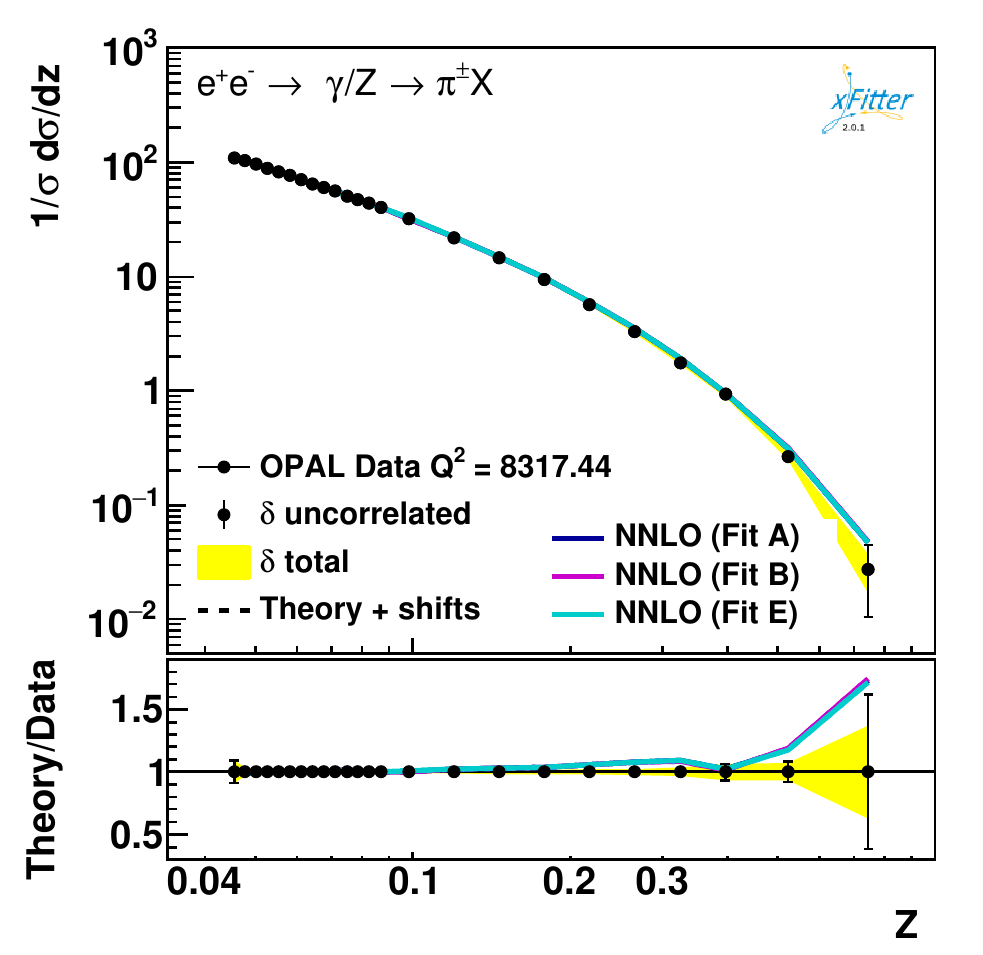}
	\includegraphics[width=0.33\textwidth]{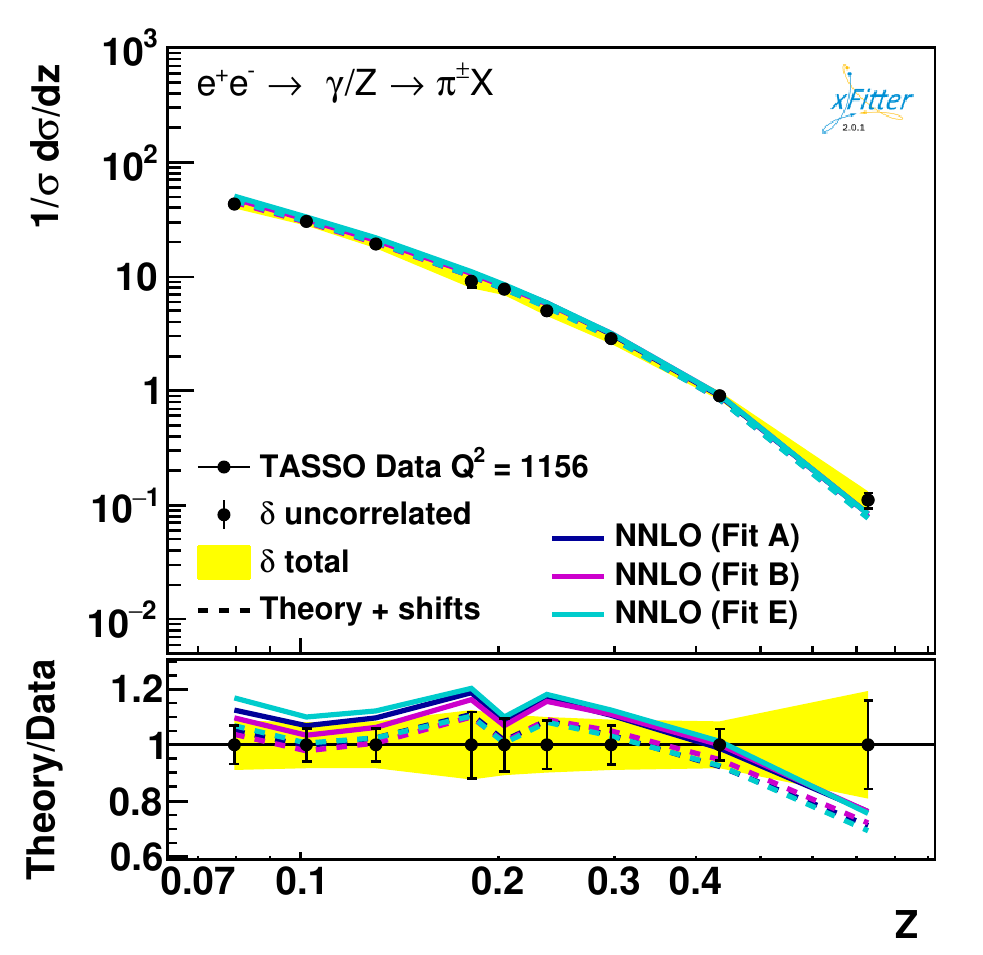}
	\\
	\includegraphics[width=0.33\textwidth]{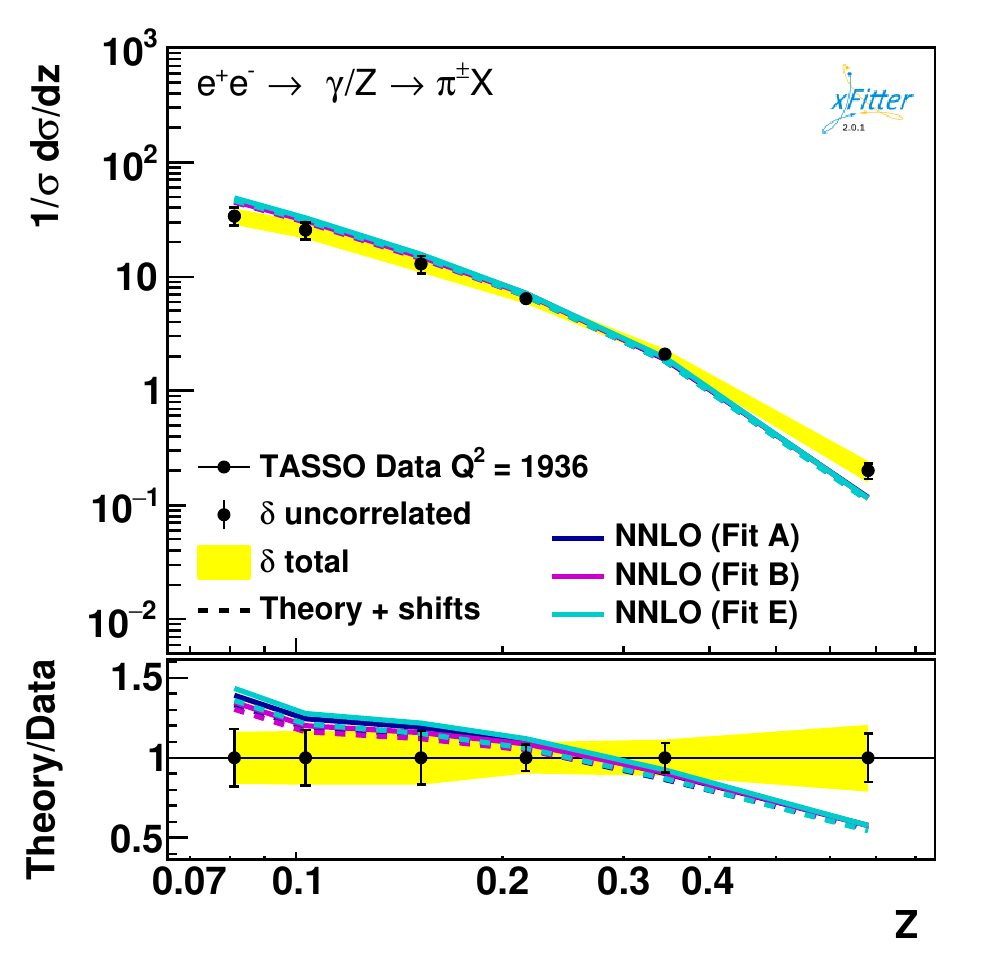}
	\includegraphics[width=0.33\textwidth]{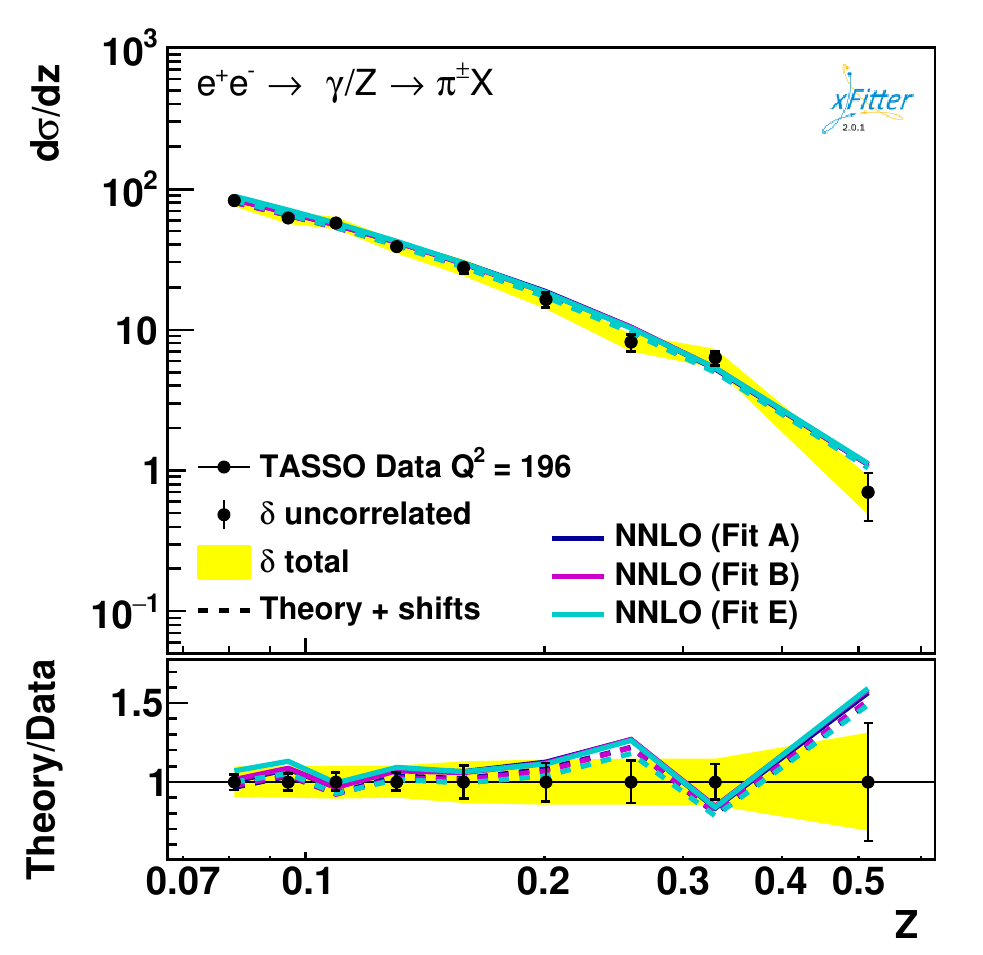}
	\includegraphics[width=0.33\textwidth]{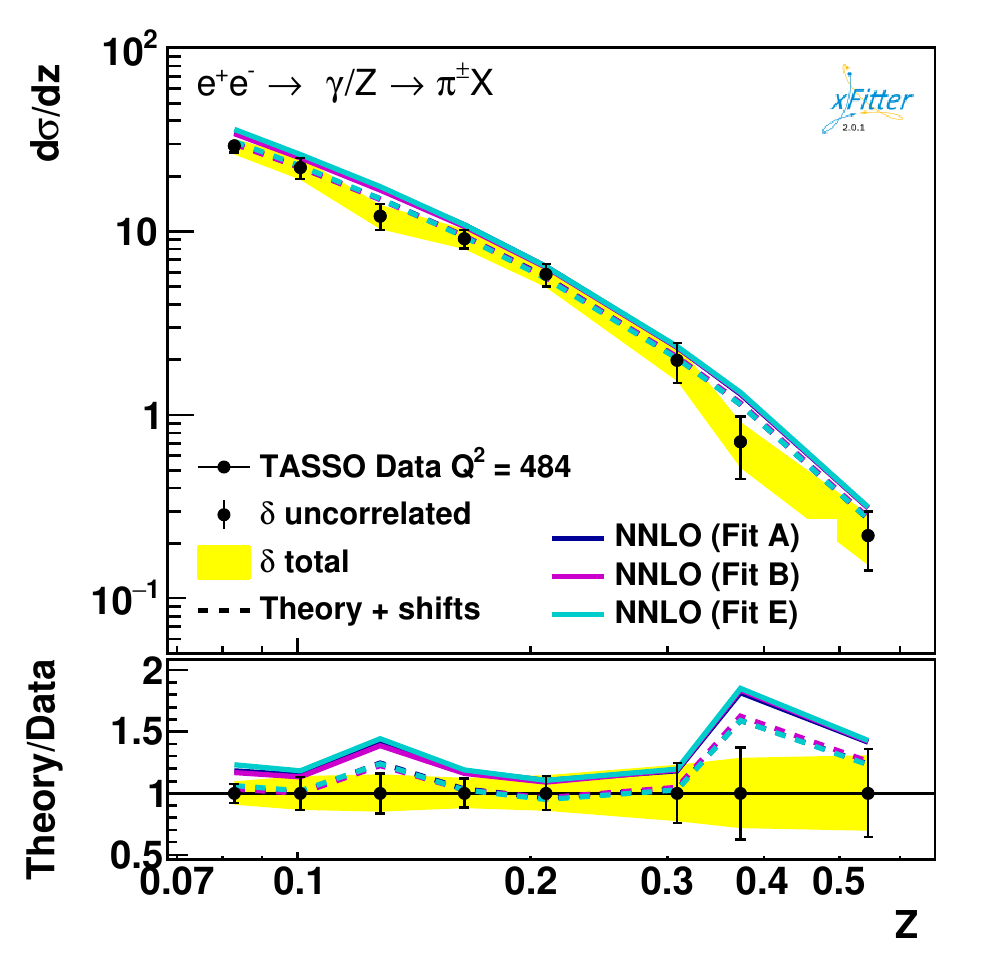}
	\\
	\includegraphics[width=0.33\textwidth]{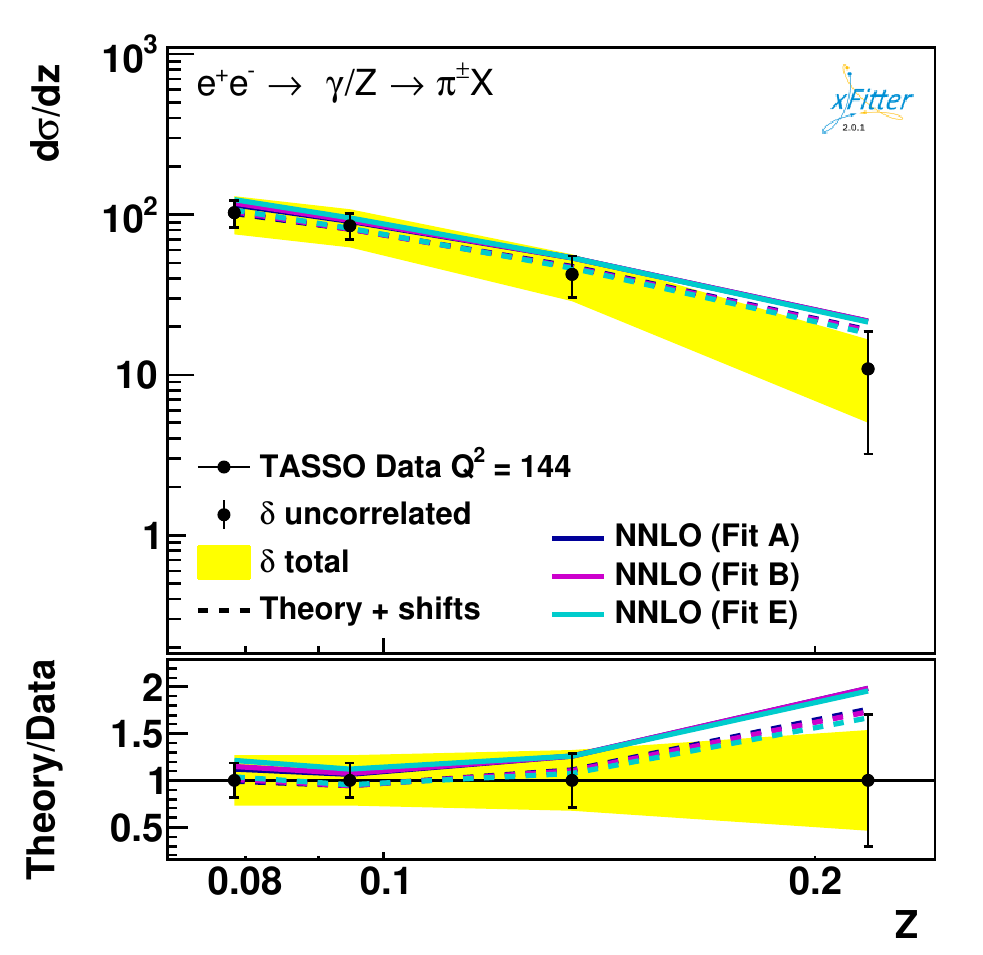}
	\includegraphics[width=0.33\textwidth]{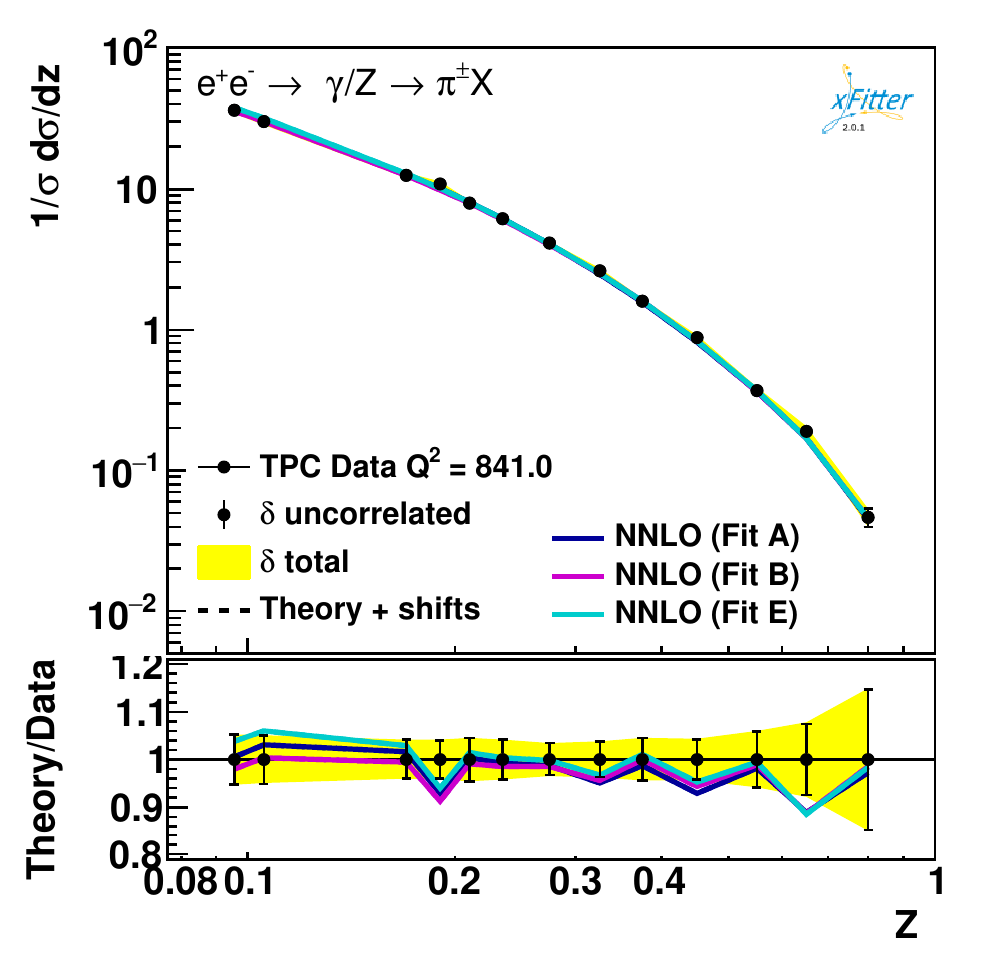}
	\includegraphics[width=0.33\textwidth]{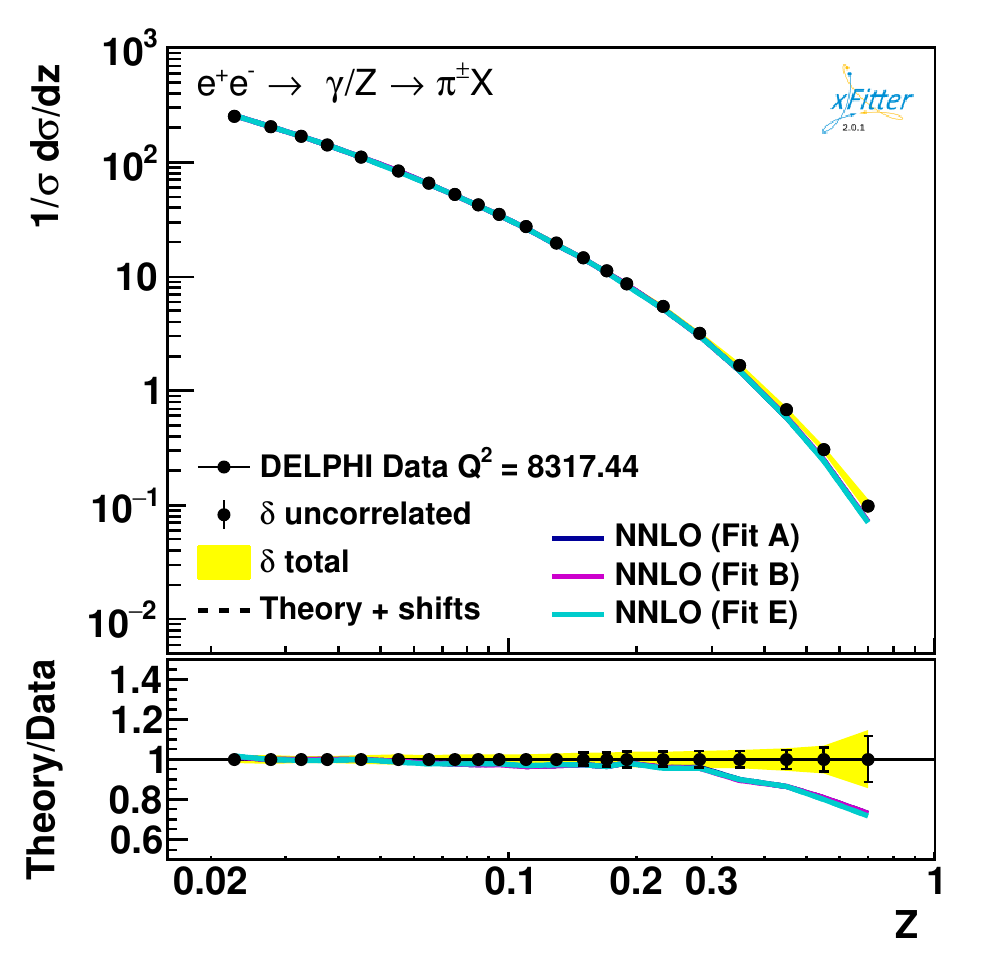}
\\
	\caption{The inclusive pion production as a function of $z$ 
	for fixed $Q^2$ values (displayed)  for  Fit~A, Fit~B and Fit~E at NNLO.
	We compare with experimental data from  
	ALEPH~\cite{Buskulic:1994ft}, 
	OPAL~\cite{Akers:1994ez}, 
	TASSO~\cite{Brandelik:1980iy,Althoff:1982dh,Braunschweig:1988hv}, 
	TPC~\cite{Aihara:1988su}, 
	DELPHI~\cite{Abreu:1998vq}, 
	SLD~\cite{Abe:2003iy}, 
	BaBar~\cite{Lees:2013rqd} 
	BELLE13~\cite{Leitgab:2013qh} 
	and 
	BELLE20~\cite{Seidl:2020mqc}.  
	We show both the absolute distributions (upper panel) 
	and the data over theory ratios (lower panel).
	The theory shifts arise from the correlated systematic overall normalization of the data points
	in the case where an individual experiment provides multiple data sets,
	such as TASSO,  DELPHI, and SLD ({\it c.f.}, Table~\ref{tab-SIA-chi2}).
	}
	\label{fig:Compare-with-SIA-1-a}
	\ContinuedFloat
\end{figure*}
}%

\def\dataFigii{
\begin{figure*}[tbh]
    \includegraphics[width=0.33\textwidth]{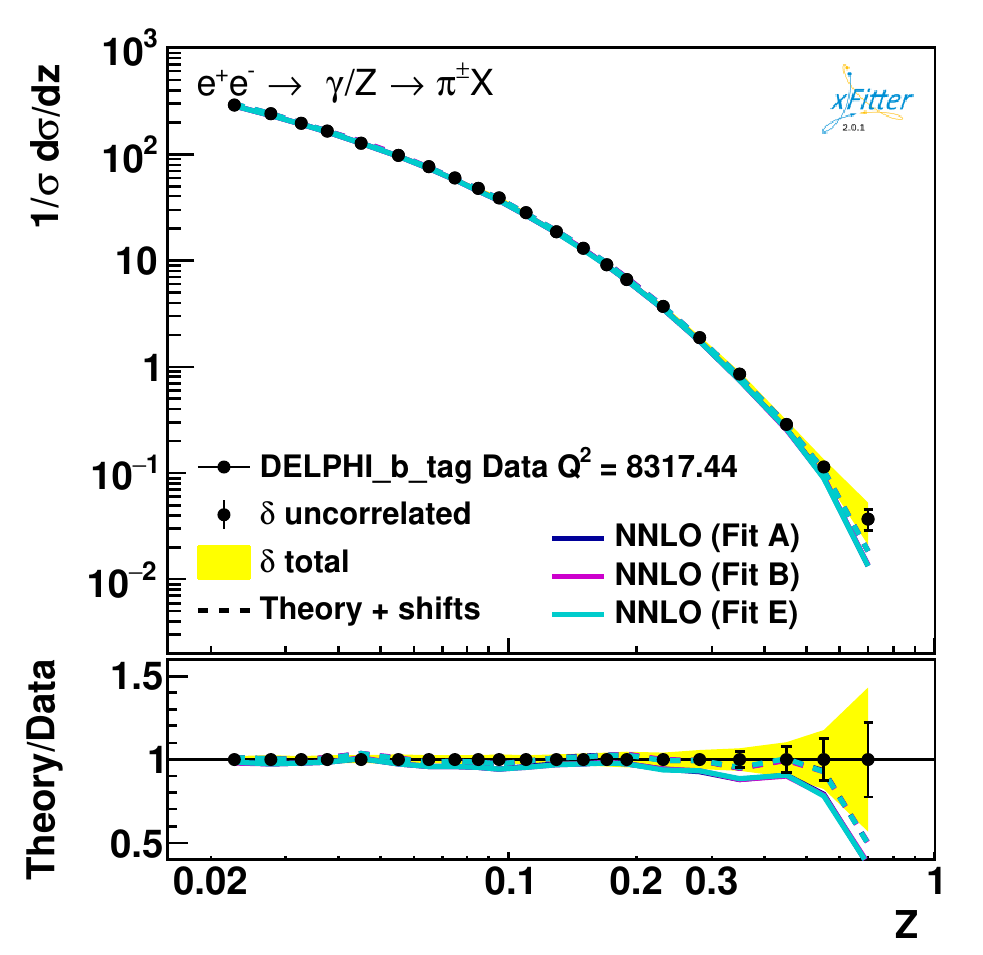}
	\includegraphics[width=0.33\textwidth]{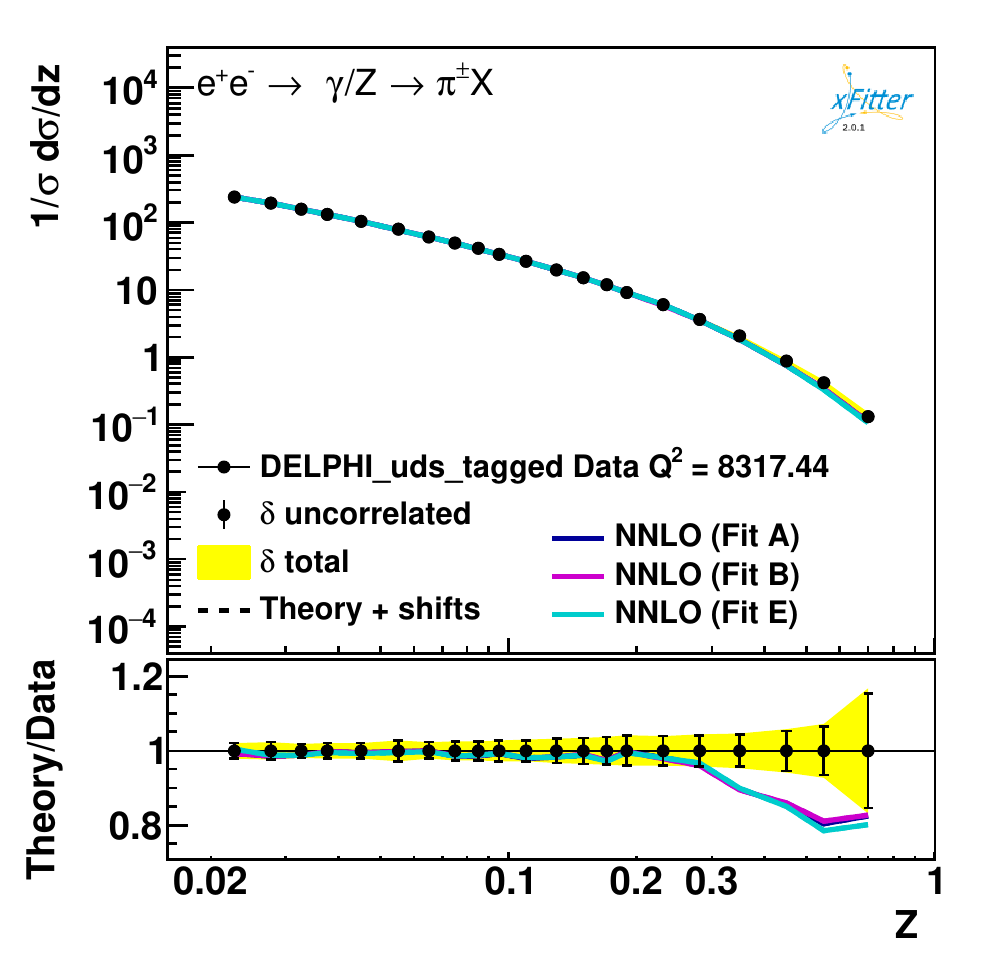}
	\includegraphics[width=0.33\textwidth]{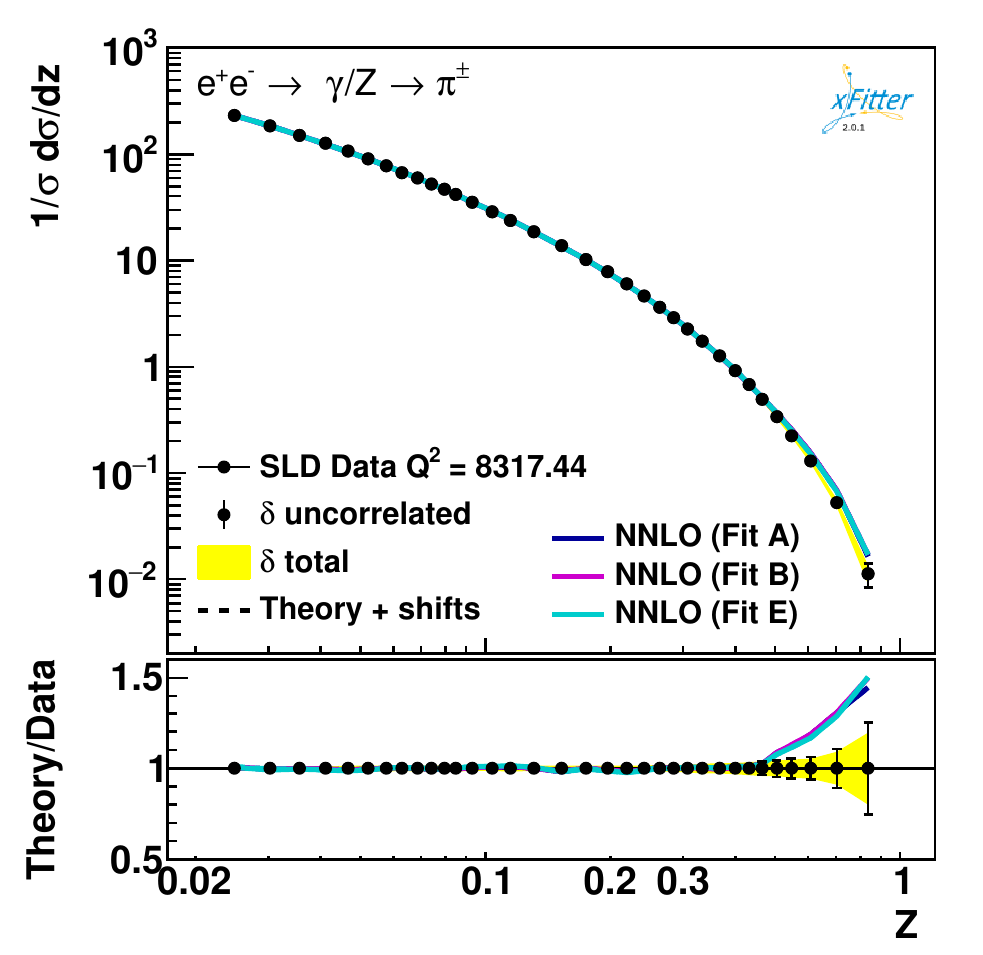}
\\
	\includegraphics[width=0.33\textwidth]{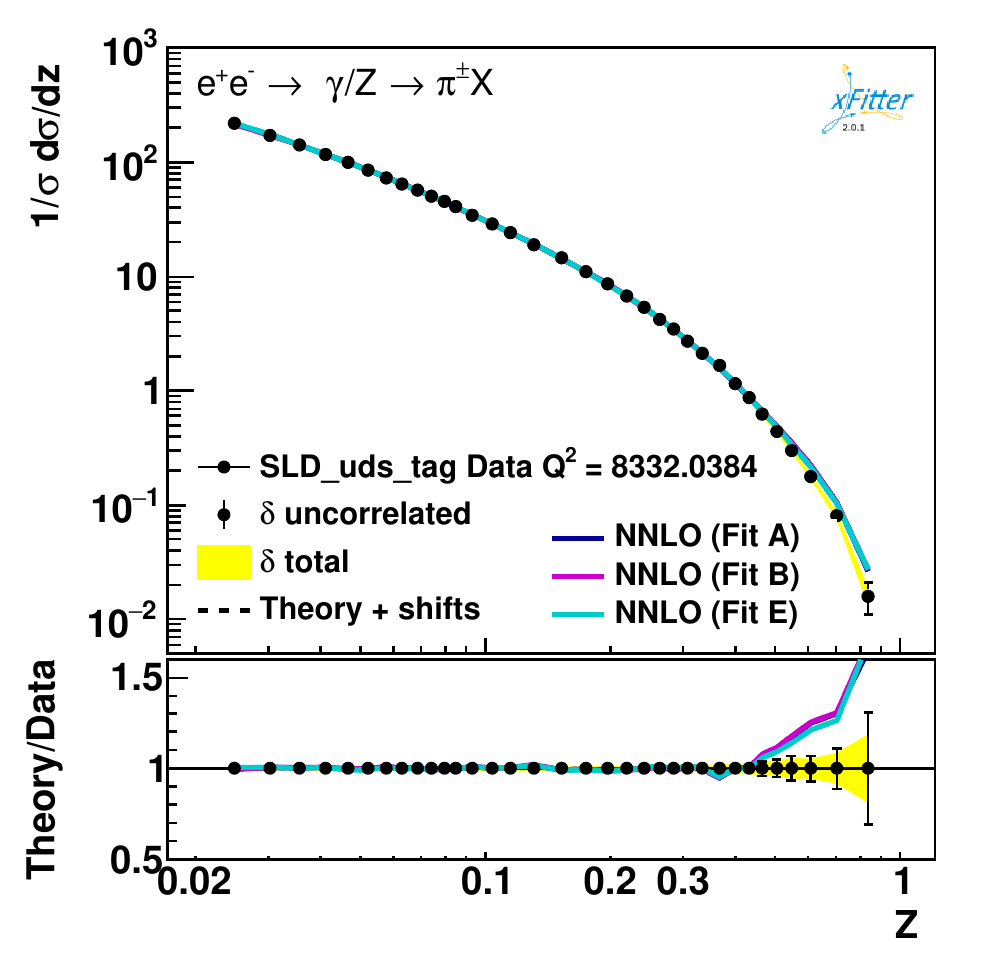}
	\includegraphics[width=0.33\textwidth]{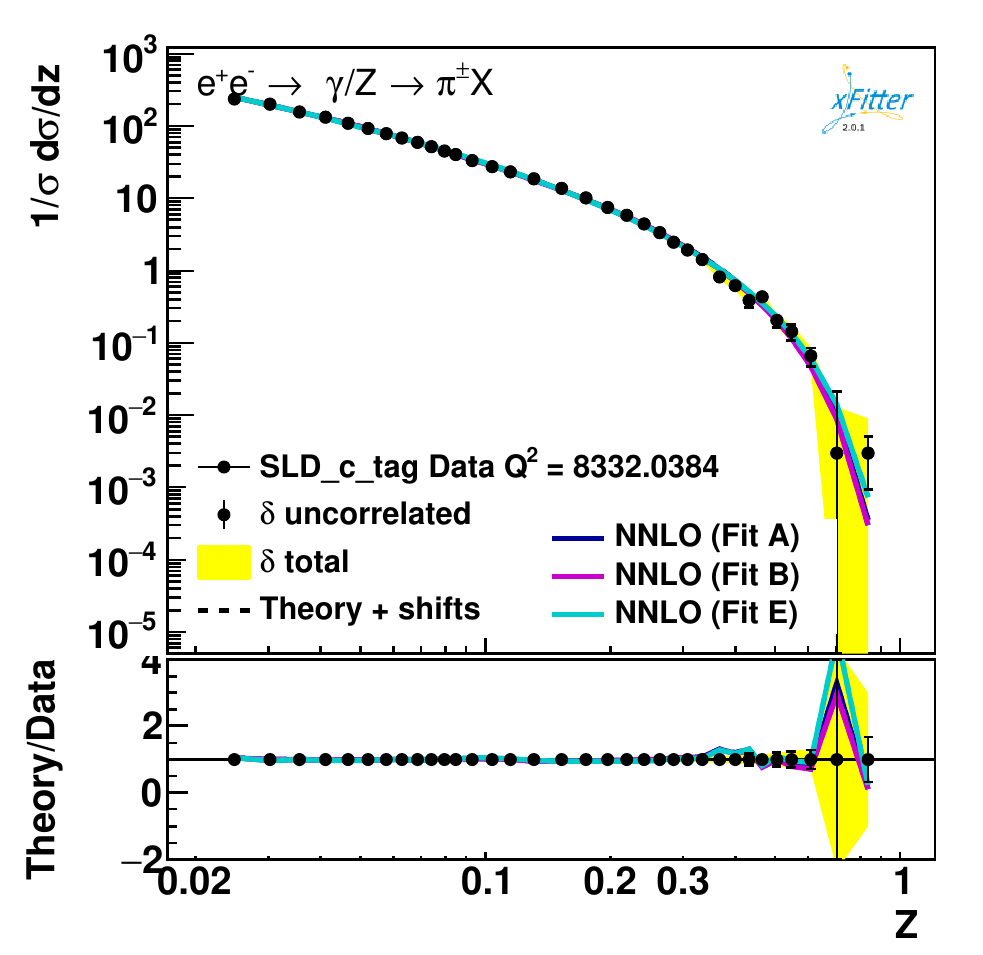}
	\includegraphics[width=0.33\textwidth]{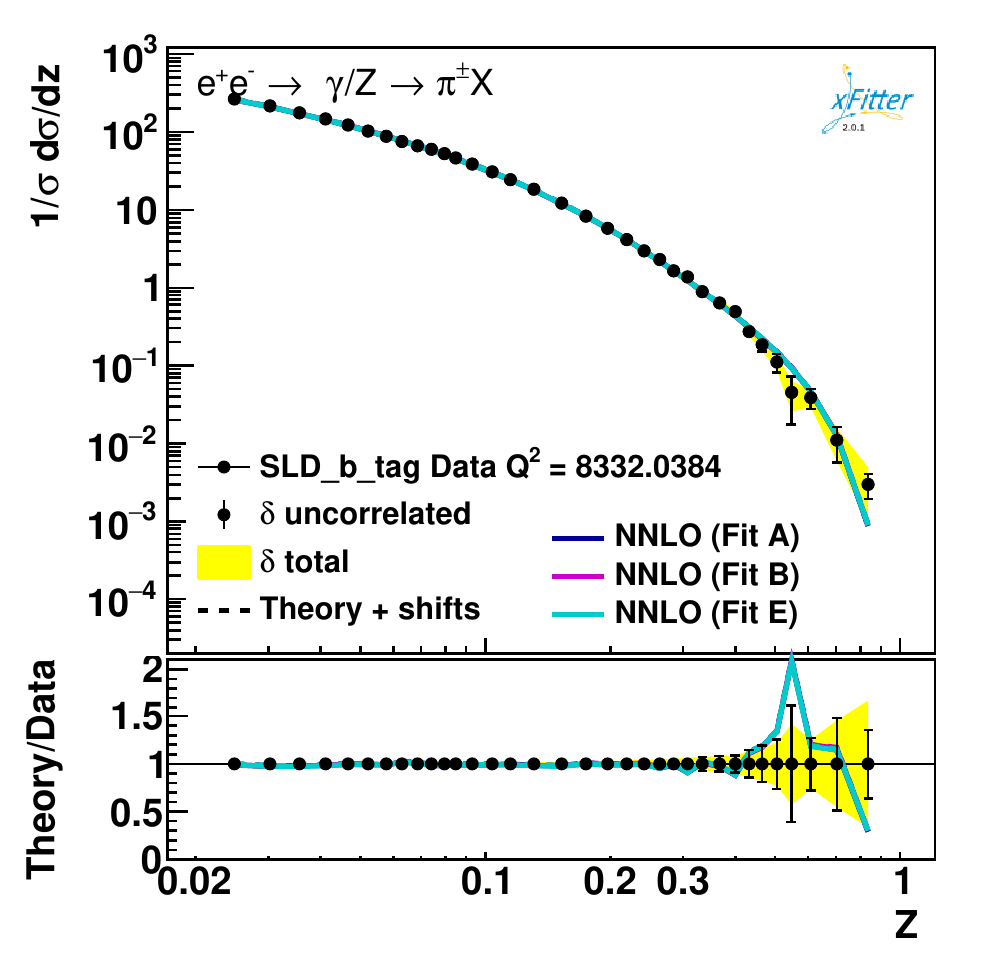}
\\	
	\includegraphics[width=0.33\textwidth,height=0.33\textwidth]{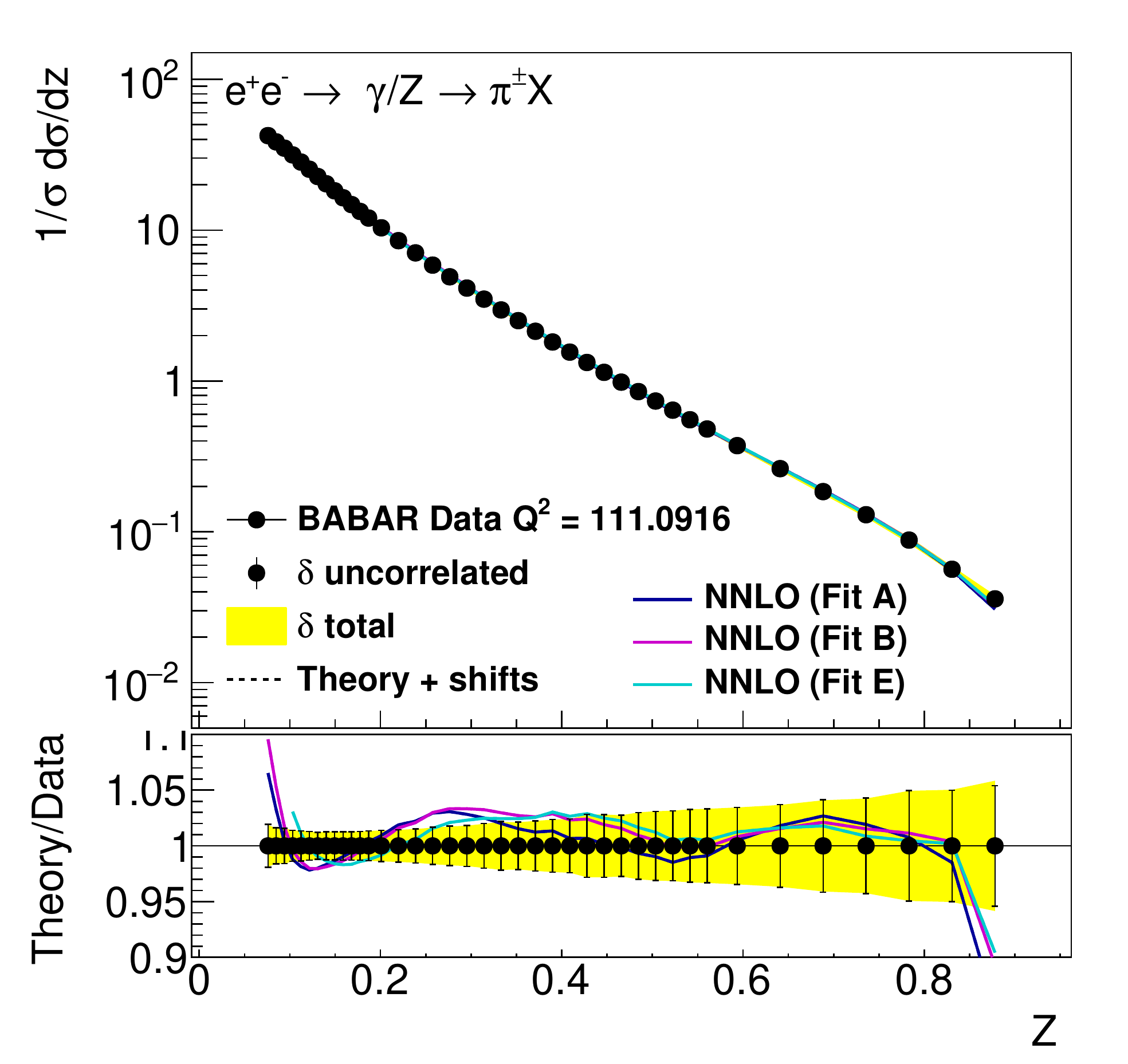}
	\includegraphics[width=0.33\textwidth,height=0.33\textwidth]{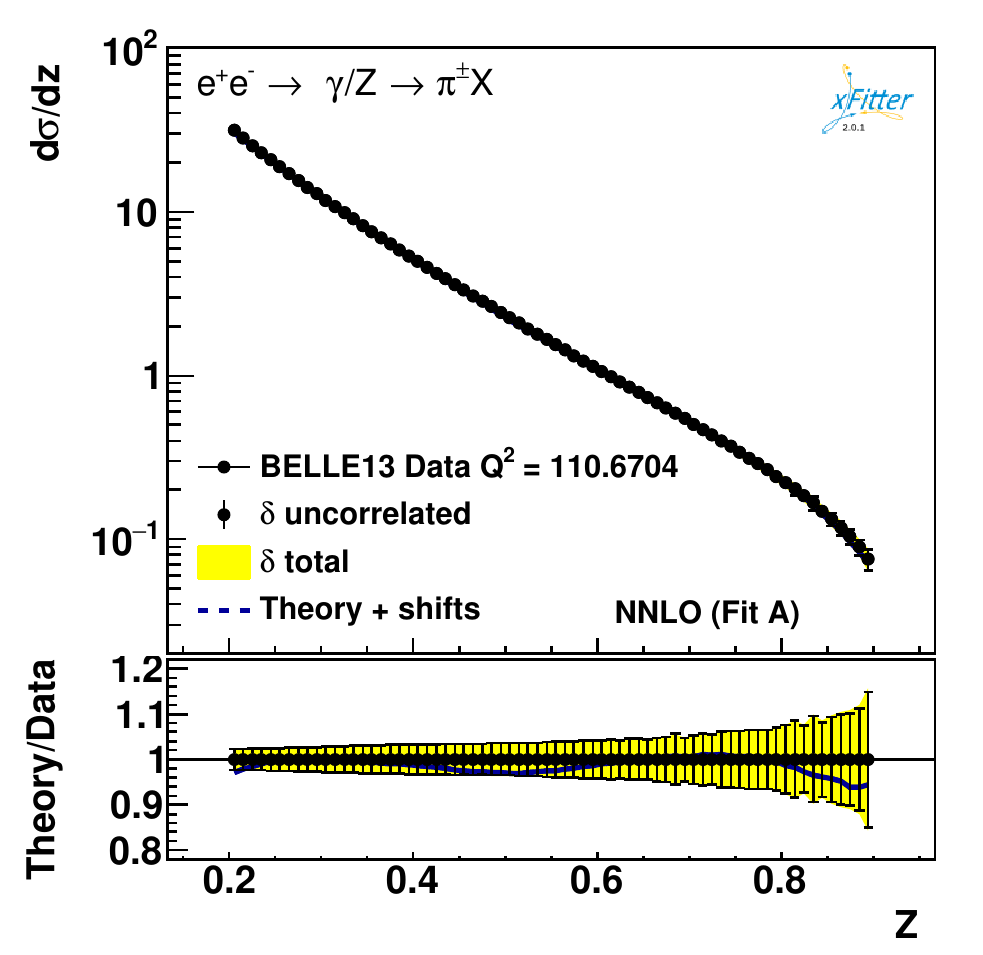}
	\includegraphics[width=0.33\textwidth,height=0.33\textwidth,]{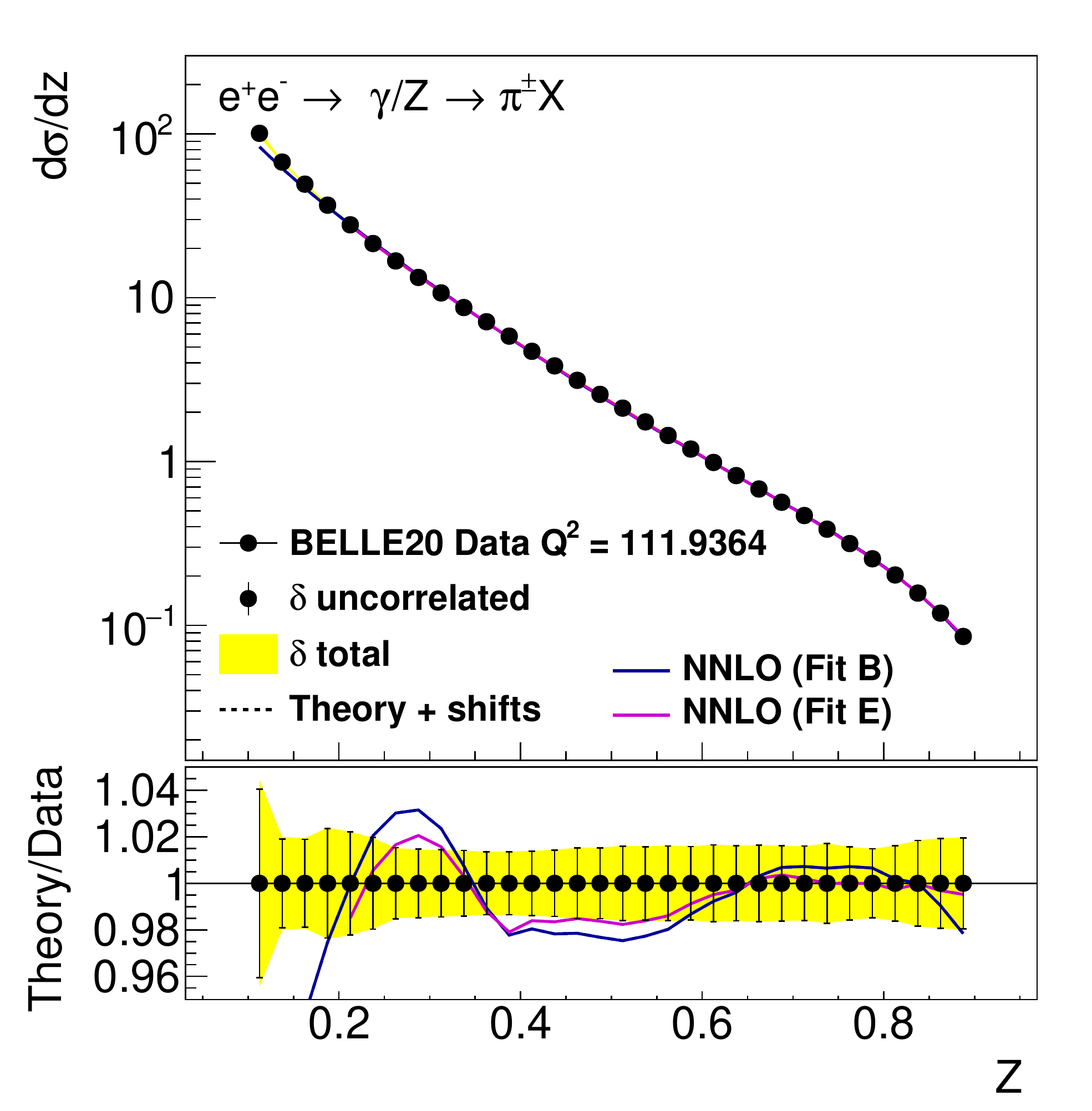}
\\
	\caption{Continuation of Fig.~\ref{fig:Compare-with-SIA-1-a}.
	}
\end{figure*}
}%

\def\tabChi{
\squeezetable
\begin{table*}[!p]
	\caption{
	The Single Inclusive electron-positron Annihilation (SIA)
		 data sets used in the   pion FFs analysis. 
		The values of $\chi^2$ per $N$ data points  for the individual SIA experiments  are shown.
		The $z$ range for each experiment is displayed in Fig.~\ref{fig:Compare-with-SIA-1-a}.
		The measured observable is also listed where $\sqrt{s}$ is the total CMS energy, 
	    $\beta=p_h/E_h$, and $z=2E_h/\sqrt{s}$.
 }
\begin{ruledtabular}
\begin{tabular}{lllcccccc}
 &  &  & \multicolumn{6}{c}{$\chi^{2}$/number of points}\\
\cline{4-9} 
Observable  & Experiment  & $\sqrt{s}$ & Fit~A  & Fit~A  & Fit~B  & Fit~C  & Fit~D  & Fit~E \\
 &  & [GeV] & (NLO)  & (NNLO)  & (NNLO) & (NNLO)  & (NNLO) & (NNLO) \\
\hline 
{\vrule height 10pt depth 0pt width 0pt}   %
$\frac{1}{\sigma_{tot}}\frac{d\sigma^{h}}{dz}$  & ${\rm SLD}$  & 91.20 & 57/34  & 41/34  & 41/34  & 48/34  & 39/34  & 45/34 \\
$\frac{1}{\sigma_{tot}}\frac{d\sigma^{h}}{dz}|_{uds}$  & ${\rm SLD_{uds}}$  & 91.20 & 66/34  & 52/34  & 56/34  & 44/34  & 43/34  & 45/34 \\
$\frac{1}{\sigma_{tot}}\frac{d\sigma^{h}}{dz}|_{c}$  & ${\rm SLD_{c}}$  & 91.20 & 35/34  & 33/34  & 32/34  & 32/34  & 32/34  & 32/34\\
$\frac{1}{\sigma_{tot}}\frac{d\sigma^{h}}{dz}|_{b}$  & ${\rm SLD_{b}}$  & 91.20 & 25/34  & 24/34  & 24/34  & 24/34  & 23/34  & 24/34\\
$\frac{1}{\sigma_{tot}}\frac{d\sigma^{h}}{dp_{h}}$  & ${\rm OPAL}$  & 91.20 & 42/24  & 41/24  & 41/24  & 39/24  & 39/24  & 39/24 \\
$\frac{1}{\sigma_{tot}}\frac{d\sigma^{h}}{dp_{h}}$  & ${\rm DELPHI}$  & 91.20 & 37/21  & 41/21  & 41/21  & 44/21  & 44/21  & 43/21 \\
$\frac{1}{\sigma_{tot}}\frac{d\sigma^{h}}{dp_{h}}|_{uds}$  & ${\rm DELPHI_{uds}}$  & 91.20 & 25/21  & 27/21  & 26/21  & 30/21  & 31/21  & 30/21 \\
$\frac{1}{\sigma_{tot}}\frac{d\sigma^{h}}{dp_{h}}|_{b}$  & ${\rm DELPHI_{b}}$  & 91.20 & 20/21  & 20/21  & 21/21  & 19/21  & 20/21  & 19/21 \\
$\frac{1}{\sigma_{tot}}\frac{d\sigma^{h}}{dz}$  & ${\rm ALEPH}$  & 91.20 & 21/23  & 14/23  & 14/23  & 11/23  & 11/23  & 12/23\\
$\frac{1}{\sigma_{tot}}\frac{d\sigma^{h}}{dz}$  & ${\rm TASSO44}$  & 44.00 & 15/6  & 17/6  & 15/6  & 18/6  & 16/6  & 18/6\\
$\frac{1}{\sigma_{tot}}\frac{d\sigma^{h}}{dz}$  & ${\rm TASSO34}$  & 34.00 & 6.8/9  & 8.0/9  & 6.8/9  & 9.3/9  & 7.3/9  & 8.3/9 \\
$\frac{1}{\beta\sigma_{tot}}\frac{d\sigma^{h}}{dz}$  & ${\rm TPC}$  & 29.00 & 6.3/13  & 11/13  & 11/13  & 11/13  & 7.1/13  & 9.2/13 \\
$\frac{s}{\beta}\frac{d\sigma^{h}}{dz}$  & ${\rm TASSO22}$  & 22.00 & 5.7/8  & 5.5/8  & 5.6/8  & 6.1/8  & 5.9/8  & 5.8/8 \\
$\frac{s}{\beta}\frac{d\sigma^{h}}{dz}$  & ${\rm TASSO14}$  & 14.00 & 11/9  & 11/9  & 11/9  & 9.9/9  & 11/9  & 9.8/9 \\
$\frac{s}{\beta}\frac{d\sigma^{h}}{dz}$  & ${\rm TASSO12}$  & 12.00 & 1.4/4  & 1.4/4  & 1.3/4  & 0.96/4  & 1.4/4  & 1.1/4 \\
$\frac{1}{\sigma_{tot}}\frac{d\sigma^{h}}{dp_{h}}$  & ${\rm BaBar}$  & 10.52 & 71/40  & 53/40  & 77/40  & -  & -  & 33/37 \\
$\frac{d\sigma^{h}}{dz}$  & ${\rm BELLE13}$  & 10.54 & 21/70  & 14/70  & -  & -  & -  & -\\
$\frac{d\sigma^{h}}{dz}$  & ${\rm BELLE20}$  & 10.58 & -  & -  & 82/32  & 32/32  & 9.2/28  & 17/28 \\[4pt]  %
\hline 
Correlated $\chi^{2}$  &  &  & 11  & 9.4  & 8.4  & 16  & 9.4  & 12 \\
Log penalty $\chi^{2}$  &  &  & +4.2  & +3.0  & +4.2  & +7.7  & +5.6  & +6.8\\
Total $\chi^{2}$/dof  &  &  & 480/386  & 427/386  & 518/348  & 404/308  & 357/304  & 410/341 \\
		\end{tabular}
		\label{tab-SIA-chi2}
	\end{ruledtabular}
\end{table*}
} %

\def\tableParm{
\begin{table*}[!p]
\caption{
		The best fit parameters with uncertainties obtained for the  pion FFs 
		 fits  at our initial scale $Q_0^2 = 25$~GeV$^2$.
		For all fits we use  $\alpha_{s}{=}0.1185$.
		}
\begin{ruledtabular}
\begin{tabular}{ l  l l l l l l}
Parameters & Fit A (NLO)& Fit A (NNLO)& Fit~B (NNLO)& Fit~C (NNLO)& Fit~D (NNLO)& Fit~E (NNLO)\\
\hline 
\hline 
$N_{g}$ & $0.559\pm0.021$& $0.564\pm0.068$& $0.43\pm0.13$& $0.346\pm0.092$ & $0.62\pm0.14$& $0.448\pm0.080$ \\
$\alpha_{g}$ & $0.91\pm0.25$& $0.22\pm0.80$& $0.8\pm3.0$& $0.26\pm0.80$ & $-0.27\pm0.43$& $-0.10\pm0.54$ \\
$\beta_{g}$ & $19.9\pm2.6$& $17.3\pm8.4$& $24\pm32$& $9.0\pm4.0$ & $8.5\pm3.2$& $9.6\pm4.0$ \\\hline 
$N_{u^{+}}$ & $1.107\pm0.021$& $1.075\pm0.093$& $1.036\pm0.082$& $1.399\pm0.100$ & $1.20\pm0.12$& $1.276\pm0.095$ \\
$\alpha_{u^{+}}$ & $-0.648\pm0.052$& $-0.69\pm0.19$& $-0.87\pm0.15$& $-1.098\pm0.088$ & $-0.99\pm0.12$& $-1.02\pm0.11$ \\
$\beta_{u^{+}}$ & $1.730\pm0.044$& $1.84\pm0.20$& $1.73\pm0.10$& $1.61\pm0.11$ & $1.61\pm0.12$& $1.64\pm0.11$\\ 
$\gamma_{u^{+}}$ & $7.37\pm0.91$& $7.4\pm2.6$& $4.0\pm2.3$& $4.1\pm1.0$ & $4.3\pm1.2$& $4.4\pm1.1$ \\
$\delta_{u^{+}}$ & $5.27\pm0.29$& $6.1\pm1.4$& $4.7\pm1.1$& $4.34\pm0.81$ & $3.87\pm0.79$& $4.58\pm0.76$ \\\hline 
$N_{s^{+}}$ & $0.348\pm0.046$& $0.45\pm0.14$& $0.66\pm0.19$& $0.22\pm0.12$ & $0.23\pm0.10$& $0.29\pm0.13$ \\
$\alpha_{s^{+}}$ & $1.03\pm0.34$& $0.75\pm0.77$& $-0.11\pm0.64$& $2.0\pm1.7$ & $1.9\pm1.3$& $1.4\pm1.1$ \\
$\beta_{s^{+}}$ & $9.89\pm0.83$& $8.5\pm2.2$& $7.8\pm3.6$& $10.7\pm3.7$ & $11.3\pm4.2$& $9.7\pm2.7$ \\\hline 
$N_{c^{+}}$ & $0.754\pm0.013$& $0.776\pm0.055$& $0.827\pm0.057$& $0.926\pm0.086$ & $0.777\pm0.088$& $0.856\pm0.070$ \\
$\alpha_{c^{+}}$ & $-0.843\pm0.037$& $-0.90\pm0.15$& $-0.93\pm0.13$& $-1.23\pm0.13$ & $-1.09\pm0.15$& $-1.14\pm0.14$ \\
$\beta_{c^{+}}$ & $4.99\pm0.18$& $5.03\pm0.65$& $5.13\pm0.62$& $4.11\pm0.58$ & $4.33\pm0.66$& $4.32\pm0.57$ \\\hline 
$N_{b^{+}}$ & $0.711\pm0.013$& $0.731\pm0.047$& $0.778\pm0.054$& $0.833\pm0.054$ & $0.722\pm0.068$& $0.788\pm0.051$\\ 
$\alpha_{b^{+}}$ & $-0.432\pm0.050$& $-0.47\pm0.19$& $-0.49\pm0.16$& $-0.75\pm0.13$ & $-0.65\pm0.17$& $-0.69\pm0.15$\\ 
$\beta_{b^{+}}$ & $4.05\pm0.24$& $4.08\pm0.78$& $4.03\pm0.84$& $3.89\pm0.68$ & $3.90\pm0.74$& $3.94\pm0.69$ \\
$\gamma_{b^{+}}$ & $12.5\pm1.7$& $12.7\pm5.2$& $13.9\pm6.8$& $8.9\pm3.3$ & $9.3\pm3.8$& $9.1\pm3.3$ \\
$\delta_{b^{+}}$ & $8.12\pm0.42$& $8.4\pm1.6$& $8.4\pm1.7$& $8.7\pm1.7$ & $8.5\pm1.8$& $8.5\pm1.7$ \\
		\end{tabular}
		\label{tab-parms}
	\end{ruledtabular}
\end{table*}
} 

\section{Introduction}\label{sec:introduction}

In the context of high-energy particle physics, Quantum Chromodynamics
(QCD) is a remarkably successful theory which governs the interactions
of strongly interacting particles in hard scattering processes.
Accurate predictions from the QCD theory are  essential to
provide a detailed description of the hadron structure.  This
information allows for incisive tests of the Standard Model (SM),
and facilitates searches for new physics.

Although the non-perturbative character of QCD imposes formidable
calculational challenges, perturbative QCD  circumvents this
difficulty by factorizing the physics into universal parton distributions
functions (PDFs) and fragmentation functions (FFs)  which encode 
the dynamical structure of the hadrons.

The PDFs and FFs must be determined phenomenologically by a global QCD
analysis of a diverse collection of hard scattering measurements
involving initial- and final-state hadrons.\footnote{%
Note that in recent years, significant advancements in Lattice QCD 
have enabled these calculations to contribute to 
constraining the PDFs and FFs in global analyses. For  recent reviews, see Refs.~\cite{Lin:2017snn,Lin:2020rut,Ji:2020ect}.
} %
The universal feature of the PDFs and FFs allows us to extract them 
in a global fit  if we have sufficient data to constrain these distributions.
Fortunately, we have extensive and precise measurements from a variety of
facilities including 
PETRA, SLC, LEP, BELLE, BaBar,
 Tevatron, HERA, and the Large Hadron Collider
(LHC).

In recent years, a large number of studies have performed global QCD
analyses of experimental data to obtain precise phenomenological
parametrizations of collinear (or integrated)
FFs~\cite{Bertone:2017tyb,deFlorian:2014xna,Sato:2016wqj,Soleymaninia:2020bsq},
as well as transverse momentum dependent FFs (TMD
FFs)~\cite{Scimemi:2019cmh,Soleymaninia:2019jqo,Jung:2021mox}.
Additionally, a recent analysis included the effects of medium
modified fragmentation functions (nFFs)~\cite{Zurita:2021kli}.
Furthermore, a simultaneous QCD analysis of both PDFs and FFs has
been performed by the {\tt JAM} Collaboration using inclusive and
semi-inclusive deep inelastic scattering (SIDIS), Drell-Yan, and
single inclusive electron-positron annihilation (SIA) experimental
data~\cite{Sato:2016wqj,Ethier:2017zbq,Sato:2019yez,Moffat:2021dji}.

Recent determinations of the pion FFs include results by several groups 
such as
{\tt HKNS}~\cite{Hirai:2007cx},
{\tt NNFF}~\cite{Bertone:2017tyb},
{\tt JAM}~\cite{Sato:2019yez,Sato:2016wqj},
{\tt SGKS}~\cite{Soleymaninia:2020bsq}, 
{\tt SGK}~\cite{Soleymaninia:2019sjo}, 
and other studies available in the literature 
including 
{\tt KKP}~\cite{Kniehl:2000hk,Kniehl:2000fe},
{\tt BKK}~\cite{Binnewies:1994ju},
and 
{\tt KLC}~\cite{Kretzer:2001pz}
These analyses differ in their choices of input functional form for the pion
FFs, the fitting methodology, the method of error estimation, the
input experimental data sets, 
and the order of the QCD perturbative expansion in the theory predictions.

The primary objective of the present study is to perform an extraction of 
the collinear
FFs for the pion (in a vacuum) using an exhaustive set of SIA data along with 
the very recent  pion production measurements by the BELLE collaboration (BELLE20)~\cite{Seidl:2020mqc}.
To help benchmark the impact of this new data set we will be especially
interested to compare the BELLE20 data with the previous 2013 BELLE measurement (BELLE13)~\cite{Leitgab:2013qh} 
as well as  with the BaBar~\cite{Lees:2013rqd} data. 
Additionally, we will  compare   our analysis at NLO and NNLO accuracy.

\xfitter{} is an open-source package that provides a framework for the
determination of the PDFs, and now FFs, for many different kinds of
analyses in QCD. The current \xfitter{} version 2.1 offers an expanded
set of tools and options, and incorporates experimental data from a
wide range of experiments~\cite{Alekhin:2014irh,Bertone:2017tig}.
\xfitter{} can analyze this data up to next-to-next-to-leading-order
(NNLO) in perturbation theory with a variety of theoretical
calculations.
While primarily based on the collinear factorization foundation,
\xfitter{} also provides facilities for fits of dipole models and
transverse-momentum dependent (TMD) PDFs.

\xfitter{} has been used for a wide range of  analyses involving different processes;
examples include:   
pion PDFs \cite{Novikov:2020snp}; 
nuclear PDFs \cite{Walt:2019slu}; 
and as well as various other studies~\cite{Alekhin:2014irh,Bertone:2017tig,Abdolmaleki:2019acd,Abdolmaleki:2019qmq,Abdolmaleki:2019tbb}.
This work   further extends the utility and capabilities of \xfitter{} 
to the  case of unpolarized collinear FFs of hadrons analysis. 
The pion FFs analysis presented in this paper represents the 
first step of a broader program to demonstrate this open-source \xfitter{} tool 
which can serve as the basis for  future analyses of both 
PDFs and FFs.

The structure of this paper is as follows. 
Sec.~\ref{sec:theoretical}  reviews 
the theoretical formalism for  $e^+ e^-$ annihilation into hadrons.
The analysis framework and methodology for the global QCD analysis is presented
in Sec.~\ref{sec:Analysis}.
Here we also  describe our methodology including 
our parametrization  and the
treatment of the FFs uncertainties. 
The SIA experimental data sets analyzed in this study are 
summarized in Sec.~\ref{sec:data_selection}. 
Sec.~\ref{sec:results} presents  the details of our results,
as well as comparisons  other  FFs   available in the literature.
Additionally, we discuss the impact of the new  BELLE20 data  on  the extracted pion FFs
and compare with results from the literature. 
Finally, we summarize our conclusions in Sec.\ref{sec:conclusion} 
and consider future applications that may prove fruitful for the \xfitter{} framework.

To facilitate further investigations of the pion FFs, both the FF grids and the code are publicly available on the \xfitter{} webpage (\texttt{xfitter.org}). 

This project is a collaborative effort between the 
Institute for Research in Fundamental Sciences (IPM) and 
 \xfitter{}; hence, we   shall identify our fit results 
as \ipmxx{}, or just \ipmx{} for short.

\section{Theoretical framework}\label{sec:theoretical}

The cross section for charged hadron production in an electron-positron 
annihilation process \mbox{$(e^{+} e^{-} \rightarrow h^{\pm} X)$} 
is typically measured as a function of the scaling variable 
\mbox{$z = 2 E_{h}/\sqrt{s}$} at a given center of mass energy of $\sqrt{s} = Q$
with hadron energy $E_h$.
The differential cross section is related to the hadronic
fragmentation function $F_2^{h^\pm}$ as\footnote{%
Here, we will follow the notation of Ref.~\cite{Bertone:2017tyb}, 
and the subscript on $F_2^{h^\pm}$ suggest an analogy with the $F_2$ DIS
hadronic structure function.
} %
\begin{eqnarray}\label{cross}
\frac{d\sigma^{h^\pm}}{dz}(z, Q) 
=
\sigma_0 \ 
F_2^{h^\pm}(z,Q) \quad ,
\end{eqnarray}
where $\sigma_0{=} 4\pi \alpha^2/Q^2$
and $\alpha$ is the electromagnetic coupling. 

The factorization theorem allows us to write the
non-perturbative hadronic fragmentation function $F_2^{h^\pm}$ as a convolution of 
a perturbative  coefficient function $C_{i}$ and
a \hbox{non-perturbative} partonic fragmentation function  $D_{i}^{h^\pm}$
given by:~\cite{Field:1977fa,Collins:1981uk,Collins:1981uw}
\begin{equation}\label{eq:structure}
F_2^{h^\pm}(z,Q)
=\sum_i C_{i} (z,\alpha_s(Q))
\otimes
D_i^{h^\pm}(z, Q)\ ,
\end{equation}
where we sum over parton flavors $i=q,\bar{q},g$.
The coefficient functions $C_{i}$
have been calculated up to the NNLO accuracy in the $\overline{\rm MS}$ 
scheme~\cite{Rijken:1996ns,Mitov:2006wy}.
The non-perturbative partonic FFs  $D_{i}^{h^\pm}$
are universal and represent the number density
for a hadron of type $h^\pm$ from  parton $i$ with momentum fraction $z$
at scale $Q$. 
It is the universal property of the FFs which will allow us to extract
these quantities by parameterizing their functional form and fitting 
to experimental data. 

To simplify the  expansion of the hard scattering cross-section, 
 we choose the renormalization scale $\mu_R$ and the  factorization scale $\mu_F$  
equal to the  center-of-mass energy;\footnote{%
To be more precise, in Eq.~(\ref{eq:structure}) 
the $\alpha_S(\mu_R)$ depends on the renormalization scale $\mu_R$, 
and the partonic fragmentation function $D_i^{h^\pm}(z, \mu_F)$ 
depends on the factorizaton scale $\mu_F$.
} %
thus, we have $\mu_R = \mu_F=\sqrt{s} \equiv  Q$.

The scale dependence of the partonic FFs is described 
by the DGLAP evolution equations,~\cite{Gribov:1972ri,Lipatov:1974qm,Altarelli:1977zs,Dokshitzer:1977sg}
\begin{equation}\label{DGLAP}
\frac{dD_i^{h^\pm}(z,Q)}
{d\ln (Q^2)}=[P_{ij}
\otimes D_{j}^{h^\pm}] 
(z,Q)\quad ,
\end{equation}
where $P_{ij}$ are the perturbative time-like splitting functions,
 and the convolution integral $\otimes$ is 
\begin{equation}\label{otimes}
[P\otimes D](z)=\int ^1_z 
\frac{dy}{y}P(y)
D\left(\frac{z}{y}\right)\quad .
\end{equation}
We solve the integro-differential DGLAP equations directly in $z$ space
using the {\tt APFEL} package~\cite{Bertone:2013vaa}
which provides NLO and NNLO accuracy.

Now that we have outlined the key elements of the SIA cross section calculation,
we next examine the framework for our analysis,
including the parameterization of the non-perturbative FFs. 

\section{Analysis framework}\label{sec:Analysis}

We will obtain the Fragmentation Functions (FFs) by
parameterizing their functional form in $z$ and then 
performing a fit by minimizing a $\chi^2$ function in comparison with 
experimental data. 
In the following,  we  detail the analysis 
framework  including the parameterization form,
the fitting procedure, and the uncertainty analysis.

\subsection{FFs parametrization and assumptions}

We  parametrize  the $z$ dependence of the FFs 
at an initial scale $Q_0 =5$~GeV which keeps us above $m_b$,
and use the DGLAP equations to evolve to arbitrary $Q$ scale.
The flexible parametric form we use is: 
\begin{eqnarray}
\label{eq:parametrization}
D^{\pi^\pm}_i(z, Q_{0})
= \frac{{\cal N}_i
z^{\alpha_i}(1 - z)
^{\beta_{i}}
[1 + \gamma_{i}
(1 - z)^{\delta_i}]}
{B[2 {+} \alpha_i, \beta_i {+} 1] {+}
  \gamma_i B[2 {+} \alpha_i, \beta_{i} {+} \delta_{i} {+} 1]
}\ ,
\nonumber \\
\end{eqnarray}
which has (maximally) five free parameters
$\{{\cal N}_i, \alpha_i,\beta_i,\gamma_i,\delta_i\}$ per parton flavor.
Here, $B[a,b]$ is the Euler beta function.
For the charged pion FFs,
we fit the flavor combinations  $i=u^+, d^+, s^+, c^+, b^+$ and $g$.
The beta functions in the denominator of Eq.~\ref{eq:parametrization} simply ensures
$\int_0^1 dz\, z D_i={\cal N}_i$.

There are a number of constraints we can impose to reduce the number of 
free parameters of the fit. 
From the  energy sum rule, we have the relation:
\begin{eqnarray}\label{eq:sumrule}
\sum _{h} \int ^1_0 dz \, z D^{h}_i
&=&
\sum_{h} {\cal N}_i^{h}
=1
\  ,
\end{eqnarray}
where $h$ sums over all possible produced hadrons.
For the pion FFs ($h=\pi^\pm$) this relation provides only an upper bound, 
but if we expect the lighter pions carry most of the parton momentum, then we have
\begin{eqnarray}\label{eq:Nsumrule}
{\cal N}_i^{\pi^\pm}&<& 1 \quad , 
\end{eqnarray}
where $i={g,q, \bar{q}}$.
Note, in Table~\ref{tab-parms} we report ${\cal N}_{u^+}$
where $u^+=u+\bar{u}$, hence the limit on this quantity is ${\cal N}_{u^+}<2$.

Thus, we will use four shape parameters  $\{\alpha_i,\beta_i,\gamma_i,\delta_i\}$ 
together with the normalization parameter ${\cal N}_i$  
to fit our FFs.

For the $\pi^+$ FFs,  we assume isospin symmetry for the favored 
($u,\bar{d}$) and unfavored ($\bar{u},d$) components~\cite{Hirai:2007cx,Bertone:2017tyb,Moffat:2021dji}:
\begin{eqnarray}\label{eq:assumm}
D^{\pi^+}_{u}
&=&
D^{\pi^+}_{\bar{d}}
\nonumber\\
D^{\pi^+}_{\bar{u}}
&=&
D^{\pi^+}_{d}\ .
\end{eqnarray}
We can also use charge conjugation to relate the
above  $\pi^+$ FFs  to the  $\pi^-$ FFs:
\begin{eqnarray}\label{eq:assumm_charge}
D^{\pi^+}_{i}=
D^{\pi^-}_{i},~~~
i=u^+, d^+, s^+, c^+, b^+,g\ .
\end{eqnarray}
Additionally, for the $s^+, c^+$ and $g$ flavors
we impose the condition $\gamma_i{=}\delta_i{=}0$.

We thus have  a total of $19$ free parameters to be determined
by a standard $\chi^2$ minimization strategy.
We now discuss the details of the fit.

\subsection{Fitting procedure}
By comparing the theoretical and experimental measurements
from our selected data sets, we determine the optimal values of the
19 independent fit parameters by minimizing the $\chi^2$ function
as implemented in the  \xfitter{} framework:~\cite{Alekhin:2014irh,James:1975dr,Abramowicz:2015mha}
\begin{equation}
\chi^2 = 
\sum_i 
\frac{ [ d_i - t_i(1- \sum_\alpha b_\alpha \delta^i_{\alpha}) ]^2 }
{\delta^2_{i, {\rm unc}}t_i^2  + \delta^2_{i, {\rm stat}}d_i t_i  } 
+\sum_{\alpha} b^2_{\alpha} 
\ ,
\label{eq:chi2}
\end{equation} 
where the sum on $i$ is over the number of data points. 
Here $t_i$ is the theory prediction, 
$d_i$ is the data measurement, 
$b_{\alpha}$ are the nuisance parameters,
and
$\delta_{i, stat}$ and $\delta_{i, unc}$ are the  statistical 
and uncorrelated systematic uncertainties.

In general, for this analysis we will add the correlated and uncorrelated
uncertainties in quadrature; while this allows a generous band of uncertainties on our FFs, we still find it challenging to describe all the data as
we detail in Sec.~\ref{sec:results}.
In the case where an individual experiment provides multiple data sets
({\it e.g.}, TASSO,  DELPHI, and SLD {\it c.f.}, Table~\ref{tab-SIA-chi2}),
the overall normalization of these data points is correlated; 
we take this into account as a systematic error. 
The effect of this correlated systematic normalization will be evident in Fig.~\ref{fig:Compare-with-SIA-1-a} which displays both the theory prediction
as well as the ``theory+shifts'' showing the impact of the correlated normalization.

\subsection{Uncertainty analysis}

In addition to a central value of the FFs and PDFs, it is important to also evaluate the 
uncertainty band of these quantities. 
For the present study, we will  apply the Hessian method to estimate the FFs uncertainty band~\cite{Alekhin:2014irh,Pumplin:2001ct,Pumplin:2000vx}.
The Hessian approach is based on a quadratic expansion of the $\chi^2$ function 
about its global minimum and can be expressed as:
\begin{eqnarray}
\chi^2 &=& \chi^2_0 + \Delta \chi^2 
= \chi^2_0 + 
\sum_{i,j} H_{ij} \, y_i \,  y_j\ ,
\end{eqnarray}
where $i,j$ run over the 19 free parameters of the fit. 
Here, $\chi^2_0$ is the global minimum, 
$\Delta \chi^2$ is the  displacement from the minimum,
and
$y_i$ are the parameter shifts about the minimum.

The Hessian matrix is constructed from the second derivatives of $\chi^2$ and defined as
$H_{ij} {=} \frac{1}{2} (\partial^2 \chi^2)/( \partial y_j  \partial y_j)$.
The Hessian  is a symmetric $n{\times} n$ matrix where $n{=}19$ is the number of free parameters,
and it will have $n$ orthogonal eigenvectors $v_i$ with eigenvalues $\epsilon_i$.

We find it convenient to transform from the $y_i$ parameter basis to the $v_i$ eigenvector basis
as the Hessian is diagonal in this basis. 
Additionally, we scale the eigenvectors by the square-root of their eigenvalue to 
introduce a scaled eigenvector 
$z_i {=}  \sqrt{\epsilon_i} \, v_i$. 
Thus, we can express the $\chi^2$ shift as: 
\begin{eqnarray}
\Delta \chi^2 &=& 
\sum_{i,j} H_{i,j} \, y_i y_j 
=
\sum_i \epsilon_i v_i^2
= 
\sum_i z_i^2 \quad . 
\end{eqnarray}
Finally, we can relate the original fitting parameters $y_i$ to the scaled eigenvectors $z_i$
via the equation: 
\begin{eqnarray}
y_i &=&
\sum_j 
V_{ij}^{-1}
\frac{z_j}{\sqrt{\epsilon_j}} \ ,
\end{eqnarray}
where $V_{ij}$ is the $n\times n$ matrix composed of the eigenvectors $v_i$
which diagonalizes the Hessian as in $V_{ii'}H_{i'j'}V_{j'j}^{-1}=\delta_{ij} \epsilon_i$.

An important consideration when constructing the FFs uncertainty range is to decide upon 
the acceptable tolerance range $T=\Delta \chi^2$ over which we will allow the fit parameters to vary. 
For a simple case of a single parameter, 
the confidence  level is 68\% for $\Delta \chi^2 {=}1$.
The present case of the pion FFs involves combining large data sets across many different measurements
with uncertainties from both experimental and theoretical sources that can only be approximated. 
Thus, determination of a   $\Delta \chi^2$ tolerance criteria is not straightforward.
For guidance on the choice of tolerance, we reviewed the analyses  used  in Refs.~\cite{Hirai:2007cx,James:1975dr,Giele:2001mr,Kovarik:2010uv,Zyla:2020zbs}.
For the present analysis we will use $\Delta \chi^2 {=} 20$ 
which was chosen based on a number of factors.
We shall present the details in the following sections, 
but  the $\Delta \chi^2 {=}20$ tolerance criteria gives us a reasonable overlap 
of the bands for our comparison of NLO and NNLO fits,
our fits comparing BELLE13 and BELLE20, 
and also our fits using different scale choices. 
With a tolerance of  $\Delta \chi^2 {=} 1$, the bands for the above comparisons 
would not overlap, and would clearly underestimate the true uncertainties.

\section{Experimental observables}\label{sec:data_selection}

The data sets used in  the present pion FFs analyses are based on 
electron-positron SIA cross-sections for pion 
($\pi ^\pm$) production. 
We incorporate   SIA experimental data from an  range of experiments
including 
SLAC 
(BaBar~\cite{Lees:2013rqd},  TPC~\cite{Aihara:1988su}, SLD~\cite{Abe:2003iy}),
CERN 
(ALEPH~\cite{Buskulic:1994ft},  OPAL~\cite{Akers:1994ez} DELPHI~\cite{Abreu:1998vq}), 
KEK 
(BELLE~\cite{Seidl:2020mqc,Leitgab:2013qh}), 
and 
DESY 
(TASSO~\cite{Brandelik:1980iy,Althoff:1982dh,Braunschweig:1988hv}).
These data sets contain inclusive and 
flavor-tagged measurements   of pion scaled energy ($z$) corresponding 
to the sum of light quarks $(u, d,s)$
as well as 
individual charm $(c)$ and bottom $(b)$ quarks.

We list all the data sets used in our fits in Table.~\ref{tab-SIA-chi2},
and indicate the experiment and the observable. 
The bulk of our data points are measured by the LEP and SLD Collaborations ($\sqrt{s} {=} M_Z$) 
and the B-factory experiments ($\sqrt{s}{\simeq} 10$).
The remaining   measurements range across  intermediate scales of energy. 

In a manner similar to Ref.~\cite{Bertone:2017tyb}, we will impose $z$~cuts on the data sets 
to avoid non-perturbative effects not described by our calculations;
such effects could arise from contributions beyond the order of our NNLO calculation,
or from non-factorizable  (e.g., higher twist) contributions.
Specifically, for the experiments with $\sqrt{s}{=}M_Z$ we take data in 
the range \mbox{$z\in[0.02,0.9]$,} and for all other sets we take  \mbox{$z\in[0.075,0.9]$.}

For all our fits we used    $\alpha_s (M_Z^2) {=} 0.1185$ which is compatible with the  world average data~\cite{Zyla:2020zbs}.
We also use $M_Z{=}91.1876$~GeV, $\sin^2\theta_W{=}0.231$,  $Q_0{=}5$~GeV, and 
pole masses for the heavy quarks of $m_c{=}1.43$~GeV and $m_b{=}4.5$~GeV.

We now describe the specific features of the B-factory measurements  from BaBar and BELLE
which are taken at the lower  scale $\sqrt{s} \simeq 10$ GeV.
As this energy is below the  B-meson pair production threshold,  
the effects of bottom quark production are not included for this data. 
These data sets are important for the fit as they have   
high precision and extend the momentum fraction $z$ close to one.

\subsection{Updated BELLE Data}

The BELLE Collaboration has recently released a new measurement  (BELLE20)~\cite{Seidl:2020mqc}
which we will compare to the previous data set  (BELLE13)~\cite{Leitgab:2013qh}.

The new data set  (BELLE20) has removed the backgrounds from $\Upsilon$ decays, $\tau$ production and 
two-photon processes, and also  applied an updated initial-state radiative (ISR)  correction.
Additionally,  the updated BELLE20 data contains all systematic uncertainties 
separated into correlated and uncorrelated contributions.
For the present analysis, these uncertainties are added in quadrature;
even with  this generous error budget, we will find this data challenging to describe with 
our theoretical calculations.
Furthermore, the number of equidistant bins in the range $0.075 {<} z {<} 0.9$ has decreased from $70$ in BELLE13 measurements to $32$ in BELLE20.

In the case of the BELLE13 experiment, the measurements have been provided in the form $d\sigma ^{H^\pm}/dz$.
To properly treat this data set consistently with the other measurements included in our analysis, 
we multiply the BELLE13 measurements~\cite{Leitgab:2013qh} by a   correction factor
as explained in Ref.~\cite{Bertone:2017tyb} to compensate for the  kinematic cut on radiative photon
events which was applied to the BELLE13 data. 
In contrast, no such correction is needed for the  BELLE20 data.

\tableParm
\tabChi

\section{Results} \label{sec:results}
We now  present the results of our fit for the charged pion FFs.
Additionally, we will also compare our resulting FFs with  recent results in the literature.

\subsection{Pion Fragmentation Function Fits}
\subsubsection{\ipmxx{} Fits}

As we studied the influence of the various data sets, 
we found it especially useful to 
compare the effects of  B-factory data sets from BELLE and BaBar. 
Thus, we perform five  sets of fits which we summarize 
in the following. 
Note, we include all the data in Table~\ref{tab-SIA-chi2} {\it except} what is indicated below. 
\begin{itemize}
{\setlength\itemindent{15pt}

\item[{\bf Fit~A:}] 
This fit focuses on the impact of the BELLE13 data set.
Thus, we {\it exclude} the BELLE20 data for a total of 386 points. 

\item[{\bf Fit~B:}] 
This fit focuses on the impact of the BELLE20 data set.
Thus, we {\it exclude} the BELLE13 data for a total of 348 points. 

\item[{\bf Fit~C:}] 
This fit focuses on the impact of the BELLE20 data \textit{without} the  BaBar set.
Thus,  we {\it exclude} the BaBar and BELLE13 data for a total of 308 points. 

\item[{\bf Fit~D:}] 
This fit focuses on the impact of cutting the \mbox{low-$z$} BELLE20 data. 
Thus,  we {\it exclude} the BaBar and BELLE13 data,
and impose a $z{>}0.2$ cut on the BELLE20 data,  
for a total of 304 points.

\item[{\bf Fit~E:}] 
This fit focuses on the impact of the BELLE20 and  BaBar sets
with cuts imposed to remove low-$z$ data.
Thus,  we {\it exclude} the BELLE13 data,
 impose a $z{>}0.2$ cut on the BELLE20 data
and 
 impose a $z{>}0.1$ cut on the BaBar data
for a total of 341 points. 

}\end{itemize}
Thus, in the above, 
Fit~A will reflect the influence of the updated BELLE13 and Fit~B will reflect BELLE20. 
We can discern the impact of the BaBar data by comparing Fits~C and~D with Fit~E.
And finally, we can see the impact of the low~$z$ cuts by comparing Fits~D and~E.

\subsubsection{Fit Parameters}

In Table~\ref{tab-parms} we present the  values of the 19 fitted parameters, Eq.~\eqref{eq:parametrization}, 
and their uncertainties for our fits as computed by Minuit~\cite{James:1975dr}.

Examining the relatively small uncertainties on the shape parameters 
(with the possible exception of $\alpha_g$ and $\alpha_{s^+}$), 
this suggests that these parameters appear to be  relatively  
well-constrained by the SIA data.

Similarly, Table~\ref{tab-SIA-chi2} displays the the $\chi^2$ values for each individual experiment
and the number of degrees of freedom, as well as the totals. The normalizations are treated as
a correlated systematic and are also reported~\cite{Alekhin:2014irh}.

\subsubsection{NLO and NNLO Comparison}

Since \xfitter{} can perform  both NLO and NNLO calculations, 
we will compare these  for the case of  Fit~A. 
The quality of our fits can be surmised by examining the total $\chi^2$ and 
$\chi^2$ per degree of freedom ($\chi^2/dof$) 
as displayed in  Table~\ref{tab-SIA-chi2}. 
For Fit~A we observe that the NNLO fit  yields a substantially improved $\chi^2/dof$
of 427/386 for NNLO \textit{vs.}\ 480/386 for NLO representing an improvement of   ${{\sim}} 11\%$.

If we look in more detail,
Table.~\ref{tab-SIA-chi2} also displays the $\chi^2$ values for each individual experiment. 
While most of the data sets show improvement of the NNLO compared to the NLO results, 
there are  some observables  where the individual $\chi^2$ does  slightly increase 
such as  the ones measured by the TASSO, TPC and DELPHI experiments; these increases are small compared 
to the overall improvement of the fit from $\chi^2/dof {\sim} 1.24$ to $1.11$. 
A similar behavior is observed in other analyses such as Ref.~\cite{Bertone:2017tyb}.

Finally, if we examine the uncertainty bands for the FFs of Fig.~\ref{fig:FitA-1},
the \textit{a posteriori} fact  the NLO and NNLO bands overlap
can be taken as an indication that perturbative uncertainties are under control 
 and this additionally suggests that our choice of tolerance $T$ is reasonable.

While we do not explicitly show details of the other NLO fits, the results 
are similar. 
Hence, the inclusion of the higher-order QCD corrections noticeably improves
the quality of our fits, and this result  is in agreement with other analyses in the literature~\cite{Bertone:2017tyb,Soleymaninia:2019sjo,Soleymaninia:2018uiv}. 

Since the NNLO yields improved results over the NLO calculation, 
we will focus only on the NNLO fits for the following analyses.

\subsection{Comparison of Data Sets}
\label{sec:belle}

\subsubsection{BELLE Data}

To  examine how well our fits describe the experimental data,
we start by comparing 
Fit~A which includes the  BELLE13 data
and Fit~B which includes the recent BELLE20 data.
This information is contained both in  Table.~\ref{tab-SIA-chi2} and also  graphically in 
Fig.\ref{fig:chi2-all}.

Examining  Table.~\ref{tab-SIA-chi2}, we see 
at NNLO order 
Fit~A yields a 
$\chi^2/dof$ of  $427/386{\sim} 1.11$
as compared to the NNLO 
Fit~B with $518/348{\sim} 1.49$. 

If we look at the $\chi^2$ values for the individual experiments, we see 
that most of the values are comparable between the Fit~A and Fit~B results, 
with the exception of the BaBar data which contributes an extra ${\sim} 24$ units of $\chi^2$ to Fit~B. 
The remainder of the difference is of course due to the BELLE data itself where 
the BELLE13 data (Fit~A)   gives $14/70{\sim} 0.2$ \textit{vs.}
the BELLE20 data (Fit~B) with $82/32{\sim} 2.6$.

In Fig.~\ref{fig:Belle20-1} we plot the FFs for both Fits~A and B.
Note that in our framework, the $u^+$ and $d^+$ distributions will be equivalent;
hence, we  display $u^+$ 
as well as the combination $d^+ {+} s^+$.
We also display the uncertainty bands for the FFs computed with our tolerance of $T=20$.
We see that the $u^+{=}d^+$, charm and bottom distributions 
show only moderate difference between  the two fits, 
and Fit~B yields a slightly softer distribution for  $u^+$ and $d^+$.
In contrast, the strange and gluon show larger difference 
which reflects, in part, their  larger uncertainties. 
Compared to Fit~A, Fit~B moderately increases $s^+, c^+, b^+$ while decreasing $g$.

\subsubsection{BELLE and BaBar Comparison}

Comparing Fits~A and B above, we observed that 
the inclusion of the  updated BELLE20 data caused a substantial increase
in the BaBar $\chi^2$. 
To explore if this is primarily a tension between the BELLE20 and BaBar data sets, 
we introduce Fit~C which fits BELLE20 but excludes the BaBar data.

Examining the results of Fit~C in  Table.~\ref{tab-SIA-chi2} we see 
the quality of the BELLE20 data $\chi^2$ improves from 82/32 (Fit~B) to 32/32 (Fit~C), 
and the overall $\chi^2$ improves from $518/348{\sim} 1.49$ to 
$404/308{\sim} 1.31$.

Additionally, 
in Fig.~\ref{fig:Compare-with-SIA-abc} we display the comparisons of our 
A,~B, and~C fits with the data
for BELLE13, BELLE20 and BaBar. 
For BELLE13 (Fit~A), the theoretical predictions are entirely consistent with the 
experimental data, but this is partly due to the larger uncertainties; 
note the vertical scale of the ``theory/data'' plot. 

Comparing BELLE20 and BaBar,
 we see that the fits yield a good description of the data   with the exception of the  low-$z$ region.
Here,  the BELLE20 and BaBar data sets appear to pull the fit in opposite directions; 
BELLE20 is above the theory, and BaBar is below the theory.

If we compare Fit~B (with BaBar) to Fit~C (without BaBar) 
we notice Fit~C shifts toward the data throughout the $z$ range, 
including the lower $z$ region,
thus reducing the $\chi^2$ for BELLE20 from 82/32 (Fit~B) to 32/32 (Fit~C).

\subsubsection{Low-\texorpdfstring{$z$}{z} Cuts}

The deviations between theory and data  in the  low-$z$ region, as observed in Fig.~\ref{fig:Compare-with-SIA-abc},  warrants a closer examination. 
Thus, we  introduce 
Fit~D which imposes a $z{>}0.2$ cut on the BELLE20 data
again does not include the  BaBar data.
As summarized in Table~\ref{tab-SIA-chi2},
the BELLE20 data set is reduced from 32 to 28 points with a significantly improved $\chi^2$ of 9.2/28, and the total  $\chi^2$ for Fit~D improves to $357/304=1.17$.

The encouraging results of imposing the low-$z$ cut leads us to 
try  Fit~E which includes both the BELLE20 and BaBar data,
but with low-$z$ cuts imposed. 
Specifically, for BELLE we impose $z{>}0.2$ which removes 4 points, 
and  for BaBar we impose $z{>}0.1$ which removes 3 points.
(Note, we could impose a more stringent cut of $z{>}0.2$ on both data sets,
but we shall find this more conservative choice already yields a notable improvement.)

We note that other analyses in the literature have also imposed 
kinematic cuts on the low~$z$ data. 
For example,  the  JAM19 focus was on SIDIS in the region  of larger~$z$;
hence, they imposed a cut of  $z\gtrsim 0.2$ on their data sets. 
Similarly,   NNFF1 used a lower kinematic cut of $z_{min} {=} 0.02$ for $Q{=}M_Z$ and $0.075$ for $Q<M_Z$.

The result of these low-$z$ cuts is dramatic. 
We obtain an improved 17/28 for BELLE20, 
33/37 for BaBar, and a good $\chi^2=410/341{\sim} 1.20$ for the total data set of Fit~E.

To see the impact of these fits in more detail, 
Fig.~\ref{fig:Compare-with-SIA-de} displays the comparisons of our fits with the data
for BELLE20 and BaBar. 
Examining the  BaBar results, 
we see that by cutting out the lowest 3 points in $z$,
the theoretical predictions are generally contained within the band of the experimental 
uncertainty. 

Similarly, for BELLE20 we see that the predictions also are generally contained within the band of the experimental uncertainty. 
We also see that including the  BaBar data  (Fit~E) increases the theory curve 
in the low-$z$ region, consistent with the effects observed in Fig.~\ref{fig:Compare-with-SIA-abc}

 To see the impact of the low-$z$ cuts on the FFs, 
 in Fig.~\ref{fig:FFs-BE} we display the results of 
 Fit~B (no $z$ cuts)
 and
Fit~E (with $z$ cuts). 
The difference is dramatic and (with the exception of the heavy charm and bottom quarks)
the uncertainty bands do not overlap throughout much of the $z$ range. 

While our tolerance criteria of $T=20$ gave  reasonable results for the comparisons
of the NLO \textit{vs.}\  NNLO results of Fig.~\ref{fig:FitA-1} and the comparison of BELLE13 and 
BELLE20 of Fig.~\ref{fig:Belle20-1}, 
Fig.~\ref{fig:FFs-BE} underscores the fact that this does not represent 
the uncertainty due to the choice of data in the low~$z$ region; 
hence, this must be separately accounted for when computing the maximal FF uncertainty. 

Regarding the impact of the $z$~cuts on the shape of   FFs,   we see that this generally increases the FFs for up, down, charm and bottom 
while decreasing the strange and gluon FFs (which have comparably larger uncertainties).

\subsubsection{Scale Variation}

Since the results presented thus far have imposed the scale choice $\mu_F=\mu_R =Q$, 
we also want to  study the  effect of varying the perturbative scales, in part, to see if this could 
have any significant effect on the partonic FF $z D^{\pi^\pm}(z,Q) $ in the low-$z$ region. 
In Fig.~\ref{fig:FFs-scale}, we display the resulting FFs for Fit~E with their uncertainties for the variation of the scale choice as a function of $z$. 
We start with a factorization scale of $\mu=Q= 10$~GeV,
and vary this up and down by a factor of~2 
to display FFs for $\mu=Q=\{5,10,20\}$~GeV.
Essentially, this reflects the influence of the DGLAP evolution in $Q$.

The variation of the uncertainty bands for the different scales is rather modest for 
all the flavors with the exception the strange FFs which shows a observable shift
in the intermediate $z$ region, and the gluon FFs which shows a shift in the low $z$ region. 
This variation can provide an indication of higher order contributions which may not 
be included in our fixed-order perturbative calculation. 
For increasing scales, the FFs generally decrease in the small $z$ region ($z{\sim}0.01$), 
and this is most evident in the gluon FF.

\subsubsection{Comparison with SIA Data}

To provide  further details  on the quality of the QCD fits, 
we now compare our theoretical predictions directly with the inclusive pion production data. 

In Fig.~\ref{fig:Compare-with-SIA-1-a} 
we display the theoretical predictions for  Fit~A, Fit~B and Fit~E
at NNLO together with the inclusive pion production data and their uncertainties.
The comparisons are given as a function of $z$ for the various energies, 
and the measured $z$ region for each experiment can be read from the plots. 
To provide detail, we show both the absolute distributions (upper panel) 
and the data over theory ratios (lower panel).

The results of  TASSO,  DELPHI, and SLD are split across multiple data sets 
({\it c.f.}, Table~\ref{tab-SIA-chi2}),
but the experiments provide the correlated systematic uncertainty 
for the  overall normalization of the data.
For these data sets, the difference between the ``theory" 
and ``theory+shifts" is due to this systematic normalization. 
For the other data sets of Fig.~\ref{fig:Compare-with-SIA-1-a}, 
there is no systematic normalizaton, so the ``theory" 
and ``theory+shifts"  curves will coincide.

The theoretical predictions are generally in good agreement with the experimental measurements 
in the central   $z$-range.
We observe some deviations in the large-$z$ region, but 
as the data uncertainties are typically larger here this does not 
significantly impact the $\chi^2$ of the fits. 
In contrast, in the low-$z$ region where the experimental uncertainties are typically smaller
we do see some deviations of the theoretical predictions which, as discussed above, 
can substantially impact both the FFs (\textit{e.g.,} Fig.~\ref{fig:FFs-BE}) and the 
$\chi^2$ values  (\textit{e.g.,} Tab.~\ref{tab-SIA-chi2}).
For the BaBar and BELLE20 plots,  
we note  Fits~A and~B  show  deviations in the low~$z$ region,
and this is in contrast to  Fit~E where these deviations are largely absent.

\subsubsection{Theoretical Effects at Low-\texorpdfstring{$z$}{z}}

What could be the source of the large $\chi^2$ contributions coming from the low-$z$ region?
One concern is that we could be missing important theoretical corrections 
such as non-factorizable higher-twist contributions, or terms beyond the order or our perturbation 
theory, some of which can be estimated using a resummation procedure. 
The original set of $z$ cuts discussed in Sec.~\ref{sec:data_selection} 
and outlined in Ref.~\cite{Bertone:2017tyb}
aim to minimize the non-perturbative contribution not described by our calculations.

Additionally,
Ref.~\cite{Anderle:2016czy} performs an  all-order resummation of logarithmically enhanced contributions at small momentum fraction $z$ 
at NNLO  logarithmic order, including the dependence on the factorization and renormalization scales.
Specifically, they examine terms  of the form $(\alpha_s^n \ln^{k} z)$ and resum these 
contributions in Mellins space to reduce the theoretical uncertainty in the low $z$ region.

This analysis suggests that although the lowest few $z$ data points of BELLE and BaBar may 
have modest corrections from resummation contributions, these alone are not sufficient 
to address the differences we observe with the current fits.  
Additionally, the fact that the BELLE and BaBar data seem to pull the theory in 
opposite directions further complicates the resolution of this situation
and may point to non-perturbative effects  not incorporated in our calculations. 

In conclusion, while we generally obtain a good quality for our fits across most of the $z$ range, 
the description of the data in the low-$z$ region remains an unresolved puzzle, 
and we must leave this for future investigation.

\subsection{Comparison to results from the literature}

We now  compare our results with recent analyses from the literature. 
While Fit~E (\ipmxx)  is our preferred final fit, we also display Fit~B to highlight the impact of the low~$z$ cuts.
We compare with the results from the {\tt NNFF1}~\cite{Bertone:2017tyb,Nocera:2017qgb,Bertone:2017xsf},
{\tt JAM19}~\cite{Sato:2019yez} and {\tt DSEHS14}~\cite{deFlorian:2014xna} collaborations
which are computed at NLO 
({\tt JAM19} and {\tt DSEHS14}) and at NNLO ({\tt NNFF1}).
These results  are displayed in Figs.~\ref{fig:FFs-other-1} and \ref{fig:FFs-other-2} 
at $Q^2 = 100$~GeV$^2$ and $Q^2 = M_Z^2$, respectively.

Discretion is necessary when interpreting the very low~$z$ region as  
the extrapolation of the FF grids may extend beyond the region fitted in the 
individual analyses of Refs.~\cite{deFlorian:2014xna,Bertone:2017tyb,Sato:2019yez}. 
For example, the {\tt JAM19} focus was on SIDIS in the region $z\gtrsim 0.2$,
{\tt NNFF1} used a lower kinematic cut of $z_{min} {=} 0.02$ for $Q{=}M_Z$ and $0.075$ for $Q<M_Z$,
and {\tt DSEHS14} also used various $z$ cuts. 
Additionally, differences can arise due to the choice of the  tolerance criteria
as well as the method for computing the error bands such as the Hessian or Monte Carlo approach. 

The {\tt DSEHS14} uses a combination of data from SIA, SIDIS, and hadron-hadron collisions
in an NLO framework; additionally, 
our parametric form of  Eq.~\ref{eq:parametrization} matches their initial FFs. 
The {\tt NNFF1} analysis  is based on electron-positron SIA cross-sections
for the sum of charged pion, charged kaon, and proton/antiproton production.
The {\tt JAM} analysis simultaneously fits both the PDFs and FFs using  DIS, SIDIS, Drell-Yan and SIA data.

Comparing our up and down FFs with {\tt NNFF1} and {\tt DSEHS14},
we see they  are generally compatible at larger $z$, but 
differ in the low-$z$ region; this is more pronounced for Fit~E which imposes cuts on some 
of the low-$z$ data. Similar conclusions apply to the case of  the $c^+$ and $b^+$ heavy quarks.
Comparing our gluon FF with {\tt NNFF1} and {\tt DSEHS14}, again 
we see our FFs are generally  compatible as our curves lie within the {\tt NNFF1}  uncertainty band, 
and have overlap with {\tt DSEHS14} in the larger $z$ region. 
For the strange quark there is more of a spread in results suggesting 
an overall increased uncertainty (note the vertical scale). 
This reflects, in part, the fact that our chosen data set has minimal constraints on the strange FF.

In contrast, the above FFs generally have  a different behavior as compared with the {\tt JAM19} 
analysis.\footnote{While the JAM parameterizes their initial distributions as  \mbox{$z^a (1-z)^b (1+c \sqrt{z} + d z)$},
it is possible to match this to our functional form quite closely at $Q_0$; hence,this is not the source of the differences.} 
The {\tt JAM19} FFs have a much steeper slope at small $z$ for the  quark
flavors as compared to both our Fit~B and Fit~E, while the gluon generally lies above our curves 
for intermediate to larger $z$ values.
In addition, much smaller error bands are obtained in {\tt JAM19} analysis in respect to our results and the {\tt NNFF1} analysis. 
As noted above, the  focus of the {\tt JAM19} study was on SIDIS in the region $z\gtrsim 0.2$,
and in this region there is closer agreement among the FFs. 
The {\tt JAM19} study also performed a separately analysis without the SIDIS data (not shown here)
and this resulted in a substantive shift of FFs. 
Specifically, this tends to increase the $u^+$ and decrease the $g$ FFs in the 
larger~$z$ region, $z{\gtrsim}0.2$.
As the {\tt JAM19} results are obtain by a combined fit to both PDFs and FFs, 
it would be interesting to investigate further to see what difference might be 
separately attributable  to this combined analysis and the choice of data sets.

\section{Summary and Conclusions} \label{sec:conclusion}

In this paper we have presented the first open source analysis 
of  unpolarized pion FFs at NNLO accuracy in pQCD. 
Our analysis is based on a comprehensive data set from SIA experiments, 
including   the most recent measurement of single pion production by the BELLE collaboration~\cite{Seidl:2020mqc}.

This analysis includes several novel features including the use of the open source \xfitter{}  framework to perform the analysis, 
as well as  to compute the Hessian 
uncertainties for the FFs. 
We have compared both  NLO and NNLO results, and also studied the impact of the 
B-factory data.

The SIA experimental data are reasonably well described by our QCD fits, 
with the exception of the low-$z$ region.
This remains as an important unresolved issue
and will clearly require additional study.

The comparisons of our pion FFs to the results in the literature are also informative.
While some of the results agree within uncertainties, 
there are a number of areas 
where the FFs differ, and thus warrants further examination. 

This paper facilitates  future investigations as these results are available within 
 the  open-source \xfitter{} program, and also 
as   pion FF grids formatted for the {\tt LHAPDF6}  library. 
Therefore,  this analysis serves as an important step toward a broader program 
to improve the determination of the  nuclear PDFs and FFs of hadrons 
as we strive to better comprehend the full character of the QCD theory.

Note added:\ 
As we completed this study, 
a related investigation was reported in Ref.~\cite{mapff1} 
using a complementary neural-network approach with different data sets;
this opens new avenues for future study.

\newpage 
\begin{acknowledgments}

The authors thank
V.~Bertone,
\mbox{A.~Cooper-Sarkar},
\mbox{P.~Duwent\"aster},
N.~Sato,
I.~Schienbein,
and
R.~Seidl
for valuable discussion and suggestions. 
The authors would like to thank the DESY IT department for the provided computing resources and for
their support of the \xfitter{} project.
H.A., M.S.\ and H.K.\  thanks the  School of Particles and Accelerators, Institute for Research in Fundamental Sciences (IPM) for financial support of this project.
H.K.\ also thanks the University of Science and Technology of Mazandaran for financial support provided for this research.
F.O.\   acknowledges  support through US DOE grant DE-SC0010129. 
This research was supported in part by the National Science Foundation under Grant No. NSF PHY-1748958.

\end{acknowledgments}

\vspace*{2cm}
\bibliographystyle{utphys}
\bibliography{refs,extra}

\figChiPlot
\figNNLO
\figAB
\figABC
\figDE
\figBElowQ
\figScale
\dataFigi
\dataFigii
\figCompLitLowQ
\figCompLitHighQ  %

\end{document}